\documentclass[a4paper,12pt,oneside]{report}
\usepackage[utf8]{inputenc}
\usepackage{graphicx}
\usepackage{tensor}
\usepackage{bm}
\usepackage{color}
\usepackage{amsmath}
\usepackage{amssymb}
\usepackage{hyperref}

\newcommand{\R}[1]{\tensor{\mathring{R}}{#1}} 
\newcommand{\G}[1]{\tensor{\mathring{G}}{#1}} 
\newcommand{\rcd}[1]{\tensor{\overset{\circ}{\nabla}}{#1}}
\newcommand{\leviconnection}[1]{\tensor{\overset{\circ}{\bm{\nabla}}}{#1}}
\newcommand{\rsconnection}[1]{\tensor{\mathring{\omega}}{#1}}

\newcommand{\connection}[1]{\tensor{\Gamma}{#1}}
\newcommand{\chr}[1]{\tensor{\mathring{\Gamma}}{#1}}
\newcommand{\contorsion}[1]{\tensor{K}{#1}}
\newcommand{\potential}[1]{\tensor{\Sigma}{#1}}
\newcommand{\pd}[1]{\tensor{\partial}{#1}}
\newcommand{\nablab}{\bm{\nabla}}
\newcommand{\T}{\bm{T}}

\newcommand{\e}[2][]{\tensor{#1{e}}{#2}}
\newcommand{\ebar}[1]{\tensor{\bar{e}}{#1}}
\newcommand{\teta}[2][]{\tensor{#1\vartheta}{#2}}
\newcommand{\tetabar}[1]{\tensor{\bar{\vartheta}}{#1}}
\newcommand{\torsione}[1]{\tensor{T}{_{\bf e}#1}}
\newcommand{\torsionbare}[1]{\tensor{\bar{T}}{_{\bf e}#1}}
\newcommand{\torsionbarbar}[1]{\tensor{\bar{T}}{_{\bf \bar{e}}#1}}
\newcommand{\lorentz}[2][]{\tensor{#1\Lambda}{#2}}
\newcommand{\nonholon}[1]{\tensor{\Omega}{#1}}
\newcommand{\nonholonbar}[1]{\tensor{\bar{\Omega}}{#1}}
\newcommand{\bracket}[1]{\left<#1\right>}
\newcommand{\torsion}[1]{\tensor{T}{#1}}
\newcommand{\torsionbar}[1]{\tensor{\bar{T}}{#1}}
\newcommand{\sconnection}[2][]{\tensor{#1\omega}{#2}}
\newcommand{\scbar}[1]{\tensor{\bar{\omega}}{#1}}
\newcommand{\scbare}[1]{\tensor{\bar{\omega}}{_{\bf e}#1}}
\newcommand{\scbarbar}[1]{\tensor{\bar{\omega}}{_{\bf \bar{e}}#1}}
\newcommand{\scebar}[1]{\tensor{\omega}{_{\bf \bar{e}}#1}}
\newcommand{\psenergy}[1]{\tensor{\epsilon}{#1}}
\newcommand{\potentialbarbar}[1]{\tensor{\bar{\Sigma}}{_{\bf \bar{e}}#1}}
\newcommand{\potentialbare}[1]{\tensor{\bar{\Sigma}}{_{\bf e}#1}}
\newcommand{\psenergybarbar}[1]{\tensor{\bar{\epsilon}}{_{\bf \bar{e}}#1}}
\newcommand{\psenergybare}[1]{\tensor{\bar{\epsilon}}{_{\bf e}#1}}
\newcommand{\psenergyebar}[1]{\tensor{\epsilon}{_{\bf \bar{e}}#1}}

\newcommand{\energy}[2][]{\tensor{#1{t}}{#2}}
\newcommand{\energybarbar}[1]{\tensor{\bar{t}}{_{\bf \bar{e}}#1}}
\newcommand{\energybare}[1]{\tensor{\bar{t}}{_{\bf e}#1}}
\newcommand{\torsionhehl}[2][]{\tensor[#1]{{\cal T}}{#2}}
\newcommand{\baseeta}[1]{\tensor{\bm{\eta}}{#1}}

\newcommand{\potentiale}[1]{\tensor{\Sigma}{_{\bf e}#1}}
\newcommand{\potentialebar}[1]{\tensor{\Sigma}{_{\bf \bar{e}}#1}}
\newcommand{\connectiontilde}[1]{\tensor{\widetilde{\Gamma}}{#1}}
\newcommand{\connectioninfinity}[1]{\tensor{\widetilde{\Gamma}}{#1}(\infty)}
\newcommand{\energye}[1]{\tensor{t}{_{\bf e}#1}}
\newcommand{\energyebar}[1]{\tensor{t}{_{\bf \bar{e}}#1}}
\newcommand{\pifactorebar}[1]{\tensor{\Pi}{_{\bf \bar{e}}#1}}
\newcommand{\pifactorbar}[1]{\tensor{\bar{\Pi}}{_{\bf e}#1}}
\newcommand{\torsionebar}[1]{\tensor{T}{_{\bf \bar{e}}#1}}
\newcommand{\E}{$\mathbb{E}$\ }
\newcommand{\Ebar}{${\bar{\mathbb{E}}}$\ }
\newcommand{\bdeltadn}[1]{\delta^{\textrm{\tiny (#1)}}}                                                  
\newcommand{\bdeltaup}[1]{\delta_{\textrm{\tiny (#1)}}}

\newcommand{\bzero}{\textrm{\tiny (0)}}
\newcommand{\bone}{\textrm{\tiny (1)}}
\newcommand{\btwo}{\textrm{\tiny (2)}}
\newcommand{\bthree}{\textrm{\tiny (3)}}

\newcommand{\bbt}{\hat{t}}
\newcommand{\bbrho}{\hat{\rho}}
\newcommand{\bbtheta}{\hat{\theta}}
\newcommand{\bbthree}{\hat{\phi}}
\newcommand{\bbz}{\hat{z}}
\newcommand{\bbs}{\hat{s}}
\newcommand{\bbphi}{\hat{\phi}}

\newtheorem{definition}{Definition}[section]
\newtheorem{theorem}{Theorem}[section]

\newtheorem{postulate}{Postulate}[section]

\title{The meaning of torsion in teleparallel theories}
\author{J. B. Formiga}
\date{}

\begin{document}

\huge In this version I have split the paper  ``The meaning of torsion in teleparallel theories'' in three parts\footnote{J. B. Formiga, jansen@fisica.ufpb.br, Departamento de F\' isica, Universidade Federal da Para\' iba, Caixa Postal 5008, 58051-970 Jo\~ ao Pessoa, Pb, Brazil}: 

\large
\begin{enumerate}
\item Revisiting the gravitational energy of the Schwarzschild spacetime with a new approach to the calculations: {\bf page \textcolor{blue}{\pageref{paper0}}} . \textcolor{red}{(This preprint has not undergone peer review or any post-submission improvements or corrections. The Version of Record of this 
article is published in Brazilian Journal of Physics, and is available online at https://doi.org/10.1007/s13538-021-00975-8.)}
\item On the teleparallel frame problem: {\bf page \textcolor{blue}{\pageref{paper1}}}
\item The meaning of torsion in teleparallel theories: {\bf page \textcolor{blue}{\pageref{paper2}}}
\end{enumerate}

\chapter{Revisiting the gravitational energy of the Schwarzschild spacetime with a new approach to the calculations}\label{paper0}

Recently, a new method to simplify calculations in teleparallel theories has been put forward. In this article, 
this method is improved and used to analyze the gravitational energy-momentum tensor (density) of the Schwarzschild solution in a frame that is arbitrarily accelerated along the $z$-direction. It is shown that, for a special type of frame, one cannot make the gravitational energy density vanish along the observer's worldline, regardless of the observer's acceleration. The role played by the frame and its relation to the observers' worldlines are investigated. It is shown that, for the aforementioned special frames, the results for the gravitational energy and angular momenta are consistent with what we would expect.

\section{Introduction}
One of the open problems in physics is the determination of a gravitational energy-momentum tensor density (EMTD). Although a lot of progress has been made since Einstein started this quest \cite{Einstein:1915by,10.2307/20488488,PhysRev.89.400,MOLLER1961118,Landaufourthv2,Weinberg:1972kfs,aldrovandi2012teleparallel,ANDP:ANDP201200272}, there is no consensus about the localization of the gravitational energy. Some authors argue that the equivalence principle prevents us from having a well-defined  energy density for gravity; however, at least in context of teleparallel theories, this problem can be circumvented \cite{ANDP:ANDP201200272}. In at least one of these theories, known as the Teleparallel Equivalent of General Relativity (TEGR), one can define an EMTD,  a gravitational energy-momentum tensor (EMT) inside a three-dimensional volume, and an angular momentum \cite{ANDP:ANDP201200272}. Problems arise, though, when one tries to calculate all these quantities. The first problem is the ambiguity of the frame where all the calculations should be done. The second problem is the long, tedious and messy calculations, mainly when evaluating the EMTD.

Teleparallel theories belong to a set of alternative theories of gravity that are  based on a geometry endowed with a special connection called  Weitzenb\"{o}ck connection, which has torsion but no curvature \cite{ANDP:ANDP201200272}. There are many different teleparallel theories \cite{PhysRevD.19.3524,Cai_2016,0264-9381-34-11-115012,Silva2016}, but most of them can be generalized as a $f(T)$ theory, where $T$ is the {\it torsion scalar} (the analogous to the Ricci scalar). The TEGR corresponds to the particular case where $f(T)\sim T$, and its field equations  are the Einstein field equations written in terms of the tetrad field $\e{^a_\mu}$. In its teleparallel form, the left-hand side of Einstein equations is the $4$-divergence of a superpotential $\potential{^a^\mu^\nu}$, while the right-hand side has the stress-energy-momentum content, including the gravitational one. In turn, this energy-momentum content obeys a conservation equation and we can assign a $4$-momentum to the gravitational field.

Because teleparallel approach is suitable to solve the gravitational energy problem, it has been used to deal with this problem in many different contexts. These applications include  thermodynamics \cite{PhysRevD.47.1407,PhysRevD.85.044050,PhysRevD.100.124040}, gravitational waves \cite{PhysRevD.78.047502,Obukhov_2009,doi:10.1002/andp.201800320}, and cosmology \cite{Xulu2000,Vargas2004,SOUSA2010,doi:10.1142/S021827181001813X,ABEDI201954}. However, a deep analysis of the gravitational energy density has not yet been made, owing to the usual lengthy calculations that are involved. In order to make these calculations neat and tidy, a hybrid approach that helps us to take advantages of the spacetime and the frame symmetries all at once was developed in Ref.~\cite{doi:10.1002/andp.201900507}. In section \ref{26122019sa} here, this approach is improved by adopting a uniform notation and providing some new identities. As an application, the gravitational energy-momentum and the gravitational energy density in a frame that accelerates along the $z$-direction is analyzed in detail (section \ref{26122019sd}). Section \ref{26052021a} is devoted to the conclusions.

Our notation is as follows: The metric components in a coordinate basis is denoted by $g_{\mu\nu}$, while in a tetrad basis is denoted by $\eta_{ab}$; the spacetime signature is $(+,-,-,-)$. Greek letters represent spacetime indices, and  Latin letters represent tangent space ones, except for Latin letters in the middle of the alphabet ($i,j,k,\ldots$), which stand for spatial coordinate indices. The components of the frame $e_a$ and the coframe $\teta{^a}$ in a coordinate basis are represented by $\tensor{e}{_a^\mu}$ and $\tensor{e}{^a_\mu}$, respectively. The tangent space indices and the coordinate indices are distinguished  by using the former between  parentheses: $\tensor{e}{_{(0)}^0}$, $\tensor{e}{_{(1)}^2}$ and so on. We also use the convention $A_{[ab]}\equiv(1/2)(A_{ab}-A_{ba})$. 

Before going to the results of this paper, we recall some basic features of the TEGR in the next section.

\section{Teleparallelism}\label{22032020aa}

The field equations of the TEGR are the Einstein equations written in the form
\begin{equation}
\pd{_\alpha}\left( e\potential{^a^\mu^\alpha}\right)=\frac{e}{4k}\left(\energy{^\mu^a}+T^{\mu a}\right), \label{29032019ka}
\end{equation}
where $k=1/(16\pi)$ in natural units,  $e=\det(\e{^a_\mu})=\pm\sqrt{-\det g}$ is the determinant of the tetrad field,  $\potential{^a^\mu^\alpha}$ is known as {\it superpotential},     $T^{\mu a}$ is the matter energy-momentum tensor (density), and $\energy{^\mu^a}$ is identified as the gravitational energy-momentum tensor density.

The superpotential is given by the expression
\begin{equation}
\potential{^\lambda^\mu^\nu}\equiv\frac{1}{4}\left( \torsion{^\lambda^\mu^\nu}+2\torsion{^{[\mu|}^\lambda^{|\nu]}}\right)+g^{\lambda[\nu}T^{\mu]}, \label{10112017la}
\end{equation} 
where $\torsion{^\lambda^\mu^\nu}$ is the torsion tensor of the Weitzenb\"{o}ck spacetime \cite{ANDP:ANDP201200272}, and  $\torsion{_a}\equiv\torsion{^b_b_a}$.  The stress-energy-momentum tensor for the gravitational field is given by
\begin{equation}
\energy{^\mu ^a}=k(4\potential{^b^c^\mu}\torsion{_b_c^a}-\e{^a^\mu}T), \label{29032019ha}
\end{equation}
where  $T\equiv\potential{^a^b^c}\torsion{_a_b_c}$ is the torsion scalar.

In the TEGR, there is always a frame where the Weitzenb\"{o}ck torsion can be written as
\begin{align}
\torsion{^a_\mu_\nu}=\pd{_\mu}\e{^a_\nu}-\pd{_\nu}\e{^a_\mu}. \label{04102019pa}
\end{align}

The superpotential is antisymmetric in the last two indices and, as a result, we have the conservation law
 $\pd{_\mu}\left[e\left(\energy{^\mu^a}+T^{\mu a}\right)\right]=0$.  In turn, the the energy-momentum contained within a three dimensional volume $V$ is \cite{ANDP:ANDP201200272}
\begin{equation}
P^a=\int_V d^3x e\energy{^0^a}+\int_V d^3x eT^{0a}, \label{10082019aa}
\end{equation}
where the first integral is the gravitational energy-momentum tensor, and  the total energy-momentum $P^a$ can be written in terms of the superpotential as
\begin{equation}
P^a=4k\oint_S dS_i e\Sigma^{a0i}.\label{22072018ga}
\end{equation}

The total angular momentum  can be defined as \cite{ANDP:ANDP201200272}
\begin{align}
L^{ab}\equiv -\int_V d^3xM^{ab},\label{30122019ca}
\end{align}
where 
\begin{align}
M^{ab}=-4ke\left(\potential{^a^0^b}-\potential{^b^0^a} \right) \label{30122019da}
\end{align}
is total angular momentum density, which is interpreted as the gravitational angular momentum density when the matter fields are absent. (The quantities $L^{(0)b}$ and $L^{(i)(j)}$ may be interpreted as the center of mass moment and the angular momentum, respectively.)

\section{Hybrid machinery}\label{26122019sa}
The evaluation of the frame-dependent quantities, such as the gravitational stress-energy momentum tensor, can be very long and tedious. In this section, the formalism that was developed in Ref.~\cite{doi:10.1002/andp.201900507} to simplify these calculations will be improved. We will see that this machinery is very useful in simplifying the calculations whenever we mix the directions of a Cartesian coordinate system with the directions of any other coordinate system. This mixing can be done with many different coordinate systems, however, here  we focus  on the spherical and  cylindrical coordinate systems.

\subsection{Mixing Cartesian unit vectors with spherical ones}\label{26122019sb}
Let  $(\rho, \theta, \phi)$ and  $(x,y,z)$ be the spherical and the Cartesian coordinate systems. The relation between them is given by $x=\rho\sin\theta\cos\phi$, $y=\rho\sin\theta\sin\phi$, and $z=\rho\cos\theta$. From these expressions, we can define the unit vectors:
\begin{align}
\hat{t}^a\equiv \bdeltaup{0}^a,\ \hat{\rho}^a\equiv\sin\theta(\cos\phi\hat{x}^a+\sin\phi\hat{y}^a)+\cos\theta\hat{z}^a,
\nonumber\\
\hat{\theta}^a\equiv\pd{_\theta}\hat{\rho}^a=\cos\theta(\cos\phi\hat{x}^a+\sin\phi\hat{y}^a)-\sin\theta\hat{z}^a,
\nonumber\\
\hat{\phi}^a\equiv\pd{_\phi}\left(\hat{\rho}^a/\sin\theta\right)=-\sin\phi\hat{x}^a+\cos\phi\hat{y}^a,
\nonumber\\
\hat{x}^a=\bdeltaup{1}^a,\ \hat{y}^a=\bdeltaup{2}^a,\ \hat{z}^a=\bdeltaup{3}^a.
 \label{10112019da}
\end{align}
Without loss of generality, we can define the act of raising and lowering the indices of these vector with the Minkowski metric: $\hat{t}_a\equiv \eta_{ab}\hat{t}^b=\bdeltadn{0}_a$,  $\hat{x}_a\equiv \eta_{ab}\hat{x}^b=-\bdeltadn{1}_a$ and so on. Note that  $\hat{t}_a=\hat{t}^a$, $\hat{\rho}_a=-\hat{\rho}^a$, etc. It is also easy to see that $\hat{t}^a\hat{t}_a=1$, $\hat{\rho}^a\hat{\rho}_a=-1$, and so on. These vectors also satisfy the relations
\begin{align}
\pd{_\theta}\hat{\theta}^a=-\hat{\rho}^a,\ \pd{_\phi}\hat{\theta}^a=\cos\theta\hat{\phi}^a,
\nonumber\\
\pd{_\theta}\hat{\phi}^a=0,\ \pd{_\phi}\hat{\phi}^a=-\sin\theta\hat{\rho}^a-\cos\theta\hat{\theta}^a,
\nonumber\\
\pd{_\mu}\hat{t}^a=0. \label{10112019qa}
\end{align}

Now, for a given curved spacetime $g_{\mu\nu}$, we can adopt a tetrad field $\e{_a^\mu}$ such that $g_{\mu\nu}=\e{_a_\mu}\e{^a_\nu}$. Once we have chosen the tetrad we want to work with, we can use the following definitions:
\begin{align}
\hat{t}^\lambda\equiv\e{_a^\lambda}\hat{t}^a,\ \hat{\rho}^\lambda\equiv \e{_a^\lambda}\hat{\rho}^a,\ \hat{\theta}^\lambda=\e{_a^\lambda}\hat{\theta}^a,\ \hat{\phi}^\lambda\equiv \e{_a^\lambda}\hat{\phi}^a, \label{10112019e}
\end{align} 
which imply $\hat{t}^\mu\hat{t}_\mu=\e{_a^\mu}\e{_b_\mu}\hat{t}^a\hat{t}^b=\hat{t}^a\hat{t}_a=1$, and of course $\hat{\rho}^\mu\hat{\rho}_\mu=\hat{\theta}^\mu\hat{\theta}_\mu=\hat{\phi}^\mu\hat{\phi}_\mu=-1$.

Expanding the tetrad field in terms of the unit vectors and using Eq.~(\ref{10112019e}), we find that
\begin{align}
\e{_a^\lambda}=\hat{t}^\lambda\hat{t}_a-\hat{\rho}^\lambda\hat{\rho}_a-\hat{\theta}^\lambda\hat{\theta}_a-\hat{\phi}^\lambda\hat{\phi}_a, \label{19112019a}
\\ 
\e{^a_\lambda}=\hat{t}_\lambda\hat{t}^a-\hat{\rho}_\lambda\hat{\rho}^a-\hat{\theta}_\lambda\hat{\theta}^a-\hat{\phi}_\lambda\hat{\phi}^a. \label{19112019b}
\end{align}
It follows from $g_{\mu\nu}=\e{^a_\mu}\e{_a_\nu}$ that
\begin{align}
g_{\mu\nu}=\hat{t}_\mu\hat{t}_\nu-\hat{\rho}_\mu\hat{\rho}_\nu-\hat{\theta}_\mu\hat{\theta}_\nu-\hat{\phi}_\mu\hat{\phi}_\nu. \label{22022020a}
\end{align}

In this formalism, it becomes easier to go from the coordinate basis to the tetrad one . For instance, from Eq.~(\ref{22022020a}) we can infer that   the components of the metric tensor in the tetrad basis can be written as
$\eta_{ab}=\hat{t}_a\hat{t}_b-\hat{\rho}_a\hat{\rho}_b-\hat{\theta}_a\hat{\theta}_b-\hat{\phi}_a\hat{\phi}_b$. This suggests that, in order to simplify the calculations,  we should write most of the objects in terms of the unit vectors $\{\hat{t}^\mu,\hat{\rho}^\mu,\hat{\theta}^\mu,\hat{\phi}^\mu\}$.

The torsion components $\torsion{^a_\mu_\nu}$ can be expanded as
\begin{align}
\torsion{^a_\mu_\nu}=\torsion{^\bzero_\mu_\nu}\hat{t}^a-\torsion{^\bbrho_\mu_\nu}\hat{\rho}^a-\torsion{^\bbtheta_\mu_\nu}\hat{\theta}^a-\torsion{^\bbthree_\mu_\nu}\hat{\phi}^a. \label{25112019x}
\end{align}

To calculate the superpotential\footnote{There is an alternative way to calculate both $\potential{^a^\mu^\nu}$ and $\energy{^\mu^\nu}$ from the Levi-Civita spin connection and the identities (55) and (57) in Ref.~\cite{formiga2020meaning}.}, it is useful to use the identity 
\begin{align}
g^{\lambda[\nu}T^{\mu]}=&\ \hat{t}^\lambda\hat{t}^{[\nu}T^{\mu]}+\hat{\rho}^\lambda\hat{\rho}^{[\mu}T^{\nu]}+\hat{\theta}^\lambda\hat{\theta}^{[\mu}T^{\nu]}
\nonumber\\
&+\hat{\phi}^\lambda\hat{\phi}^{[\mu}T^{\nu]}.\label{25112019xx}
\end{align}
In turn, to calculate $\energy{^\mu^\nu}$, we can use
\begin{align}
-k\e{^a^\lambda}T=kT(-\hat{t}^\lambda\hat{t}^a+\hat{\rho}^\lambda\hat{\rho}^a+\hat{\theta}^\lambda\hat{\theta}^a+\hat{\phi}^\lambda\hat{\phi}^a). \label{25112019xy}
\end{align}

So far we have not fixed the coordinate system: Although we have used the unit vectors of the Cartesian and spherical coordinate systems, we can write $\{\hat{t}^\mu,\hat{\rho}^\mu,\hat{\theta}^\mu,\hat{\phi}^\mu\}$ in any coordinate system we want, with or without performing a change of the coordinate basis. For example, we can write $\hat{t}=\hat{t}^t(u,v,w,q)\pd{_t}+\hat{t}^\rho(u,v,w,q)\pd{_\rho}+\hat{t}^\theta(u,v,w,q)\pd{_\theta}+\hat{t}^\phi(u,v,w,q)\pd{_\phi}$, where $(u,v,w,q)$ is some coordinate system; or  $\hat{t}=\hat{t}^u(u,v,w,q)\pd{_u}+\hat{t}^v(u,v,w,q)\pd{_v}+\hat{t}^w(u,v,w,q)\pd{_w}+\hat{t}^q(u,v,w,q)\pd{_q}$ and so on. This freedom gives us the opportunity to perform the calculations in the most convenient way. 

If we choose to work only with spherical coordinates, then by substituting Eq.~(\ref{19112019b}) into (\ref{04102019pa}) we find that the components in Eq.~(\ref{25112019x}) will be given by

\begin{align}
\torsion{^\bzero_\mu_\nu}=2\partial_{[\mu}\hat{t}_{\nu]}, \label{25112019c}
\\ 
\torsion{^\bbrho_\mu_\nu}=2\partial_{[\mu}\hat{\rho}_{\nu]}-2\delta^2_{[\mu}\hat{\theta}_{\nu]}-2\sin\theta\delta^3_{[\mu}\hat{\phi}_{\nu]}, \label{25112019d}
\\ 
\torsion{^\bbtheta_\mu_\nu}=2\partial_{[\mu}\hat{\theta}_{\nu]}+2\delta^2_{[\mu}\hat{\rho}_{\nu]}-2\cos\theta\delta^3_{[\mu}\hat{\phi}_{\nu]}, \label{25112019dd}
\\ 
\torsion{^\bbthree_\mu_\nu}=2\partial_{[\mu}\hat{\phi}_{\nu]}+2\sin\theta\delta^3_{[\mu}\hat{\rho}_{\nu]}+2\cos\theta\delta^3_{[\mu}\hat{\theta}_{\nu]}. \label{25112019e}
\end{align}

Again, if we have a term like $\torsion{^a_\mu_\nu}=()\hat{t}^a\hat{t}_{[\mu}\hat{\rho}_{\nu]}$, we know immediately that   $\torsion{^a_b_c}=()\hat{t}^a\hat{t}_{[b}\hat{\rho}_{c]}$, and $\torsion{^\lambda_b_\nu}=()\hat{t}^\lambda\hat{t}_{[b}\hat{\rho}_{\nu]}$ etc. In the last example, we have $\hat{t}_{[b}\hat{\rho}_{\nu]}\equiv(1/2)(\hat{t}_b\hat{\rho}_\nu-\hat{t}_\nu\hat{\rho}_b)$. In some cases, we can use the definition $\hat{t}_{[b}\hat{\rho}^{\nu]}\equiv (1/2)(\hat{t}_b\hat{\rho}^\nu-\hat{t}^\nu\hat{\rho}_b)$.

Equations (\ref{25112019c})-(\ref{25112019e}) are particularly useful when we are working with a radial motion and the spacetime is spherically symmetric.

The whole formalism shown in this section is compatible with that of MAG (for more details, see Ref.~\cite{formiga2020meaning}). It is possible that, by using them together, we could improve the calculations even more.

\subsubsection{Changing the signature}
In case the reader is willing to work with the signature $(-+++)$, the hybrid machinery can be easily adapted to this signature. Equations (\ref{10112019da}), (\ref{10112019qa}), (\ref{10112019e}) and (\ref{25112019d})-(\ref{25112019e}) do not change. In the case of Eq.~(\ref{25112019c}), the expression will also be the same if we write it in terms of $\torsion{^\bbt_\mu_\nu}$, instead of $\torsion{^\bzero_\mu_\nu}$, i.e, we will still have $\torsion{^\bbt_\mu_\nu}=2\partial_{[\mu}\hat{t}_{\nu]}$. 

On the other hand, we now have $\hat{t}_a=-\delta_a^\bzero$, $\hat{x}_a=\delta_a^\bone$, and so on. Of course, we also have $\hat{t}^a\hat{t}_a=\hat{t}^\mu\hat{t}_\mu=-1$, $\hat{\rho}^a\hat{\rho}_a=\hat{\rho}^\mu\hat{\rho}_\mu=1$ and so on. Expressions (\ref{19112019a})-(\ref{22022020a}) can be adapted to the new signature by changing the sign of the right-hand sides:  $\e{_a^\lambda}=-\hat{t}^\lambda\hat{t}_a+\hat{\rho}^\lambda\hat{\rho}_a+\hat{\theta}^\lambda\hat{\theta}_a+\hat{\phi}^\lambda\hat{\phi}_a$ and $g_{\mu\nu}=-\hat{t}_\mu\hat{t}_\nu+\hat{\rho}_\mu\hat{\rho}_\nu+\hat{\theta}_\mu\hat{\theta}_\nu+\hat{\phi}_\mu\hat{\phi}_\nu$. Equation (\ref{25112019x}) can also be adapted in this way if we use $\torsion{^\bbt_\mu_\nu}$ rather than $\torsion{^\bzero_\mu_\nu}$. In this case, the expression becomes $\torsion{^a_\mu_\nu}=-\torsion{^\bbt_\mu_\nu}\hat{t}^a+\torsion{^\bbrho_\mu_\nu}\hat{\rho}^a+\torsion{^\bbtheta_\mu_\nu}\hat{\theta}^a+\torsion{^\bbthree_\mu_\nu}\hat{\phi}^a$.

We continue to work with the signature $(+---)$ in the next section. Nevertheless, the signature can be changed to $(-+++)$ following the same logic as explained above. 

\subsection{Mixing Cartesian and cylindrical unit vectors with spherical coordinates}\label{26122019sc}
Sometimes it is more convenient to work with the unit vectors of the cylindrical coordinate system because the frame used possess this symmetry. (This is exactly the case of Sec.~\ref{26122019sd}.)

For $x=s\cos\phi$, $y=s\sin\phi$, and $z=z$, we have:
\begin{align}
\hat{s}^a\equiv \cos\phi\hat{x}^a+\sin\phi\hat{y}^a,\ \hat{\phi}^a\equiv \pd{_\phi}\hat{s}^a,\ \pd{_\phi}\hat{\phi}^a=-\hat{s}^a. \label{14122019a}
\end{align}
The tetrad field can be written as
\begin{align} 
\e{_a^\lambda}=\hat{t}^\lambda\hat{t}_a-\hat{s}^\lambda\hat{s}_a-\hat{\phi}^\lambda\hat{\phi}_a-\hat{z}^\lambda\hat{z}_a, \label{14122019d}
\\
\e{^a_\lambda}=\hat{t}_\lambda\hat{t}^a-\hat{s}_\lambda\hat{s}^a-\hat{\phi}_\lambda\hat{\phi}^a-\hat{z}_\lambda\hat{z}^a, \label{14122019e}
\end{align}
while the metric becomes
\begin{align}
g_{\mu\nu}=\hat{t}_\mu\hat{t}_\nu-\hat{s}_\mu\hat{s}_\nu-\hat{\phi}_\mu\hat{\phi}_\nu-\hat{z}_\mu\hat{z}_\nu. \label{14122019f}
\end{align}

Since we are going to apply this formalism to the Schwarzschild spacetime, which is spherically symmetric, we will use the spherical coordinate system $x^\mu=(t,\rho,\theta,\phi)$ and its coordinate basis $(\pd{_t},\pd{_\rho},\pd{_\theta},\pd{_\phi})$. 

Applying $\pd{_\mu}$ to Eq.~(\ref{14122019a}) and using Eq.~(\ref{10112019da}), we obtain
\begin{align}
 \pd{_\mu}\hat{s}^a=\delta_\mu^3\hat{\phi}^a,\  \pd{_\mu}\hat{\phi}^a=-\delta_\mu^3\hat{s}^a. \label{22022020b}
\end{align}

Like $\torsion{^a_\mu_\nu}$ in  Eq.~(\ref{25112019x}), the torsion components here can be written in the form 
\begin{align}
\torsion{^a_\mu_\nu}=\torsion{^\bzero_\mu_\nu}\hat{t}^a-\torsion{^\bbs_\mu_\nu}\hat{s}^a-\torsion{^\bbphi_\mu_\nu}\hat{\phi}^a-\torsion{^\bbz_\mu_\nu}\hat{z}^a. \label{15122019a}
\end{align} 
Now, using Eqs.~(\ref{14122019e}) and (\ref{22022020b}) into Eq.~(\ref{04102019pa}), and comparing the result with Eq.~(\ref{15122019a}), we find that
\begin{align}
\torsion{^\bzero_\mu_\nu}=2\partial_{[\mu}\hat{t}_{\nu]}, \label{15122019b}
\\ 
\torsion{^\bbs_\mu_\nu}\equiv 2\partial_{[\mu}\hat{s}_{\nu]}-2\delta^3_{[\mu}\hat{\phi}_{\nu]}, \label{15122019c}
\\ 
\torsion{^\bbphi_\mu_\nu}\equiv 2\partial_{[\mu}\hat{\phi}_{\nu]}+2\delta^3_{[\mu}\hat{s}_{\nu]}, \label{15122019d}
\\ 
\torsion{^\bbz_\mu_\nu}\equiv 2\partial_{[\mu}\hat{z}_{\nu]}. \label{15122019e}
\end{align}

The equivalents of Eqs.~(\ref{25112019xx}) and (\ref{25112019xy}) are
\begin{align}
g^{\lambda [\nu}T^{\mu]}=&-\hat{t}^\lambda\hat{t}^{[\mu}T^{\nu]}+\hat{s}^\lambda\hat{s}^{[\mu}T^{\nu]}+\hat{\phi}^\lambda\hat{\phi}^{[\mu}T^{\nu]}
\nonumber\\
&+\hat{z}^\lambda\hat{z}^{[\mu}T^{\nu]}, \label{14122019g}
\end{align}
\begin{align}
-\e{^a^\lambda}T=T(-\hat{t}^\lambda\hat{t}^a+\hat{s}^\lambda\hat{s}^a+\hat{\phi}^\lambda\hat{\phi}^a+\hat{z}^\lambda\hat{z}^a). \label{15122019f}
\end{align}

We could go on and define the equivalents of Eqs. ~(\ref{15122019a})-(\ref{15122019e}) to other coordinate systems, but the procedure would be basically the same. 

\section{Accelerated observers in Schwarzschild spacetime}\label{26122019sd}
Let us now apply the hybrid machinery to a tetrad field adapted to accelerated observers moving along the $z$-direction in Schwarzschild spacetime. 

In isotropic coordinates, we have
\begin{align}
ds^2=& A^2dt^2-B^2(dx^2+dy^2+dz^2)
\nonumber\\
=& A^2dt^2-B^2(d\rho^2+\rho^2d\theta^2+\rho^2\sin^2\theta d\phi^2), \label{10112019a}
\end{align}
where $B=[1+r_0/(4\rho)]^2$,  $r_0=2m$, and $A=2/B^{(1/2)}-1$.

In the Cartesian basis, the static tetrad can be written in the form
\begin{align}
\e[\bar]{_a}=\frac{1}{A}\hat{t}_a\pd{_t}-\frac{1}{B}\left(\hat{x}_a\pd{_x}+\hat{y}_a\pd{_y}+\hat{z}_a\pd{_z} \right), \label{10112019f}
\end{align}
where $\hat{x}^a$, $\hat{y}^a$, $\hat{z}^a$ are given by Eq.~(\ref{10112019da}). 

We can build a frame  adapted to arbitrarily accelerated  (along the $z$-direction)  observers by applying the local Lorentz transformation
\begin{align}
\lorentz{_a^b}=(f\hat{t}_a-g\hat{z}_a)\hat{t}^b+(g\hat{t}_a-f\hat{z}_a)\hat{z}^b-\hat{x}_a\hat{x}^b-\hat{y}_a\hat{y}^b, \label{10112019g}
\end{align}
where $f=f(t,z)$ (Lorentz factor) and $g=g(t,z)=\pm\sqrt{f^2-1}$. Note that $f$ depends on $z$, too. We will perform the calculations assuming this dependence to show that it may produce some undesirable effects.

Using Eq.~(\ref{10112019g}) and $\e{_a}=\lorentz{_a^b}\e[\bar]{_b}$, we find that
\begin{align}
\e{_a^{\bar{\lambda}}}=&\ (\frac{f}{A}\delta^{\bar{\lambda}}_0+\frac{g}{B}\delta^{\bar{\lambda}}_3)\hat{t}_a-\frac{1}{B}\delta^{\bar{\lambda}}_1\hat{x}_a-\frac{1}{B}\delta^{\bar{\lambda}}_2\hat{y}_a
\nonumber\\
&-(\frac{g}{A}\delta^{\bar{\lambda}}_0+\frac{f}{B}\delta^{\bar{\lambda}}_3)\hat{z}_a, \label{10112019h}
\end{align}
where the overbar denotes the ``Lorentzian'' coordinate system $\bar{x}^\lambda=(t,x,y,z)$. Changing these components to the spherical coordinate system and using Eq.~(\ref{14122019a}), we find that
 \begin{align}
\hat{t}^\lambda=\frac{f}{A}\delta^\lambda_0+\frac{g}{B}(\cos\theta\delta^\lambda_1-\frac{\sin\theta}{\rho}\delta^\lambda_2), \label{15122019g}
\\
\hat{s}^\lambda=\frac{\sin\theta}{B}\delta^\lambda_1+\frac{\cos\theta}{\rho B}\delta^\lambda_2, \label{15122019h}
\\ 
\hat{\phi}^\lambda=\frac{\delta^\lambda_3}{\rho B\sin\theta}, \label{15122019j}
\\ 
\hat{z}^\lambda=\frac{g}{A}\delta^\lambda_0+\frac{f}{B}\cos\theta\delta^\lambda_1-\frac{f}{\rho B}\sin\theta\delta^\lambda_2. \label{15122019i}
\end{align}
Lowering the indices, we obtain
\begin{align}
\hat{t}_\lambda=fA\delta_\lambda^0-Bg(\cos\theta\delta_\lambda^1-\rho \sin\theta\delta_\lambda^2), \label{15122019l}
\\ 
\hat{s}_\lambda=-B\sin\theta\delta^1_\lambda-\rho B\cos\theta\delta^2_\lambda, \label{15122019m}
\\ 
\hat{\phi}_\lambda=-\rho B\sin\theta\delta_\lambda^3, \label{15122019o}
\\ 
\hat{z}_\lambda=gA\delta^0_\lambda-fB\cos\theta\delta^1_\lambda+f\rho B\sin\theta\delta^2_\lambda. \label{15122019n}
\end{align}
The next step is to invert these expression so as to write the ``deltas'' in terms of $\{\hat{t}_\lambda,\hat{s}_\lambda,\hat{\phi}_\lambda,\hat{z}_\lambda\}$. Once we have the ``deltas'' written in this basis, then we can write everything else in terms of it. The remaining calculations can be found in appendix \ref{03092019f}.

\subsection{Gravitational angular momentum density}
From Eq.~(\ref{27122019e}) in appendix \ref{03092019f} and Eq.~(\ref{30122019da}), we find that
\begin{align}
M^{ab}=\frac{8ke h_6}{A}\left( f\sin\theta\hat{t}^{[a}\hat{s}^{b]}+\cos\theta\hat{t}^{[a}\hat{z}^{b]}
+g\sin\theta\hat{s}^{[a}\hat{z}^{b]}\right), \label{23122019a}
\end{align}
where we have used the identity  $ gh_{12}-fh_{14}=-h_6\cos\theta$ [ see Eq.~(\ref{15122019p})]; the determinant of the tetrad field is $e=AB^3\rho^2\sin\theta$, while $h_6$ is given by  
\begin{align}
h_6=-\frac{r_0}{2\rho^2B^{(3/2)}}. \label{15032020a}
\end{align}
Note that, since  $h_6$ vanishes when $r_0=0$,  the density $M^{ab}$ vanishes in the absence of gravity.

Substituting Eq.~(\ref{23122019a}) into Eq.~(\ref{30122019ca}) and integrating  over a spherical volume, we get
\begin{align}
L^{ab}=&-16\pi k\left(\int_0^r B^3\rho^2h_6 d\rho \right)
\nonumber\\
&\times \left(\int_0^{\pi}d\theta\sin\theta\cos\theta \right)\hat{t}^{[a}\hat{z}^{b]}=0, \label{17032020a}
\end{align}
where we have used $\int_0^{2\pi}\hat{s}^ad\phi=0$. The integral in $\rho$ diverges, though. (This might be a problem for frames that are not axially symmetric.)

As expected, the total angular momentum of the gravitational field vanishes. (The observers' trajectories are symmetric.) With respect to the density, we see in Eq.~(\ref{23122019a}) that the contribution to $L^{(i)(j)}$ comes from the third term, which is associated to rotations of the gravitational field in the plane $(\hat{s}^a,\hat{z}^a)$; using Eq.~(\ref{15032020a}) we can write this contribution as $M^{(i)(j)}=4kgr_0B^{(3/2)}\sin^2\theta \hat{z}^{[(i)}\hat{s}^{(j)]}$, which corresponds to an angular momentum density in the direction of $-\hat{\phi}^a$ for $g>0$ (recall that $L \sim-M$). This density vanishes at $\theta=0,\pi$ and reaches its maximum when $\theta=\pi/2$. These properties are consistent with our intuition.

It is worth noting that Eq.~(\ref{23122019a}) yields $M^{ab}=(8keh_6/A)\hat{t}^{[a}\hat{\rho}^{b]}$ for static observers ($f=1$ and $g=0$), where $\hat{\rho}$ is given by Eq.~(\ref{10112019da}).  Since $M^{(i)(j)}$ vanishes, the angular momentum density also vanishes, as expected; the density associated to the gravitational center of mass moment, however, does not vanish (we have rotations in the $(t,\rho)$ plane). These results are consistent with those of Ref.~\cite{ANDP:ANDP201200272} 

It is also interesting to note that $M^{ab}=(8keh_6/A)\hat{t}^{[a}\hat{\rho}^{b]}$ does not vanish along the worldline of any static observer (except for $r\to\infty$). So, in this sense, this density can be localized by the observers that are not at spatial infinity.

\subsection{Gravitational energy}\label{29122019a}
Let us now calculate the gravitational energy within a sphere of radius $\rho$. By exchanging the letter $b$ (tangent space index) for $1$ (the coordinate index associated with $\rho$) in Eq.~(\ref{27122019e}) and using Eqs.~(\ref{15122019g})-(\ref{15122019i}), we get 
\begin{align}
\potential{^a^0^1}=\frac{\partial_\rho B}{AB^3}\left[ -f(t,z)\hat{t}^a+g(t,z)\hat{z}^a\right]+\frac{\sin\theta}{2fAB^2}(\pd{_z}g)\hat{s}^a, \label{05012020ab}
\end{align}
where we have used the identities $ h_{14}\cos\theta+fh_6\sin^2\theta=fh_6$ and $ h_{12}\cos\theta+gh_6\sin^2\theta=gh_6$ [see Eq.~(\ref{15122019p})].

Substituting Eq.~(\ref{05012020ab}) into  Eq.~(\ref{22072018ga}) and simplifying, we obtain
\begin{align}
P^a=\frac{E_s}{2\rho}\int_{-\rho}^\rho dz\left[f(t,z)\hat{t}^a-g(t,z)\hat{z}^a\right], \label{10122019bb}
\end{align}
where $E_s=-16\pi k\rho^2\partial_\rho B=8\pi k r_0[1+r_0/(4\rho)]$ is the energy measured by the static observers, and the integral in $\theta$ has been changed to an integral in $z$. [The value of $E_s$ in the Schwarzschild coordinate system can be found in Ref.~\cite{doi:10.1002/andp.201900507}, Eq.~(45) with $f=1$.]

It is clear in Eq.~(\ref{10122019bb}) that $P^a$ is zero when $r_0=0$ (no gravity). For $r_0\neq 0$, however, things get more complicated. As discussed in section VI C of Ref.~\cite{formiga2020meaning}, to have a meaningful calculation of the energy of a field, we need uniformity. This uniformity can be achieved by taking $f=f(t)$ and $g=g(t)$, in which case  we have $P^a=f(t)E_s\hat{t}^a-g(t)E_s\hat{z}^a$. This is exactly what we should expect: a ``particle'' of mass $E_s$ moving with velocity $v=-g(t)/\sqrt{1+g(t)^2}$.

If we let the observers' velocity depend on $z$, then we will have to impose some restrictions on $f$ and $g$. The first one is to assume that they are even functions of $z$ (for more details, see the discussion in section VI C of Ref.~\cite{formiga2020meaning}). The second one is to take the asymptotic limit. (We will see that the first restriction is necessarily only asymptotically.)

We can recast the first integral in Eq.~(\ref{10122019bb}) as  $\int_{-\rho}^\rho dz f(t,z)=\int_{-\rho}^{-a}f(t,z)dz+\int_{-a}^{a}f(t,z)dz+\int_a^\rho f(t,z)dz$, where $a$ is an arbitrary positive constant. Because of the factor $1/\rho$ in Eq.~(\ref{10122019bb}), if  the second integral above  does not diverge at $\pm a$, then this integral will not contribute to $P^a$ at spatial infinity. Besides, the gravitational momentum will not depend on $a$. In this case, we can write $\int_{-\rho}^\rho dz f(t,z)=2\int_a^\rho f(t,z)dz$, where we have used $f(t,z)=f(t,-z)$. The same holds for the second integral in Eq.~(\ref{10122019bb}). Hence, the limit of Eq.~(\ref{10122019bb}) as $\rho$ goes to infinity is
\begin{align}
P^a_\infty=\lim_{\rho\to\infty}E_s\left(\frac{\hat{t}^a}{\rho}\int_a^\rho f(t,z)dz-\frac{\hat{z}^a}{\rho}\int_a^\rho g(t,z)dz \right). \label{12122019a}
\end{align}

\begin{figure}[h]
\includegraphics[scale=0.25]{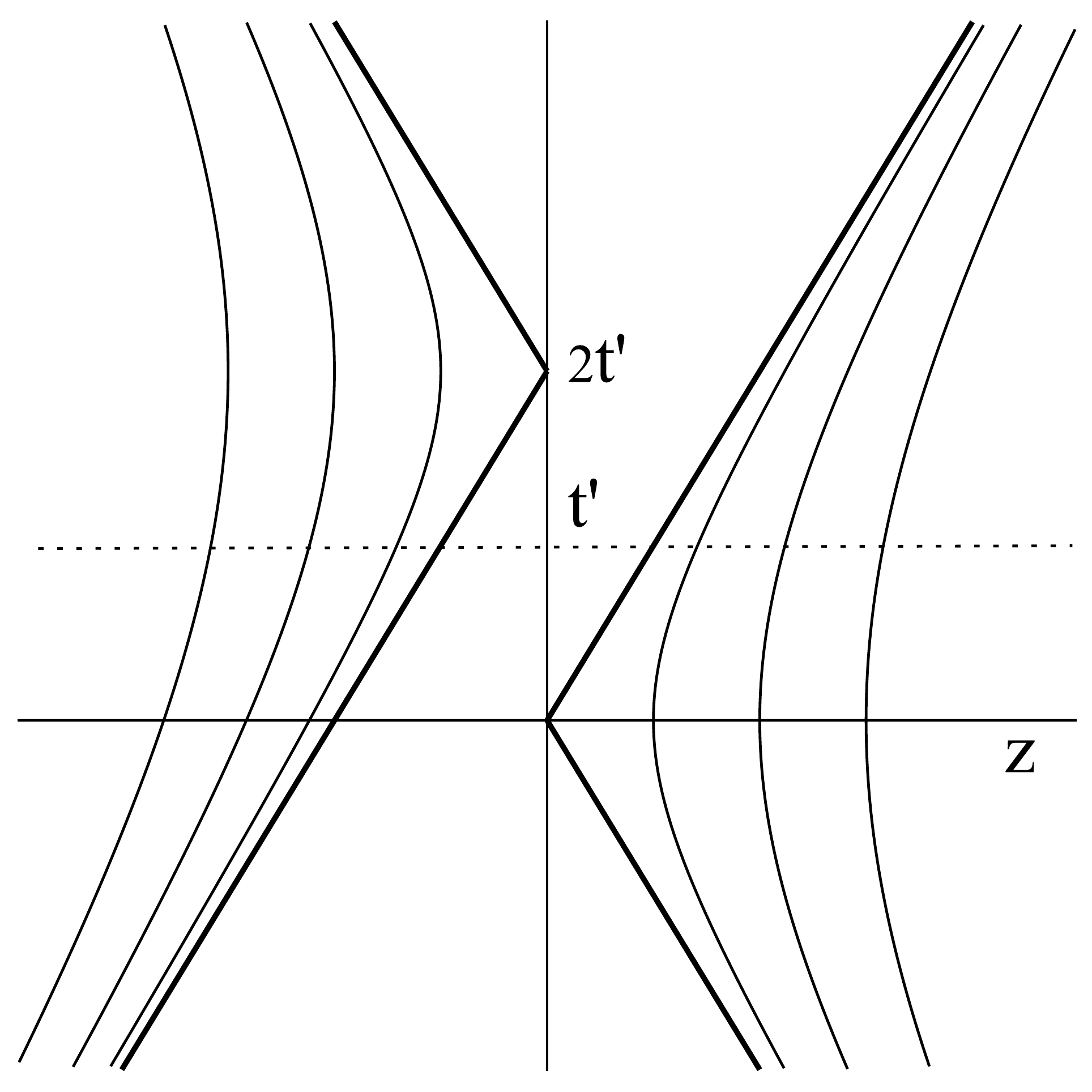}
\caption{ In this figure we show how to connect the observers in the region $z>0$ with those in $z<0$ so as to have $f(t',-z)=f(t',z)$ and $g(t',-z)=g(t',z)$ at a instant $t'$. While the curves on the right are $f(t,z)=|z|/\sqrt{z^2-t^2}$ and $g(t,z)=t/\sqrt{z^2-t^2}$, those on the left are given by $f(t,z)=|z|/\sqrt{z^2-(2t'-t)^2}$ and $g(t,z)=(2t'-t)/\sqrt{z^2-(2t'-t)^2}$. This ensures that all the observers are moving towards  the right and in a symmetric way.}
\label{10012020a}
\end{figure}

It is easy to calculate these integrals because we need only $f$ and $g$ in the limit $z\to \infty$. If $\lim_{z\to\infty}f(t,z)=\gamma(t)$ and $\lim_{z\to\infty}g(t,z)=\beta(t)\gamma(t)$, then we will obtain the $4$-momentum of a particle in special relativity: $P_\infty^a=\gamma(t)m\hat{t}^a-\beta(t)\gamma(t)m\hat{z}^a$, where $m$ is the black hole mass. As an example, let us take the Lorentz factor of the Rindler observers, namely, $f(t,z)=|z|/\sqrt{z^2-t^2}$ and $g(t,z)=t/\sqrt{z^2-t^2}$. To guarantee that the trajectories are symmetric, we synchronize the observers in the region $z<0$ with those in region $z>0$ as shown in Fig.~\ref{10012020a}. For constant $t$ and $z\gg t$, we see that $f(t,z)\approx 1$ and $g(t,z)\approx t/|z|$, which gives $P_\infty^a=E_s\hat{t}^a$ (the observers are at rest at infinity). Another interesting example, which we will discuss later, is a frame with $f(t,z)=\gamma(t)/\sqrt{B(|z|)}$. For $z\gg r_0$, we get $f\approx\gamma(t)$ and $g\approx\sqrt{\gamma(t)^2-1}$. Then, for this case, we have $P_\infty^a=\gamma(t)m\hat{t}^a-\beta(t)\gamma(t)m\hat{z}^a$.

The above examples show that the mess caused by the non-uniformity in the frame cancels out at infinity, if the observers' velocity are finite and, of course, uniform there.

\subsection{Acceleration}
The antisymmetric acceleration tensor can be written in the form [see, e.g., Eq.~(9) of Ref.~\cite{ANDP:ANDP201200272}]
\begin{align}
\phi_{ab}=\frac{1}{2}(\torsion{_\bzero_a_b}+\torsion{_a_\bzero_b}-\torsion{_b_\bzero_a}). \label{23122019b}
\end{align}
From Eq.~(\ref{27122019d}), one can check that $ 2\torsion{^{[\mu|}^\lambda^{|\nu]}}=\torsion{^\lambda^\mu^\nu}$. Hence, the above expression reduces to $\phi_{ab}=\torsion{_\bzero_a_b}$. Substituting  Eq.~(\ref{23122019c}) into this expression, we obtain
\begin{align}
\phi_{ab}=2h_7\hat{t}_{[a}\hat{s}_{b]}+2h_8\hat{t}_{[a}\hat{z}_{b]}+2h_9\hat{s}_{[a}\hat{z}_{b]}, \label{23122019f}
\end{align}
where the functions $h_7$, $h_8$, $h_9$ can be found in Eq.~(\ref{15122019p}). So, the frame acceleration is given by
\begin{align}
a^b=h_7\hat{s}^b+h_8\hat{z}^b. \label{23122019g}
\end{align}

The components $h_7$ and $h_8$ are accelerations along $\hat{s}$ and $\hat{z}$, respectively. (The component $h_7$ is a result of an external force that prevents the observers from moving towards the black hole.) The function $h_9$ is associated to a rotation of the spatial frame about an certain direction in the tangent plane formed by $\pd{_x}$ e $\pd{_y}$: $\phi_{\bone\bthree}=h_9\cos\phi$ and $\phi_{\btwo\bthree}=h_9\sin\phi$.

\subsection{Energy density}
It is desirable to have a theory where the gravitational energy density cannot be made to vanish along the observer's worldline. Here we show that, for the tetrad given by Eq.~(\ref{10112019h}), we cannot cancel this density if  $f=f(t)$. 

We can take $\nu=0$ and exchange the coordinate index $\alpha $ for the  tangent space one $(0)$ in Eq.~(\ref{20122019c}), then  use Eqs.~(\ref{10112019da}), (\ref{14122019a}),  and (\ref{15122019g})-(\ref{15122019i}) to simplify the expression and substitute the result into Eq.~(\ref{29032019ha}). To calculate the second term of Eq.~(\ref{29032019ha}), we can use Eqs.~(\ref{20122019d}), (\ref{14122019d}),  and procedure in a similar way. Doing this, we obtain
\begin{align}
\energy{^0^{(0)}}=&\ k\bigl[(2h_9^2+2h_7h_{15}+4h_8h_{14}-4h_{12}^2-8h_{10}h_{12}
\nonumber\\
&-T)\hat{t}^0+(4h_{12}h_{18}-2h_9h_{13})\hat{z}^0 \bigr]. \label{24122019a}
\end{align}
 Now, using Eq.~(\ref{15122019p}) and the identity $\partial_\rho f(t,z)=\cos\theta\partial_zf(t,z)$, we can write this component as
\begin{align}
\energy{^0^{(0)}}=\frac{kr_0}{\rho^2AB^{(7/2)}}\left[\partial_\rho\left(fB\right)+B\partial_\rho f \right]. \label{24122019b}
\end{align} 

The energy density will vanish if $\partial_\rho\left(fB\right)+B\partial_\rho f=0$. The solution for this equation is  $f(t,z)=\gamma(t,\theta)/\sqrt{B(\rho)}$, where $\gamma(t,\theta)$ is an integration constant. However, since $\sqrt{B}=(1+r_0/(4\rho))$, the dependence on $\theta$ cannot disappear. Thus, we can make the energy density vanish only along the $z$ axis ($\theta=0,\pi$). In this case, we have  $f(t,z)=\gamma(t)/\sqrt{B(|z|)}$. This case corresponds to the example given in Sec.~\ref{29122019a} [see text below  Eq.~(\ref{12122019a})], where we have extended this Lorentz factor to all observers and have shown that the value of $P^a$ at infinity is consistent.  

Despite the fact that $P^a$ for  $f(t,z)=\gamma(t)/\sqrt{B(|z|)}$ is consistent at infinity, the curve that makes (\ref{24122019b}) vanish is not the curve of one particular observer, but rather the collection of many observers: for $t$ constant, we have different observers for different values of $z$, each with a Lorentz factor given by $f(t,z)=\gamma(t)/\sqrt{B(|z|)}$. 

By removing the dependence on $z$ and coming back to Eq.~(\ref{24122019b}), we see that $\energy{^0^{(0)}}\sim f(t)\partial_\rho B$, which cannot vanish. Therefore, if we limit ourselves to frames where $f$ is a function of $t$ only, the gravitation energy density will not  vanish at  any finite point of an accelerated observer whose trajectory is in the $z$ direction of the isotropic coordinate system.

\section{Conclusions}\label{26052021a}
We have improved the hybrid machinery of Ref.~\cite{doi:10.1002/andp.201900507} and used it to analyze the gravitational energy problem in a accelerated frame, moving along the $z$-direction. We have found that when the Lorentz factor depends only on $t$, the gravitational energy and momentum takes the usual form of a particle moving with a velocity $v$ in Minkowski spacetime; however, when this factor depends on $z$, the expression of the energy-momentum tensor $P^a$ becomes more involved: we have to demand $f(t,z)=f(t,-z)$ to have a meaningful result. (A deeper discussion of this point can be found in Ref.~\cite{formiga2020meaning}.)

By imposing $f=f(t)$, we have also been able to show that the gravitational energy density cannot be made to vanish in any point of the observers' trajectory, which is consistent with an energy that can be localized.

\chapter{On the teleparallel frame problem}\label{paper1}

The role played by the congruence in the prediction of the gravitational energy in the teleparallel equivalent of general relativity is discussed. It is shown that some congruences yield unphysical predictions. It is also shown that the energy-momentum tensor density predicted by the proper reference frame of an arbitrary accelerated observer vanishes along the observer's worldline, regardless of its acceleration. The latter result is discussed and  arguments  both against and in favor of the use of  this type of frame to describe the gravitational energy are presented; arguments against the belief that the principle of equivalence is incompatible with the localization of the gravitational energy are also presented. A set of constraints is imposed on the teleparallel frame and its consistence is discussed: using three different tetrads for the $pp$-wave spacetimes, it is shown that the most consistent prediction comes from the tetrad that satisfies these constraints. The possibility of having a well-defined concept of absolute vacuum is also discussed.

\section{Introduction}
Teleparallel theories are based on an absolute parallelism. In this theories, the torsion tensor is used  to establish parallelism by means of a connection known as the Weitzenb\"{o}ck connection, which has vanishing curvature \cite{ANDP:ANDP201200272}. This framework allows us to formulate a theory that yields the same field equations as general relativity (GR) does, the so-called teleparallel equivalent of general relativity (TERG). Although the field equations are the same, the TEGR has the advantage of being a gauge theory of the translation group \cite{PhysRevD.14.2521,PhysRevD.14.3335,HAYASHI1977441,PhysRevD.19.3524}, and a natural place to approach the problem of the gravitational energy. However, like GR, the TEGR also faces  problems of consistence when predicting the gravitational energy. In this paper we address this problem from the perspective of observers and their frames. We also provide some constrains that can possibly lead to the solution of the energy problem.

The TEGR can be seen as a tetrad formulation of GR where the field equations  takes a peculiar form. In this form, the right-hand side is expected to represent the full energy-momentum content of the spacetime, including the gravitational contribution. The quantity that is interpreted as the gravitational energy-momentum tensor density appears on the same footing as the matter one, except for the way it depends on the teleparallel frame. This density  transforms covariantly under both general coordinate transformations and global Lorentz transformations, and is invariant under local gauge translation of the tangent-space coordinate \cite{PhysRevLett.84.4533}. An object that can be interpreted as the angular momentum density can also be defined. In turn,  one can integrate these densities on a hypersurface with constant time and defines the energy and angular momenta inside this region.

As argued by Maluf \cite{ANDP:ANDP201200272}, there is no problem with the fact that the gravitational energy in the TEGR depends on the frame, because energy is always a frame-dependent quantity. The problem with all approaches to the gravitational energy is that all of them yield objects that depend on artificial properties of either the coordinate system or the tetrad field. In the metrical formulation of GR, they all depend on the coordinate system, even on coordinate transformations of the three-dimensional space. Although some coordinate transformations may be indirectly linked to a change that has physical significance, because the new coordinates might be adapted to a new set of observers with different velocities and accelerations, there are transformations that are clearly not related to any change in the state of motion of the frame; the latter transformations are, in this sense, artificial. (An example of an artificial change is to change Cartesian coordinates to spherical ones.) To remove the artificial properties of these transformations (only asymptotically), one has to restrict the coordinate system to a particular set of the asymptotically rectangular coordinate systems, as described in \cite{Arnowitt2008}. In the TEGR, on the other hand, the artificial properties are encoded in the tetrad field. Since there is no hope of localizing the gravitational energy in the metrical formulation of GR, the analysis of these spurious effects in GR is limited to the behavior of the energy on a surface integral at spatial infinity, which limits the analysis to asymptotically flat spacetimes, or to the so-called ``quasilocal'' approach \cite{0264-9381-9-7-009,PhysRevD.47.1407,PhysRevLett.83.1897}. In the TEGR, however, one hopes to localize the gravitational energy and be able to evaluate it at any point of the spacetime. Hence, the analysis of the spurious features of the tetrad fields at finite regions of the spacetime, including the study of the behavior of the gravitation energy density along the observers' worldline, is an important task.

The paper is organized as follows. The following section is devoted to a brief description of some of the basic properties of the TEGR. In Sec.~\ref{25022020a}, we discuss the causes of inconsistent predictions for the gravitational energy-momentum (including the densities). In general, one assumes that the problem is only with rotations of the spatial triad. However,  in this article we will prove that a bad choice of the congruence yields inconsistent results, too. We will also prove that both the gravitational energy-momentum and angular momentum densities vanish along the curve of any arbitrarily accelerated observer if the proper reference frame (PRF) is used as the teleparallel frame. The possible meanings of this result and the limited role that the principle of equivalence should have in a teleparallel theory is discussed. In Sec.~\ref{24022020a}, we present a discussion on the definition of the energy of a field  to find the restrictions that need to be imposed on the possible sets of observers (timelike congruences). It turns out that, by postulating a frame that eliminates all artificial properties, we can have a reasonable restriction on the set of observers used to calculate the gravitational energy.

To show the consistence of the aforementioned restrictions, we revisit the gravitational energy-momentum  density of $pp$-waves in Sec.~\ref{08082019b}. We compare the results from three incompatible  predictions, those of Refs.~\cite{PhysRevD.78.047502,Obukhov_2009,doi:10.1002/andp.201800320}, and show that the most consistent results come from the frame that satisfies these restrictions.

We devote Sec.~\ref{23032020a} to a general discussion about the definition of vacuum in classical theories. There, we postulate that the TEGR must have a well-defined concept of absolute vacuum (the total absence of energy and momentum).

Throughout this paper, we use the signature $(+,-,-,-)$.  Greek letters represent spacetime indices, and  Latin letters represent tangent space ones, except for Latin letters in the middle of the alphabet ($i,j,k,\ldots$), which stand for spatial coordinate indices. The components of the frame $e_a$ and the coframe $\teta{^a}$ in a coordinate basis are represented by $\tensor{e}{_a^\mu}$ and $\tensor{e}{^a_\mu}$, respectively. We distinguish the tangent space indices from the coordinate ones  by using the former between  parentheses: $\tensor{e}{_{(0)}^0}$, $\tensor{e}{_{(1)}^2}$ etc.

\section{Teleparallelism}\label{22032020ab}
In teleparallel theories, there exist a tetrad field $\e{_a}$ and an affine connection $\nablab$ such that the covariant derivative of $\e{_a}$ vanishes everywhere
\begin{equation}
\nablab_{\mu}\e{_a}=0\quad (\sconnection{^a_b_c}=0), \label{14022019bb}
\end{equation}
where $\sconnection{^a_b_c}$ are the coefficients of this connection expanded in terms of $\e{_a}$. This connection is known as the Weitzenb\"{o}ck connection and has vanishing curvature, as can be easily verified from Eq.~(\ref{14022019bb}).

Since Eq.~(\ref{14022019bb}) is not invariant under local Lorentz transformations, not all frames can satisfy this condition simultaneously. Furthermore, in the pure-tetrad formulation, which is the one used here, we must fix (up to global Lorentz transformation) the frame that satisfies (\ref{14022019bb}). In principle, many types of frames can be chosen to satisfy it, but teleparallel theories give no reason for choosing one over another. The frame that satisfies Eq.~(\ref{14022019bb}) has been called, for obvious reasons, the teleparallel frame (TF).

In a TF, the components of the torsion tensor can be written in the form
\begin{align}
\torsion{^a_\mu_\nu}=\pd{_\mu}\e{^a_\nu}-\pd{_\nu}\e{^a_\mu}. \label{04102019pb}
\end{align}
In other words, in this gauge, the torsion is nothing but the structure functions, also known as the object of anholonomity.

From the torsion tensor, one defines the torsion scalar as $T\equiv\potential{^a^b^c}\torsion{_a_b_c}$, where
\begin{equation}
\potential{^\lambda^\mu^\nu}\equiv\frac{1}{4}\left( \torsion{^\lambda^\mu^\nu}+2\torsion{^{[\mu|}^\lambda^{|\nu]}}\right)+g^{\lambda[\nu}T^{\mu]} \label{10112017lb}
\end{equation}
is called {\it superpotential}, and  $\torsion{_a}\equiv\torsion{^b_b_a}$. (Note that $T$ depends on the choice of the TF. This dependence will be discussed in more details in a future paper.) 

From $T$, one can define the so-called $f(T)$ teleparallel gravity \cite{Cai_2016}. The most appealing $f(T)$ theory is the TEGR, which is characterized by $f(T)\sim T$. The only difference between the TEGR and GR actions is a surface term. As a result, taking variations with respect to the tetrad field gives the Einstein field equations written in term of the tetrad field.

In the metrical formulation of GR, one writes these equation as $\G{^\mu^\nu}=\left(1/(2k)\right)T^{\mu\nu}$, where $\G{^\mu^\nu}$ is the Einstein tensor in terms of the Levi-Civita connection, $T^{\mu\nu}$ is the matter energy-momentum tensor density, and $k=1/(16\pi)$ in natural units. On the other hand, in the TEGR, one writes
\begin{equation}
\pd{_\alpha}\left( e\potential{^a^\mu^\alpha}\right)=\frac{e}{4k}\left(\energy{^\mu^a}+T^{\mu a}\right), \label{29032019kb}
\end{equation}
where
\begin{equation}
\energy{^\mu ^a}=k(4\potential{^b^c^\mu}\torsion{_b_c^a}-\e{^a^\mu}T) \label{29032019hb}
\end{equation}
is interpreted as the gravitational energy-momentum tensor density, and $e=\det(\e{^a_\mu})=\pm\sqrt{-\det g}$ is the determinant of the tetrad field. Since the superpotential is antisymmetric in the last two indices, we see that  $\energy{^\mu^a}$ satisfies the conservation law $\pd{_\mu}\left[e\left(\energy{^\mu^a}+T^{\mu a}\right)\right]=0$.

In the pure-tetrad version of teleparallelism,  the superpotential and the gravitational stress-energy tensor density can be written in the alternative forms\footnote{These expressions are also valid for a teleparallel theory with an arbitrary affine connection as long as we use the TF.} \cite{formiga2020meaning}
\begin{equation}
\potential{_a_b_c}=\frac{1}{2}\rsconnection{_c_a_b}+\rsconnection{^d_d_{[c}}\tensor{\eta}{_{b]a}},\label{20052019fb}
\end{equation}
\begin{align}
\energy{^b_a}=&\ 2k\biggl(2\rsconnection{^c_{[ad]}}\rsconnection{^b_c^d}-2\rsconnection{^b_{[ad]}}\rsconnection{^c_c^d}
\nonumber\\
&-\rsconnection{^c_c_a}\rsconnection{^d_d^b}+\delta^{b}_{a}\rsconnection{^c_{[c|f}}\rsconnection{^d_{|d]}^f}\biggr), \label{20052019hb}
\end{align}
where $\rsconnection{^a_b_c}$ is the Levi-Civita spin connection, which can be given in the form
\begin{equation}
\rsconnection{^a_b_c}=\frac{1}{2}\left(\torsion{_b_c^a}+\torsion{_c_b^a}-\torsion{^a_b_c} \right). \label{16062021a}
\end{equation}
[Note that Eq.~(\ref{16062021a}) holds only in the TF. This is so because the torsion components  coincide with the object of anholonomity only in the TF.]

From the right-hand side of Eq.~(\ref{29032019kb}), one defines the total energy-momentum contained within a three dimensional volume $V$ as \cite{ANDP:ANDP201200272}
\begin{equation}
P^a=\int_V d^3x e\energy{^0^a}+\int_V d^3x eT^{0a}, \label{10082019ab}
\end{equation}
where the first integral on the right-hand side is identified as being the gravitational energy-momentum  and, of course, the second one is the matter energy-momentum. On the other hand, from the left-hand side of Eq.~(\ref{29032019kb}), one finds that
\begin{equation}
P^a=4k\oint_S dS_i e\Sigma^{a0i}.\label{22072018gb1}
\end{equation}

Many authors work with the following definition for the angular momentum
\begin{align}
L^{ab}\equiv -\int_V d^3xM^{ab},\label{30122019cb}
\end{align}
where 
\begin{align}
M^{ab}=-4ke\left(\potential{^a^0^b}-\potential{^b^0^a} \right) \label{30122019db}
\end{align}
is interpreted as the angular momentum density. (In ``vacuum'', it is interpreted as the gravitational angular momentum density; the question whether or not it can still be interpreted as such in the presence of a matter field will not be discussed here.). The quantity $L^{(0)b}$ may be interpreted as the center of mass moment, while $L^{(i)(j)}$ is the angular momentum per se \cite{ANDP:ANDP201200272}. Note that, using Eq.~(\ref{20052019fb}), one can recast Eq.~(\ref{30122019db}) in the alternative form
\begin{align}
M^{ab}=-4ke\left(\rsconnection{^{[ba]}_c}\e{^c^0}+\rsconnection{^d_d^{[b}}\e{^{a]}^0} \right). \label{30122019bb}
\end{align}

The reader may have noticed that, in the pure-tetrad formulation of the TEGR, everything is uniquely determined by the tetrad field and no reference to the Weitzenb\"{o}ck connection is really necessary. In other words, any ambiguity lies in the TF. No one should be surprised by this. After all, TEGR is a tetrad formulation of GR.

The fact that $\energy{^\mu^a}$ can be written in the form given by Eq.~(\ref{20052019hb}) might raise some questions over its  transformation properties. Like the acceleration tensor, $\energy{^\mu^a}$ depends on the frame field, but not on the coordinate system. A deeper discussion about this issue will be published elsewhere. (It can be found in Ref.~\cite{formiga2020meaning}.)

In relation to the field equations written in form given by Eq.~(\ref{29032019kb}), we would like to make the following point. Since many solutions to the equation $\G{^\mu^\nu}=0$ has energy, one can see that the energy-momentum content is not completely on the right-hand side of Einstein equations in the metrical formulation. In Eq.~(\ref{29032019kb}), however, we may hope that the whole energy content can be put on the right-hand side. 

The purpose of this paper is to search for the constraints that fix the TF and make sure that the whole energy-momentum content is on the right-hand side of Eq.~(\ref{29032019kb}). Of course, in a frame that satisfies these constraints, the quantities given by Eqs.~(\ref{29032019hb}) and (\ref{30122019db}) must yield consistent results. In the next section we will discuss under which conditions a frame can be considered an ideal frame to predict the gravitational energy.  

Before we go to the next section, let us clarify what we mean by ``fixing the TF''. Some authors may argue that the TEGR does not distinguish the tetrad fields because Eq.~(\ref{29032019kb}) is covariant under local Lorentz transformations; therefore, an addition principle to distinguish them would be incompatible with TEGR.  However, there is no problem in adding a new principle to the theory when analyzing the gravitational energy. This does not mean that the theory has a preferred frame; it means that some frames are better for the understanding of the gravitational energy, and thus its determination, than others. So, the principle that fixes the TF should be applied only when we are interested in analyzing the gravitational energy and similar quantities.

\section{The ideal frames}\label{25022020a}
An ideal TF is certainly one that allows for a physical interpretation and yields consistent results. They are not necessarily inertial. In fact, if one wishes to describe the world as it is, then noninertial frames of references are the only ones available. However, in teleparallelism, many authors have generally assumed   that the ideal TF should not introduce inertial effects. As we will see, this assumption may not be necessary.

\subsection{Problematic frames versus accelerated ones}\label{08082019a}
Before discussing the ideal frame to interpret quantities that depend on the choice of the TF, let us see an example of the opposite. Suppose that the spacetime metric is $ds^2=dt^2-dx^2-dy^2-dz^2$ everywhere (Minkowski). If we choose as the TF 
\begin{align}
\teta{^a}=(dt,dr,rd\theta,r\sin\theta d\phi), 
\nonumber\\
\e{_a}=(\pd{_t},\pd{_r},\frac{1}{r}\pd{_\theta},\frac{1}{r\sin\theta}\pd{_\phi}),\label{30072019a}
\end{align}
then we will have 
nonvanishing torsion components and energy: $\torsion{^{(2)}_{(1)(2)}}=\torsion{^{(3)}_{(1)(3)}}=1/r$,  $\torsion{^{(3)}_{(2)(3)}}=\cos\theta/(r\sin\theta)$, $\potential{^{(0)01}}=-1/r$, $E=-r$ (within a sphere of radius $r$), etc. It is hard to believe that these results have any physical meaning. Hence, we are led to conclude that this frame is not appropriated to interpret quantities that depend on the choice of the teleparallel frame.

The above result does not mean that the TEGR predicts a nonvanishing gravitational energy when the metric is that of Minkowski. In fact, depending on the TF, the TEGR predicts the opposite.
\begin{theorem}
Let $x^\mu(\tau)$ represent the worldline of a general accelerated observer with proper time $\tau$. Assume that $\e{_{(0)}^\mu}(\tau)=dx^\mu/d\tau$ is the observer's $4$-velocity and $\e{_{(i)}}(\tau)$ is its spatial triad. Then, parallel transport (via Levi-Civita) its frame $\e{_a}$ to all neighboring points on the spacelike hyperplane orthogonal to $\e{_{(0)}}(\tau)$, thus obtaining the reference frame $\e{_a}(\bar{x}^\mu)$, where $\bar{x}^\mu$ are the local coordinates for the observer. The gravitational energy-momentum density  vanishes if $\e{_a}(\bar{x}^\mu)$ is taken as the TF.
\end{theorem}
{\it Proof.} The frame described above can be written in the form \cite{PhysRevD.42.2045,JansenRevistaMexicana}
\begin{align}
\e{_{(0)}}=&\ f\Bigl[\pd{_\tau}+\bigl(\omega_\zeta(\tau)\chi-\omega_\chi(\tau)\zeta \bigr)\pd{_\xi}
\nonumber\\
&+\bigl(\omega_\xi(\tau)\zeta-\omega_\zeta(\tau)\xi \bigr)\pd{_\chi}+\bigl(\omega_\chi(\tau)\xi-\omega_\xi(\tau)\chi \bigr)\pd{_\zeta} \Bigr],
\nonumber\\
 \e{_{(1)}}=&\ \pd{_\xi},\ \e{_{(2)}}=\pd{_\chi},\, \e{_{(3)}}= \pd{_\zeta},
\nonumber\\
f=&\ 1/[1+a^\xi(\tau)\xi+a^\chi(\tau)\chi+a^\zeta(\tau)\zeta], \label{30072019c}
\end{align}
where the ``$a$s'' are the components of the $4$-acceleration of the observer at the origin ($\xi=\chi=\zeta=0$)  and the $\omega$s are the  rotations of its spatial triad $\e{_{(i)}}$ with respect to an inertial frame. The coframe associated with Eq.~(\ref{30072019c}) is
\begin{align}
\teta{^{(0)}}=f^{-1}d\tau,
\nonumber\\
\teta{^{(1)}}=[\omega_\chi(\tau)\zeta-\omega_\zeta(\tau)\chi]d\tau+d\xi,
\nonumber\\
\teta{^{(2)}}=[\omega_\zeta(\tau)\xi-\omega_\xi(\tau)\zeta]d\tau+d\chi,
\nonumber\\
\teta{^{(3)}}=[\omega_\xi(\tau)\chi-\omega_\chi(\tau)\xi]d\tau+d\zeta. \label{30072019b}
\end{align}
Using Eqs.~(\ref{30072019c})-(\ref{30072019b}) in Eq.~(\ref{04102019pb}), we find that 
\begin{align}
\torsion{_{(0)(0)(i)}}=\pd{_i}\ln f,
\nonumber\\
\torsion{_{(1)(0)(2)}}=-\torsion{_{(2)(0)(1)}}=-\omega_\zeta f,
\nonumber\\
\torsion{_{(1)(0)(3)}}=-\torsion{_{(3)(0)(1)}}=\omega_\chi f,
\nonumber\\
\torsion{_{(2)(0)(3)}}=-\torsion{_{(3)(0)(2)}}=-\omega_\xi f, \label{30072019d}
\end{align}
where $\pd{_i}=(\pd{_\xi},\pd{_\chi},\pd{_\zeta})$. Substituting these components into Eq.~(\ref{10112017lb}) gives
\begin{align}
\potential{^{(j)(j)(i)}}=-(1/2)\pd{_i}\ln f \quad \textrm{($i\neq j$)},
\nonumber\\
\potential{^{(0)(1)(2)}}=-(1/2)f\omega_\zeta ,\ \potential{^{(0)(1)(3)}}=(1/2)f\omega_\chi,
\nonumber\\
\potential{^{(0)(2)(3)}}=-(1/2)f\omega_\xi, \label{30072019e}
\end{align}
where the indices $(j)(j)$ above are not summed.

Since all the nonvanishing torsion components have at least one $(0)$ in the second or third indices and  $\potential{^a^{(0)}^c}=-\potential{^a^c^{(0)}}=0$, we conclude that $\torsion{}=\potential{^a^b^c}\torsion{_a_b_c}=0$.
The same argument holds for $\potential{^b^c^d}\torsion{_b_c^{(i)}}$. So, we are left with $\energy{^{(j)(0)}}=4k\potential{^b^c^{(j)}}\torsion{_b_c^{(0)}}$. It is easy to check that this expression also vanishes. Hence, we have $\energy{^\lambda^a}=0$. It is also straightforward to verify that Eq.~(\ref{30122019db}) vanishes, too.

It was proved in Ref.~\cite{PhysRevD.80.064043} that a necessary condition for the superpotential to vanish is that  the spatial rotations also vanish, which is in agreement with Eq.~(\ref{30072019e}). However, at least in Minkowski spacetime, this is not necessary to obtain a vanishing energy.

Why does the frame given by Eq.~(\ref{30072019a}) yield a nonvanishing energy? The answer to this question lies in the orientation of the spatial triad, as pointed out in Ref.~\cite{PhysRevD.99.024022}. For a constant $t$, the spatial vectors change their orientation from point to point. Note, for example, that the observer at $(r,0,0)$ uses $\e{_{(3)}}=\pd{_y}$, while the one at $(r,\pi/2,\pi/2)$ uses $\e{_{(3)}}=-\pd{_x}$. Therefore, two observers at different points use spatial frames that do not have the same orientation for a fixed value of $t$, and the associated rotation has nothing to do with their motion (is an artificial rotation). They are, in this sense, incoherent. On the other hand, when we use the frame given by Eq.~(\ref{30072019c}) we have a set of observers that, for a constant $\tau$, use spatial vectors $\e{_{(i)}}$ that do not have these spurious rotations, since their spatial frame is constructed out of the parallel transport of the spatial frame of the observer at $\xi=\chi=\zeta=0$. Thus, we can say that these observers are coherent accelerated observers (see definition \ref{24092019d}), while the ones that use Eq.~(\ref{30072019a}) are incoherent static observers.

A remark is in order here regarding the so-called inertial connection. It has been customary to state  that the connection given by $\scbare{^a_b_c}=\lorentz{_d^a}\ebar{_b^\alpha}\pd{_\alpha}\lorentz{^d_c}$, where $\ebar{_b^\alpha}$ is related to $\e{_b^\alpha}$ through the Lorentz transformation $\lorentz{^a_b}$,  represents only inertial effects \cite{aldrovandi2012teleparallel,0264-9381-34-14-145013,Krssak2017,PhysRevD.99.024022}. However, the observers who use the frame given by Eq.~(\ref{30072019a}) are at rest in an inertial frame. Hence, the transformation connecting them with the static observers which use $\e{_a}=(\pd{_t},\pd{_x},\pd{_y},\pd{_z})$ has nothing to do with motion, but rather with the incoherence in which the frame (\ref{30072019a}) is  attached to each observer\footnote{In fact, the observers that use the frame given by Eq.~(\ref{30072019a}) are basically the same set of static observers which use $\e{_a}=(\pd{_t},\pd{_x},\pd{_y},\pd{_z})$.}.  This incoherence is due to spatial rotations that are not a result of any motion.  Since inertial effects are related to changes in the state of motion, we must distinguish spatial rotations that are linked to  the motion of an observer and that are shared with all other family members, such as those of Eq.~(\ref{30072019c}), from  rotations that are not related to any motion and  are not shared with the family members. Therefore, it is important to emphasize that the inertial connection can also manifest this incoherence that has nothing to do with motion.

\subsubsection{Spurious static  rotations of the $3$-velocity}\label{19032020a}

The problems related to spurious static rotations of the $3$-velocity of a congruence of curves have never been discussed before. This is an important issue because the gravitational energy-momentum tensor is also sensitive to these rotations. To give an example, consider a  Rindler observer that has an acceleration $a$ in the $x$ direction. We can built a congruence of curves in the plane $(t,x)$ so that the coordinates $t=(1/a+\xi)\sinh(a\tau)$ and $x=(1/a+\xi)\cosh(a\tau)$ represent a set of Rindler observers with accelerations $a/(1+a\xi)$ for each constant value of $\xi$.   For these observers, we have
\begin{align}
\e{_{(0)}}=\frac{x}{\sqrt{x^2-t^2}}\pd{_t}+\frac{t}{\sqrt{x^2-t^2}}\pd{_x},
\nonumber\\
\e{_{(1)}}=\frac{t}{\sqrt{x^2-t^2}}\pd{_t}+\frac{x}{\sqrt{x^2-t^2}}\pd{_x}. \label{24092019a}
\end{align}
So far so good. Now the problem is how to extend this congruence to other directions. There are two very intuitive ways:  dragging them to $y$ and $z$ (the $3$-velocity is parallel transported along $\pd{_y}$ and $\pd{_z}$) or rotating them in such a way that they become radially accelerated observers. In the former case, the observers continue to be accelerated in the $x$-direction. In the latter, however, the $3$-velocity of the observers are rotated in such a way that the congruence acquires spurious static rotations.

The extension of the Rindler congruence and tetrad in the plane $(t,x)$ to the whole spacetime without adding rotations is achieved by adding the coordinates $y,z$ and taking  $\e{_{(2)}}=\pd{_y}$ and $\e{_{(3)}}=\pd{_z}$. This congruence is a particular case of Eq.~(\ref{30072019c}). Thus, the gravitation energy-momentum tensor vanishes. 

On the other hand, the radially accelerated frame is not a particular case of (\ref{30072019c}) and does not yield a vanishing energy. To see this, let us use the spherical coordinate system $x=\rho\sin\theta\cos\phi,\ y=\rho\sin\theta\sin\phi,\ z=\rho\cos\theta$. The radial extension of Eq.~(\ref{24092019a}) is obtained by exchanging  $x$ for $\rho$. This gives
\begin{align}
\e[\widetilde]{_{(0)}}=f(t,\rho)\pd{_t}+g(t,\rho)\pd{_\rho},
\nonumber\\
\e[\widetilde]{_{(1)}}=g(t,\rho)\pd{_t}+f(t,\rho)\pd{_\rho},
\nonumber\\
\e[\widetilde]{_{(2)}}=\frac{1}{\rho}\pd{_\theta},\, \e[\widetilde]{_{(3)}}=\frac{1}{\rho\sin\theta}\pd{_\phi}, \label{24092019b}
\end{align}
where $f(t,\rho)=\rho/\sqrt{\rho^2-t^2}$ and $g(t,\rho)=t/\sqrt{\rho^2-t^2}$; note that we have added $\e[\widetilde]{_{(2)}}$ and  $\e[\widetilde]{_{(3)}}$.

 Now, for convenience, we use the approach of Ref.~\cite{Formiga2021Braz} to simplify the calculations. In this approach, we work with the unit-vector components:
 \begin{align}
\hat{t}^a\equiv \bdeltaup{0}^a,\ \hat{\rho}^a\equiv\sin\theta(\cos\phi\hat{x}^a+\sin\phi\hat{y}^a)+\cos\theta\hat{z}^a,
\nonumber\\
\hat{\theta}^a\equiv\pd{_\theta}\hat{\rho}^a=\cos\theta(\cos\phi\hat{x}^a+\sin\phi\hat{y}^a)-\sin\theta\hat{z}^a,
\nonumber\\
\hat{\phi}^a\equiv\pd{_\phi}\left(\hat{\rho}^a/\sin\theta\right)=-\sin\phi\hat{x}^a+\cos\phi\hat{y}^a,
\nonumber\\
\hat{x}^a=\bdeltaup{1}^a,\ \hat{y}^a=\bdeltaup{2}^a,\ \hat{z}^a=\bdeltaup{3}^a,
 \label{10112019db}
\end{align}
where we define the act of raising and lowering the indices of these components with the Minkowski metric: $\hat{t}_a\equiv \eta_{ab}\hat{t}^b=\bdeltadn{0}_a$,  $\hat{x}_a\equiv \eta_{ab}\hat{x}^b=-\bdeltadn{1}_a$ and so on. Note that  $\hat{t}_a=\hat{t}^a$, $\hat{\rho}_a=-\hat{\rho}^a$, etc. It is also easy to see that $\hat{t}^a\hat{t}_a=1$, $\hat{\rho}^a\hat{\rho}_a=-1$, and so on. 

From Eq.~(\ref{10112019db}), we see that
\begin{align}
\pd{_\theta}\hat{\theta}^a=-\hat{\rho}^a,\ \pd{_\phi}\hat{\theta}^a=\cos\theta\hat{\phi}^a,
\nonumber\\
\pd{_\theta}\hat{\phi}^a=0,\ \pd{_\phi}\hat{\phi}^a=-\sin\theta\hat{\rho}^a-\cos\theta\hat{\theta}^a,
\nonumber\\
\pd{_\mu}\hat{t}^a=0. \label{10112019qb}
\end{align}

 From Eq.~(\ref{10112019db}), we see that Eq.~(\ref{24092019b}) can be written in the form $\e[\widetilde]{_a}=\hat{t}_a\left(f\pd{_t}+g\pd{_\rho} \right)-\hat{x}_a\left(g\pd{_t}+f\pd{_\rho} \right)-(\hat{y}_a/\rho)\pd{_\theta}-[\hat{z}_a/(\rho\sin\theta)]\pd{_\phi}$. In turn, the Lorentz matrix that is used to ``align'' (it does not always work) the spatial triad of a frame with the $x$, $y$, and $z$ directions can be written as
\begin{equation}
\lorentz{_a^b}=\hat{t}_a\hat{t}^b-\hat{\rho}_a\hat{x}^b-\hat{\theta}_a\hat{y}^b-\hat{\phi}_a\hat{z}^b.
\end{equation}
Defining the new frame as $\e[\hat]{_a}=\lorentz{_a^b}\e[\widetilde]{_b}$, we obtain
\begin{align}
\e[\hat]{_a}=&\ \hat{t}_a\left[f(t,\rho)\pd{_t}+g(t,\rho)\pd{_\rho}\right]
\nonumber\\
&-\hat{\rho}_a\left[g(t,\rho)\pd{_t}+f(t,\rho)\pd{_\rho}\right]
\nonumber\\
&-\frac{\hat{\theta}_a}{\rho}\pd{_\theta}-\frac{\hat{\phi}_a}{\rho\sin\theta}\pd{_\phi}. \label{17022020a}
\end{align}

The spatial triad of this frame is not aligned with the Cartesian direction because of the Lorentz contraction along the $\rho$ direction. But, here we point out other issue with this frame: the $3$-velocity $g(t,\rho)\pd{_\rho}$  rotates artificially  when we arbitrarily change the values of $\theta$ and $\phi$. (These rotations have nothing to do with the motion of the accelerated observers.) This set of observers also gives fictitious effects because of this feature. One of these effects is a false energy for gravity when the metric is that of Minkowski and, at the same time, a vanishing momentum. To see that this is the case, note that the frame (\ref{17022020a}) is the same as that of Eq.~(18) in Ref.~\cite{doi:10.1002/andp.201900507} when the black hole mass vanishes. From Eq.~(43) there, we find that $E=16\pi k \rho\left[1-f(t,\rho)\right]$, which vanishes only for $g=0$ ($f=1$); the momentum $P^{(i)}$ is zero only because $f(t,\rho)$ is spherically symmetric. The vanishing of $P^{(i)}$ is not a consequence of the absence of gravity, but rather a consequence of the spherical symmetry of $g(t,\rho)\pd{_\rho}$. If we remove this symmetry, then an artificial gravitational momentum would appear. (We will discuss this issue in more detail in Sec.~\ref{24022020a}.)

The previous explanation motivates the following definition.

\begin{definition} \label{24092019d}
A set of observers whose spatial triad and $3$-velocity do not have spurious static rotations will be called coherent observers.
\end{definition}

This definition includes the observers given by Eq.~(\ref{30072019c}) as a particular case, which means that it does not exclude the type of rotations that their triad has. Are these rotations artificial? There is a 
subtlety regarding the rotation $\vec{\omega}$. If we pretend that the frame is just a mathematical abstraction, as we normally do, we would say that the rotations  produced by $\vec{\omega}$ are artificial. However, frames are space filling system that cares energy and momentum. We might neglect their impact on the background geometry in most cases, but we can never neglect their physical reality in the dynamics of the frame: the origin propagates a long a timelike curve and the triad cannot rotate arbitrarily, otherwise $g_{\tau\tau}$ could change its sign. There is another subtlety worth mentioning. Frames with nonvanishing acceleration (and or rotations) possess non-gravitational interaction embedded in it. Do those interactions get mixed with the gravitation one, preventing $\energy{^\mu^a}$ from giving a consistent prediction for the gravitational energy? If so, is that the reason why we can cancel gravity locally? If the answer to the first question is in the affirmative, then we will have to use frames that are freely falling everywhere, otherwise $\energy{^\mu^a}$ would not properly account for the gravitation energy. Thus, it is not clear yet how  the ideal frame should be defined. But it is certainly a frame adapted to a subset of coherent observers. We need a principle to find this subset. We discuss this principle in the next subsections.

\subsection{The principle of equivalence} \label{24112021a}
In this section, we will discuss the gravitational energy and the ideal frame to analyze it from the perspective of Pauli's version of the principle of equivalence. (Pauli's version basically states that we can cancel gravity locally.) 

 In Ref.~\cite{ANDP:ANDP201200272}  Maluf  states that the vanishing of the gravitational energy-momentum, $P^a$, in a freely falling frame in the Schwarzschild case is in agreement with the principle of equivalence. This is true, however, it is not a necessary condition for this principle to hold. As was  pointed out in \cite{doi:10.1002/andp.201700175}, a freely falling frame is not necessarily a local inertial reference frame. In fact, the vanishing of both $\Pi^{ab}$ and $\energy{^a^b}$ along a curve will depend on the chosen congruence of curves: we can find congruences where these tensors does not vanishing along a freely falling frame (see, e.g., Refs.~\cite{doi:10.1002/andp.201800320} and \cite{PhysRevD.80.064043}) and, more importantly, we can find a congruence that makes them vanish along a particular curve regardless of the acceleration. As far as we are aware, the latter fact  has not been proved yet. Let us state it as a theorem\footnote{This theorem generalizes the statement made below Eq.~(38) in Ref.~\cite{doi:10.1002/andp.201700175}.}.
 
\begin{theorem}\label{30122019a}
 Let $\e{_a}$ be  the proper reference frame (PRF) of an accelerated observer $\cal{O}$. Then, if $\gamma$ is the observer's worldline, we will have $\Pi^{ab}|_\gamma=\energy{^a^b}|_\gamma=0$. 
\end{theorem}
Here, we prove this theorem in two different ways:  Since the frame given in Eq.~(\ref{30072019c}) reproduces the metric of an arbitrary curved spacetime (Levi-Civita curvature) to first order in the observer's proper coordinates [see, e.g., Eq.~(13.71) of Ref.~\cite{Gravitation}], then the vanishing result for both $\Pi^{ab}$ and $\energy{^a^b}$ when $g_{\mu\nu}$ is the Minkowski metric implies that they also vanish along the observer's worldline in a curved spacetime (again, Levi-Civita curvature). A second way of proving this is to write the Levi-Civita affine connection coefficients given by Eqs.~(13.69a)-(13.69b) in Ref.~\cite{Gravitation} in the form $\rsconnection{^a_b_c}|_\gamma=\bdeltadn{0}_b\left( a^a\bdeltadn{0}_c-a_c\bdeltaup{0}^a+\omega^d\varepsilon_{\bzero d e c}\eta^{ea}\right)$, where $a^a$ ($a^{(0)}=0$) is the observer's acceleration, $\omega^d$ the rotations, and $\varepsilon_{abcd}$ the Levi-Civita tensor. Then, by direct substitution into Eq.~(\ref{20052019hb}), we find that $\energy{^a^b}|_\gamma=0$. To check that Eq.~(\ref{30122019bb}) also vanishes, we can use $\e{_a^\mu}|_\gamma=\delta_a^\mu$.

Some comments are in order:
\begin{enumerate}
\item Let  $\energy{^\lambda^a}$ be calculated in the PRF of an accelerated observer.   The vanishing of $\energy{^\lambda^a}$ along this observer's worldline does not mean that the observer's  acceleration  does not affect $\energy{^\lambda^a}$. It does affect, but only outside the observer's worldline:  $\energy{^\lambda^a}$ in the PRF is not  insensitive to the fiducial observer's accelerations to any order.
\item The fact that $\energy{^\lambda^a}$ (when adapted to a PRF) is disconnected from inertial effects locally has two different aspects. On the one hand, it circumvents the criticism that one can conveniently  ``remove'' the gravitational energy density along an observer's worldline by just transforming an arbitrary metric tensor to a Minkowski one locally. (The gravitational energy would already be ``absent'' if the PRF were the ideal frame.). On the other hand,  it prevent us from  simulating a gravitational energy density by means of an accelerated observer. 
\item In principle, there is no need to demand that $\energy{^\lambda^a}$ satisfy the equivalence principle. After all, this will not change the fact that, at the classical level, the TEGR is equivalent to GR. (Particles follow the geodesic of the Levi-Civita connection.) Thus, we can interpret the equivalence principle as the possibility, but not the necessity, of making  $\energy{^\lambda^a}$ vanish in a  local inertial reference frame for a peculiar congruence. (This congruence does not have to be the ``ideal one''.)
\item \label{19022020a} {\bf Argument against the PRFs.} Taking the PRF of a particular observer as the ideal frame does not seem to be a good idea because, from theorem \ref{30122019a}, we see that we can make $\Pi^{ab}$ and $\energy{^a^b}$ vanish along any curve we want. This seems to  make these densities meaningless. In addition, all the consistent calculations of $P^a$ in the literature and in Sec.~\ref{08082019b} here have been made with tetrad fields that are not necessarily PRFs.
\item {\bf Argument in favor of the PRFs.} It is natural to think that, in a Levi-Civita curved spacetime, all physical quantities should be defined locally. Furthermore, only proper coordinates have a real physical meaning. Hence, a physical frame is (or could be) a frame built up with the proper coordinates of one particular observer. In this case, we expect that global quantities such as (\ref{22072018gb1}) and (\ref{30122019cb}) will always give consistent results if they are calculated with the PRFs. With respect to the vanishing of  $\Pi^{ab}$ and $\energy{^a^b}$ along the arbitrary curve $\gamma$, we could argue that this arbitrariness is equivalent to that of the zero-potential-energy configuration, where we are free to choose the zero at any point where the potential does not diverge. (The difference here of course is that the zero is at the origin of the frame, where the fiducial observer is.)
\item \label{22112021a} As was mentioned in the last paragraph of Sec.~(\ref{19032020a}), the presence of non-gravitational interactions translates itself into a non-gravitational force that accelerates and rotates the frame. These interactions may interfere with the gravitational field. It may be the case that only coordinate systems whose labels are adapted to systems with non-gravitational interactions, such as Fermi normal coordinates, which are adapted to rigid rulers that are freely falling only at the origin, can cancel gravity locally\footnote{For examples that prove that gravity cannot always be neglected locally, see Ref.~\cite{doi:10.1119/1.10744}.}. An indication that this may be the case is the impossibility of obtaining the Minkowski metric when using the transverse-traceless gauge of gravitational waves, which is adapted to freely falling test particles, i.e., it is adapted to a frame that is free from non-gravitational interactions and freely falling everywhere, not only at the origin of the coordinates. (See, e.g., pages 19 - 20 of Ref.~\cite{MicheleMaggiore1}.) Theorem \ref{30122019a} may perhaps be the tetrad version of this idea: Since the PRF generalizes the one that is adapted to Fermi normal coordinates to the case where the origin is not freely falling, i.e, the  PRF is also a ``rigid'' fame, we obtain a vanishing gravitational energy along the origin of the PRF. On the other hand, as in the case where coordinates whose positions are marked by freely falling particles do not turn $g_{\mu\nu}$ into $\eta_{\mu\nu}$, a tetrad that is freely falling everywhere and has no artificial properties may never yield a vanishing $\energy{^0^{0}}$ in a Levi-Civita curved spacetime. An example that corroborates this view will be given in section \ref{08082019b}. (If this conjecture turns out to be true, then a frame that is freely falling everywhere with no artificial properties will have a great advantage over the PRF to evaluate and interpret the gravitational energy and momentum.) 
\end{enumerate}

It is worth noting here that the frame used by Maluf et. al. in Ref.~\cite{Maluf_2007} to obtain $E=0$ is not the PRF of any of the freely falling observers that are not at spatial infinity. (This is clear because $\energy{^a^b}$ vanishes only when $r \to \infty$ [see, e.g., Eq.~(58) of Ref.~\cite{doi:10.1002/andp.201900507}].) Although this frame is a freely falling frame everywhere, it has an explicit rotation in the $3$-velocity and the time coordinate is not adapted to the observers' point of view. These problems were circumvented in Ref.~\cite{Gonalves2021} by using a frame that is freely falling everywhere and adapted to the Novikov coordinates. Hence, the result obtained by Maluf et. al. cannot be used to disregard either the PRF or frames that are freely falling everywhere.

\subsection{The calculation of energy and momentum}\label{24022020a}
Throughout this paper, we will assume that the PRF of a particular observer is not the ideal frame to interpret the gravitational energy. (We leave the opposite assumption for a future work.) Hence, we have to find a different way to approach this problem. Since our focus is on energy and  momentum, we now discuss how these quantities are calculated.

In general, to calculate the energy of a system of particles or fields, we  consider a set of observers whose velocities are the same: they are either at rest or share the same velocity. For example, the energy of an electron that is moving with respect to a set of inertial observers is the famous expression $\gamma mc^2$, where $\gamma$ and $m$ are the Lorentz factor and the electron mass, respectively. In turn, in the rest frame, we have $mc^2$. The common factor between these two frames is the uniformity of the observer's velocity. It would not make sense to say that the energy of this electron with respect to a set of observers whose velocities are not the same is $\gamma(v)mc^2$, because there would not be a unique (or effective) velocity $\vec{v}$. This becomes even more pronounced when we are calculating the energy of a field that fills the whole spacetime, where, in principle, we  need to include the ``measurements'' of every single observer.

Let's assume, for example, that we have a set of  observers that moves in one direction and another set that moves in the opposite direction with respect to a static gravitational field; one set covers half of the spacetime and the other covers the other half. In this case, if we mixed the calculation of the gravitational momentum, we would get a result that is hard to interpret because the momentum density measured by one set would be negative while the one measured by the other set would be positive; the net effect could even be zero. [see, e.g., Eq.~(50) in Ref.~\cite{Formiga2021Braz}; see also Eq.~(43) in Ref.~\cite{doi:10.1002/andp.201900507}.] It is clear, then, that we cannot use two observers that have arbitrarily different velocities in the same congruence. 

These arguments suggest that we should take a congruence where all the curves had the same constant velocity with respect to a special type of frame that we do not know yet. The problem with that idea is that it would exclude accelerated observers. We could relax this requirement by allowing the velocities of the observers to change only in time, but remain the same for all curves at a fixed coordinate time (or perhaps proper time), as is done by Maluf in Ref.~\cite{Maluf:2004vc}. (It is worth noting that this choice excludes interesting observers such as the Rindler ones\footnote{These observers can be seen as a rigid rod that accelerates in a fixed direction in Minkowski spacetime.}.) 

It is reasonable to think that the frame that is free from non-gravitational effects in a curved spacetime is the closest we can get to a global Lorentz frame. In such a frame, at least in principle, the non-uniformity of the frame is caused by the gravitational energy and momentum only. Therefore, if this frame exists, then it gives the right prediction for a pure gravitational energy, with no interference of other types of energies.

We can say with a fair degree of confidence that if a frame $\e{_a}$ is the closest we can get to a inertial frame, then when all  physical parameters (mass, electric charge etc) vanish, we must have $\rsconnection{^a_b_c}=0$ [see, e.g., Eqs.~(\ref{20052019fb})-(\ref{20052019hb})]. This is  equivalent to demanding that the frame be a holonomic frame in Minkowski: Let $\tensor{\Omega}{^a_b_c}$ be the object of anholonomity.  From $\rsconnection{^a_b_c}=(1/2)\left(\tensor{\Omega}{_b_c^a}+\tensor{\Omega}{_c_b^a}-\tensor{\Omega}{^a_b_c} \right)$, we see that if $\nonholon{^a_b_c}=0$, then $\rsconnection{^a_b_c}=0$. On the other hand, using $\nonholon{^a_b_c}=\rsconnection{^a_c_b}-\rsconnection{^a_b_c}$, we see that $\nonholon{^a_b_c}$ also vanishes when $\rsconnection{^a_b_c}=0$. This type of tetrad as been named {\it proper tetrad} (PT) by Tiago {\it et al} \cite{PhysRevD.80.064043}. (The word ``proper'' here has nothing to do with either ``proper coordinates'' or a PRF.) The motivation for this tetrad is the following theorem. 

\begin{theorem}\label{12032020a}
Let $\e{_a}$ be a tetrad field. This tetrad is holonomic if and only if it is a global  inertial frame of reference.
\end{theorem}
{\it Proof.} Taking $\nonholon{^a_b_c}=0$ ($\rsconnection{^a_\mu_{b}}=0$), we immediately see that, for any curve $x^\mu(\tau)$, we have $\mathring{D}\e{_b}/d\tau=(dx^\mu/d\tau)\rsconnection{^a_\mu_{b}}\e{_a}=0$. Furthermore, since the frame is holonomic, there must exist a coordinate system  ($t,x,y,z$) where the coframe is $\teta{^a}=(dt,dx,dy,dz)$ everywhere. On the other hand, if we assume that the frame is inertial, we will have $\rsconnection{^a_b_c}=0 \to \nonholon{^a_b_c}=0$. This proof is similar to that of Ref.~\cite{Mashhoonbook2017}, p. 83. Hence, a PT becomes a global inertial frame of reference when gravity is absent\footnote{Note that we are adopting the standard view, due to Synge, that gravity is the curvature of the Levi-Civita connection. For an interesting discussion of the difference between Einstein's and Synge's view, see section 11 of Ref.~\cite{NORTON1985203}. (See also section 7.3 of Ref.~\cite{Norton_1993}.)}.

As is well known, the PT is not uniquely defined by the metric, and the description of the gravitational energy momentum  changes as we change the PT. This leads to different predictions of energy, as exemplified in Sec.~\ref{08082019b}. 

Some authors argue that this frame does not include inertial effects (see, e.g., Sec. IV B of Ref.~\cite{PhysRevD.80.064043}). However, this is not true. As discussed in Ref.~\cite{doi:10.1002/andp.201900507}, many types of accelerated frames can be set to be a PT: we just have to change the external forces  that accelerate them in a way that the $4$-acceleration vanishes in the absence of gravity (vanishing of the Levi-Civita curvature).  

To remove this  ambiguity, it is necessary to choose only one of the possible PTs. Here, we  demand that the PT  satisfy the conditions\footnote{Equation (\ref{01062019bb}) is known as ``time gauge'' \cite{PhysRev.130.1253}.  }
\begin{equation}
\e{_{(i)}^0}=0, \label{01062019bb}
\end{equation}
\begin{equation}
\e{_{(i)j}}=\e{_{(j)i}},\ (i,j=1,2,3)  \label{01062019b2}
\end{equation}
for a particular coordinate system. 

It is very likely that these conditions do not define a unique coordinate system, but we will assume that they do. In case they do not fix the tetrad, we must search for  the closest tetrad to a global Lorentz frame.

It is worth noting that, in Ref.~\cite{PhysRevD.65.124001}, Eqs.~(4.4a)-(4.4b), Maluf et al. present these conditions demanding that the components of the tetrad be written in a Cartesian coordinate system. However, not all spacetimes admit such a coordinate system, and some admit more than one coordinate system that looks like Cartesian. That is the reason why here we avoid saying that the components given by Eq.~(\ref{01062019b2}) are written in a Cartesian basis.

 Maluf et al.  also state that this frame ``establish a unique reference space-time that is neither related by a boost transformation, nor rotating with respect to the physical space-time.'' If this is true for the PT satisfying the conditions given by Eqs.~(\ref{01062019bb}) and (\ref{01062019b2}), then this frame will be the closest we can get to a global Lorentz frame in a curved spacetime. In addition, we could interpret it as the only frame that carries only gravitational effects. 
 
The following definitions will be of great value for the next sections. 
\begin{definition}\label{15022020a}
 A PT satisfying the conditions given by Eqs.~(\ref{01062019bb}) and (\ref{01062019b2}) will be called the fundamental frame (FF). 
\end{definition}  
\begin{definition}\label{15022020b1}
Let $e{_a}$ be the FF of a certain spacetime. The frame given by $\lorentz{_a^b}e{_b}$ (or $\lorentz{^b_a}e{_b}$) will be called an ideal frame\footnote{This is not a preferred frame to solve the field equations. It is only supposed to be an ideal frame to predict the gravitational energy. } (IF) if   the transformation $\lorentz{^b_a}$  does not include artificial rotations and depends only on the time coordinate of the coordinate system that allows us to write the FF in the form given by Eqs.~(\ref{01062019bb}) and (\ref{01062019b2}).
\end{definition}

It is clear that the  FF is an IF, but it is not clear if we should fix it as the TF to interpret the gravitational energy.  The discussion whether the FF should be the TF can be found in Sec.~ IX A of Ref.~\cite{formiga2020meaning}. 

It is also clear that, although the definition \ref{15022020b1} restricts the possible congruences, an IF can still be adapted to any accelerated worldline: there is no restriction on the fiducial observer, only on the congruence. An example of an IF that accelerates along the direction defined by a Cartesian coordinate in the $pp$-wave spacetime is given in section X B of Ref.~\cite{formiga2020meaning}.

For some time, the idea of a reference frame  was ambiguous because it was mixed with the idea of a coordinate system. Nowadays, there is a clear distinction between them: coordinates are used to assign four numbers to events in spacetime, while a frame of reference is seen as a space filling system of hypothetical instruments with arbitrary velocities\footnote{For more details, see section 6.3 of Ref.~\cite{Norton_1993}}. There is no restriction either on their spatial triads or on the congruence. This allows one to take any congruence of timelike curves as a reference frame without worrying  about the orientation of their spatial frames or $3$-velocity. Since in GR the focus is only on quantities that does not depend on the choice of the frame, this definition is not a problem for GR. However, for quantities and theories that do depend on frames, it seems that we need a more restricted definition. The definition \ref{15022020b1} is an attempt to refine the idea  of a physical frame of reference.

\section{Gravitational waves}\label{08082019b} 
In this section we will see the consistence of the FF by comparing the predictions of three different gravitational energy-momenta for $pp$-wave spacetimes.

Plane-fronted gravitational waves (pp-waves) traveling along the z direction can be described by the line element
\begin{align}
ds^2=&\ (1+H)dT^2-dX^2-dY^2-(1-H)dZ^2
\nonumber\\
&-2HdT dZ, \label{01062019a}
\end{align}
where $H=H(u,X,Y)$,  $u=T-Z$, and $\frac{\pd{^2}H}{\pd{}X^2}+\frac{\pd{^2}H}{\pd{}Y^2}=0$. Here, the conventions are slightly different from those used by Maluf and Ulhoa in Ref.~\cite{PhysRevD.78.047502}: the signature is different, $u$ is also different, and $H$ here is $-1/2$ times that of Ref.~\cite{PhysRevD.78.047502}. 

To calculate $P^a$, Maluf and Ulhoa used  
\begin{equation}
\teta{^a}=(AdT+BdZ,dX,dY,CdZ),\label{02062019b}
\end{equation}
where $A=(1+H)^{1/2}$, $AB=-H$, and $AC=1$. Their result was 
\begin{equation}
P^{(0)}=P^{(3)}=-\frac{k}{2}\int_Vd^3x\frac{(\pd{_i}H)^2}{(g_{00})^{3/2}}, \label{01062019c}
\end{equation}
where $(\pd{_i}H)^2\equiv (\pd{_X}H)^2+(\pd{_Y}H)^2$. Although this expression does not seem to be problematic, it cannot reproduce the energy of linearized gravitation waves predicted in  the context of GR (see Ref.~\cite{doi:10.1002/andp.201800320} for more details). As we will see later, the frame (\ref{02062019b}) is not an IF.

On the other hand, in Ref.~\cite{Obukhov_2009}, Obukhov  et al. used\footnote{In adapting the notation of Ref.~\cite{Obukhov_2009} to the one used here, we have made the following changes from their notation to ours: $h\to 4H$, $\sigma\to u/2$, $\rho\to (T+Z)/2$, $z\to X$, and $y\to Y$.}  
\begin{align}
\teta[\widehat]{^{(0)}}=(1+\frac{H}{2})dT-\frac{H}{2}dZ,
\nonumber\\
\teta[\widehat]{^{(1)}}=-\frac{H}{2}dT-(1-\frac{H}{2})dZ,
\nonumber\\
\teta[\widehat]{^{(2)}}=dY,\ \teta[\widehat]{^{(3)}}=dX \label{02062019c}
\end{align}
 to obtain $\tensor{\widehat t}{^a_b}=0$, which yields ${\widehat P}^a=0$. This prediction is inconsistent because we know  that gravitational waves do have energy and momentum. As in the previous case, the frame ($\ref{02062019c}$) is not an IF either.

A third prediction for the gravitational energy and momentum of GWs was made in Ref.~\cite{doi:10.1002/andp.201800320}. There, the frame used was
\begin{equation}
\tetabar{^a}=(dt,f(u)dx,g(u)dy,dz),\label{02062019d1}
\end{equation}
where in this case, the $pp$-waves has only the $+$ polarization, which means that the corresponding spacetime $ds^2=dt^2-f(u)^2dx^2-g(u)^2dy^2-dz^2$ is a particular case of Eq.~(\ref{01062019a}) for $H=h(u)(X^2-Y^2)$, where $h(u)=-(1/f)(d^2f/du^2)=(1/g)(d^2g/du^2)$; the coordinates $(T,X,Y,Z)$ are related to $(t,x,y,z)$ through
\begin{align}
T=t+\frac{1}{2}\left[x^2f(u)\frac{df}{du}+y^2g(u)\frac{dg}{du}\right],
\nonumber\\
Z=z+\frac{1}{2}\left[x^2f(u)\frac{df}{du}+y^2g(u)\frac{dg}{du}\right],
\nonumber\\
X=f(u)x,\ Y=g(u)y. \label{02062019e}
\end{align}
Note that $u=T-Z=t-z$. The gravitational energy-momentum density predicted by the tetrad given by Eq.~(\ref{02062019d1}) is
\begin{equation}
 \energy[\bar]{^0^{(0)}}=\energy[\bar]{^3^{(0)}}=\energy[\bar]{^0^{(3)}}=\energy[\bar]{^3^{(3)}}=-\frac{c^2}{4\pi G fg}\frac{df}{du}\frac{dg}{du}. \label{02062019f}
\end{equation}

For $f(u)\approx (1-h_{11})^{1/2}$ and $g(u)\approx (1+h_{11})^{1/2}$ with $h_{11}$ small, we get $\energy[\bar]{^0^{(0)}}\approx c^2\left(dh_{11}/du\right)^2/(16\pi G)$. This prediction gives a nonzero and positive energy density, which is clearly incompatible with Eq.~(\ref{01062019c}) and the vanishing result predicted by Obukhov et al..  In short, we have three incomparable predictions. Since the expression (\ref{02062019f}) reproduces the expression for GWs  in GR \cite{doi:10.1002/andp.201800320}, it is plausible to assume that it is the right description of the gravitational energy. 

The incompatibility of these predictions is in the assumption that they describe the same physical quantity, namely, the gravitational energy of $pp$-waves. However, if they describe different systems, then they might be reconciled. But even in this case, Eq.~(\ref{02062019f}) is more likely to be reproducing the pure energy of $pp$-waves, since the frame (\ref{02062019d1}) is a system free from non-gravitational interactions, a system of freely falling particles (see Sec.~\ref{241120221a}).

Comparing Eqs.~(\ref{02062019b}), (\ref{02062019c}), and (\ref{02062019d1}) with Eqs.~(\ref{01062019bb}) and Eq.~(\ref{01062019b2}), one can easily check  that the only PT that satisfies the conditions in Eqs.~(\ref{01062019bb}) and (\ref{01062019b2}) is\footnote{Note in Eq.~(18) of Ref.~\cite{PhysRevD.78.047502} that $\e{_\bthree^0}\neq 0$. From Eq.~(\ref{02062019c}) one can also see that $\e{_\bone^0}\neq 0$.} $\ebar{_a}$. To be more precise, $\ebar{_a}$ is the FF of $pp$-waves with $+$ polarization. (Actually, we need to impose an additional constraint, as discussed below.)

It is important to say that the frame $\ebar{_a}$ is not a FF for any possible values of $f$ and $g$. This happens because the coordinate system where the conditions given by Eqs~(\ref{01062019bb}) and (\ref{01062019b2}) are satisfied is not fixed yet (we still have a gauge freedom). To be a FF, the frame $\ebar{_a}$ must also be a PT (for more details, see the paragraph before theorem~\ref{12032020a}). This happens only if $f$ and $g$ are constant when $H=0$. To see this, we need to remember that $h(u)=-(1/f)(d^2f/du^2)=(1/g)(d^2g/du^2)$, where $h(u)$ is the amplitude of the gravitational wave. Since  the curvature tensor of the Levi-Civita connection vanishes when $h=0$ (see, e.g., Chap. 4 of Ref.~\cite{griffiths1991colliding}), the absence of gravity (Synge's view) implies  that $f=c_1u+c_2$ and $g=c_3u+c_4$. However, by substituting these expressions     into Eq.~(\ref{02062019d1}), we obtain a frame that is not a PT for $c_1,c_3\neq 0$. Therefore, we must also demand that $f$ and $g$ be constant when $h=0$ ($H=0$). [Note that, if $f$ and $g$ were not constant in this case, Eq.~(\ref{02062019f}) would give a false gravitational energy.] 

As is clear from Refs.~\cite{PhysRevD.78.047502,Obukhov_2009,doi:10.1002/andp.201800320}, the frames~(\ref{02062019b}), (\ref{02062019c}), and (\ref{02062019d1}) (with $f$ and $g$ constant for $H=0$) satisfy the condition $\rsconnection{^a_b_c}=0$ when $H=0$, i.e., they are PTs. Hence, demanding that the TF be a PT is not enough for a suitable description of the gravitational energy. We need additional conditions to obtain a frame that will identify the gravitational energy in a consistent way. We hope that  Eqs~(\ref{01062019bb})-(\ref{01062019b2}) can fulfill this requirement.

\subsection{The problems with $\e[\widehat]{_a}$ and $\e{_a}$}
Neither  $\e[\widehat]{_a}$ nor $\e{_a}$ satisfy the time gauge, Eq.~(\ref{01062019bb}). Hence  they are not the FF of the spacetime given by Eq.~(\ref{01062019a}). However, this is not sufficient to exclude them because they could be IFs or at least something close to that. To see that they are not IFs, we need to write them in the form  $\e[\widehat]{_a}= \lorentz{^b_a}\ebar{_b}$ and show that $\lorentz{^b_a}$  is not uniform in space.

Let us work only with the $+$ polarization. From Eq.~(\ref{02062019e}), we see that
\begin{align}
dT=(1+F)dt+xff' dx+ygg' dy-Fdz,
\nonumber\\ 
dZ=Fdt+xff'dx+ygg' dy+(1-F)dz, \label{06012020f}
\end{align} 
 where $ F\equiv (1/2)\left\{x^2\left[ (f')^2+ff'' \right]+y^2\left[ (g')^2+gg'' \right] \right\}$ and $f'=df/du$. Substituting Eq.~(\ref{06012020f}) into (\ref{02062019c}), we obtain $ \e[\widehat]{^\bzero_\mu}=(1+H/2+F)\delta^0_\mu+xff'\delta^1_\mu+ygg'\delta^2_\mu-(H/2+F)\delta^3_\mu$, which is the component of  $\teta[\widehat]{^{(0)}}$ written in the coordinate basis $(dt,dx,dy,dz)$. Lowering the index $(0)$,  raising $\mu$ with  $ g^{\mu\nu}=\delta^\mu_0\delta^\nu_0-(1/f^2)\delta^\mu_1\delta^\nu_1-(1/g^2)\delta^\mu_2\delta^\nu_2-\delta^\mu_3\delta^\nu_3$, and contracting the resultant components with $\pd{_\nu}=(\pd{_t},\pd{_x},\pd{_y},\pd{_z})$, we arrive at
 \begin{align}
 \e[\widehat]{_\bzero}=(1+\frac{H}{2}+F)\ebar{_\bzero}-xf'\ebar{_\bone}-yg'\ebar{_\btwo}+(\frac{H}{2}+F)\ebar{_\bthree}, \label{06012020g}
 \end{align}
where $\ebar{_a}$ is the FF, whose coframe is given by  Eq.~(\ref{02062019d1}). It is clear that $\e[\widehat]{_a}\neq \lorentz{^b_a}(t)\ebar{_b}$, i.e., the frame $\e[\widehat]{_a}$ is not an ideal frame. Besides, it is hard to believe that the vector (\ref{06012020g}) is adapted to a congruence that has  some sort of coherence.

A similar procedure yields 
\begin{align}
\e{_\bzero}=&\ \left[ A+(A+B)F\right]\ebar{_\bzero}-(A+B)\left(xf'\ebar{_\bone}+yg'\ebar{_\btwo} \right)
\nonumber\\
&-\left[B-(A+B)F \right]\ebar{_\bthree}, \label{06012020h}
\end{align} 
which suffers  from the same problems as $\e[\widehat]{_a}$.

\subsection{Freely falling frame}\label{241120221a}
It is interesting to note that the fundamental frame $\ebar{_a}$ is a freely ``falling'' frame. To see this, let us recall that the acceleration tensor $\phi_{ab}$ and the Levi-Civita spin connection $\rsconnection{^a_b_c}$ are related to each other through $\phi_{ab}= \rsconnection{_b_\bzero_a}$. From Eq.~(22) in Ref.~\cite{doi:10.1002/andp.201800320} and Eq.~(\ref{10112019db}) here, we see that the torsion components in the basis (\ref{02062019d1}) can be written as $ \torsionbar{_a_b_c}=-2\left[(f'/f)\hat{x}_a\hat{x}_{[c}+(g'/g)\hat{y}_a\hat{y}_{[c}\right]\left(\hat{t}_{b]}+\hat{z}_{b]}\right)$. (Recall that the prime denotes the derivative with respect to $u$.)   
 Substituting these components into Eq.~(\ref{16062021a}) and using $\phi_{ab}= \rsconnection{_b_\bzero_a}$ gives $\phi_{ab}=0$.

Since $\ebar{_\bzero^\mu}=(1,0,0,0)$ and $\phi_{ab}=0$, the FF frame of the $pp$-waves (for $+$ polarization) is  adapted to a sort of ``absolute'' static observes.  Furthermore, since $t$ is the proper time of these observers, their clocks are synchronized all the time. It is an interesting fact that a spacetime that is not static admits such a type of frame, and this happens to be the FF. 
 
An important point to emphasize here is that the frame $\ebar{_a}$ is a system free from non-gravitational interactions: it is a system of freely falling test particles,  not a system of rigid rulers, such as the PRF. Thus, it is a pure gravitational system with only gravitational energy, and whose energy density cannot be zero unless the curvature tensor of the Levi-Civita connection vanishes. This gives support to the conjecture that only frames with non-gravitational interactions can cancel the gravitational energy locally, as discussed in comment \ref{22112021a} of section \ref{24112021a}.  

\section{Absolute vacuum}\label{23032020a}
The modern view of vacuum is that of quantum mechanics. In this view the vacuum is not empty, not unique, not stable, and can be seen as the remains that cannot be removed from space. On the other hand, in the classical point of view, there are two types of vacuum: a relative vacuum (RV) and an absolute vacuum (AV). The relative concept is related to the field equations of a particular theory. For instance, the vacuum of Maxwell's equations is different from that of Einstein's equations. (The former is characterized by the absence of electric charges, while the latter is characterized by the absence of any form of energy that is not gravitational.) With respect to the AV, we have the idealized concept of ``nothing'', i.e., the absence  of any form of energy and momentum. Here, we will not discuss which point of view is right. Clearly, the quantum view is incompatible with an AV. However, since the TEGR is a classical theory, we will work with the classical view here.

It would be interesting to have a gravitational theory with an AV, at least at the classical level. However, the main theory of gravity, GR, admits solutions with a curved spacetime even in its vacuum  \cite{StephaniBook}. The reason why this happens is clear: the gravitational field  does have energy and momentum on its own. So, to have a well defined AV, we need to include the gravitational energy-momentum density into the concept of vacuum.

Since the TEGR gives us the ability to identify the gravitational energy-momentum density, it is natural to speculate whether the Minkowski spacetime is the only possible solution of Eq.~(\ref{29032019kb})  when its right-hand side vanishes. This leads us to the following postulate:

\begin{postulate}\label{10032020a}
Let $\e{_a}$ be  a FF (see definition \ref{15022020a}). The metric tensor must be that of Minkowski spacetime if and only if  $\energy{^\mu^a}=T^{\mu a}=0$. 
\end{postulate}

It is clear from definition \ref{15022020a} that if $\e{_a}$ is the FF of the Minkowski spacetime, then $\rsconnection{^a_b_c}=0$. Furthermore, from Eqs.~(\ref{20052019fb}) and (\ref{20052019hb}), we also have $\pd{_\alpha}\left( e\potential{^a^\mu^\alpha}\right)=\energy{^\mu^a}=0$ ($T^{\mu a}=0$, of course). What is not clear is if $\energy{^\mu^a}=T^{\mu a}=0$ implies $g_{\mu\nu}=\eta_{\mu\nu}$. To see that this is not as simple as it may appear, notice that for the frame given by Eq.~(\ref{02062019c}), which is not a FF, we have $\energy{^\mu^a}=T^{\mu a}=0$ in a Levi-Civita curved spacetime (a spacetime with energy).

In the early days of GR, Einstein believed that in the absence of ``matter'' the right theory of gravity should not have any solution \cite{AbrahamPais}.  Initially, he thought that the problem\footnote{It is a problem only if one believes in Mach's principle.} with the vacuum solutions in the absence of matter could be solved by adding the so-called cosmological constant. However, de Sitter solution showed that the problem persists. Perhaps, in the context of the TEGR, Mach's principle might be reconciled with solutions such as the de Sitter spacetime because such spacetimes are not empty, they have gravitational energy-momentum densities.

\section{Discussion and conclusions}
We have discussed the role played by the teleparallel frame in calculating the energy of the gravitational field. It was shown that, in addition to the problems with the rotations of the spatial triad, the artificial rotations of the congruence can also yield spurious results for the gravitational energy, momentum, and angular momentum. Therefore, not all set of observers can be used to interpret these kind of objects. In search for an ideal set of frames to interpret these objects, we must impose some constraints on the timelike congruence. Note, however, that these constraints are not necessarily a restriction on the worldline of the fiducial observer at the origin of the frame, but rather a restriction on the collection of observers. In other words,  we cannot use two arbitrarily accelerated observers in the same set, but any observer can be used to build an acceptable congruence. 
 
We have proved that the proper reference frame yields a vanishing gravitational energy-momentum density (also the angular momentum density) along the fiducial observer's worldline, regardless of the accelerations of the frame. (This result, theorem~\ref{30122019a}, generalizes that of Ref.~\cite{doi:10.1002/andp.201700175}). An immediate consequence of this theorem is that we can make both $\energy{^\mu^\nu}$ and $M^{ab}$ vanish along any timelike curve we want, we just have to take the PRF adapted to this curve as the TF. This ambiguity was used as an argument  against the PRF, although we have presented some arguments in favor of it as well. We have also argued against the requirement that $\energy{^\mu^\nu}$ and $M^{ab}$ vanish in a local inertial reference frame because the principle of equivalence is connected to the motion of particles; it is not strictly connected to the energy of the gravitational field. After all, in the TEGR, particles will follow the geodesics of the Levi-Civita connection even if $\energy{^\mu^\nu}$ and $M^{ab}$ do not vanish in the particles' local inertial frames.

From the discussion in Sec.~\ref{24022020a}, it is clear that the measurement/calculation of the energy of a field demands a sense of uniformity. (In general, the observers share the same velocity with respect to a global inertial frame.) Since a global inertial frame cannot be realized in a curved spacetime (Levi-Civita curvature), we have postulated that the PT (see theorem \ref{12032020a} and the text below it)  satisfying Eqs.~(\ref{01062019bb})-(\ref{01062019b2}), which we call the fundamental frame, has only gravitational effects, i.e., its lack of uniformity is due to gravity only. (It is possible that only a frame adapted to a system of freely falling test particles can yield a pure gravitational energy, in which case the conditions given by Eqs.~(\ref{01062019bb})-(\ref{01062019b2}) may not always be possible.) Then, we have implicitly defined a uniform set of observers in  a curved spacetime as being the set of all observers whose velocities with respect to the local FF is the same at an instant $t$, where $t$ is the time coordinate where Eqs.~(\ref{01062019bb})-(\ref{01062019b2})  hold. 

To be able to describe the gravitational energy along the curve of an arbitrary accelerated observer, we have defined the IFs (see definition~\ref{15022020b1}), which are adapted to the uniform observers.  (In this definition, we have excluded all artificial rotations.) We have also discussed the meaning of the rotations in Eq.~(\ref{30072019c}) and came to the conclusion that we should treat them as non-artificial.

To show the consistence of the FF, the gravitational energy predicted by tree different frames were compared. The only one that yields a nonvanishing gravitational energy that can reproduce the well-known positive energy density of pp-waves in the weak field approximation is the FF. It is also possible to build an IF that accelerates along the same direction as the wave and gives consistent results. (This demonstration will be published elsewhere.)  For more  consistent predictions of IFs, see Refs.~\cite{formiga2020meaning}.

We have seen that, unlike GR, the TEGR may have a well-defined concept of absolute vacuum.  This is a important issue because it might help us to understand the relation between the structure of spacetime and the concept of energy and momentum.

This article is not the final version of the ``right'' approach to the teleparallel frame problem, because there are still some issues open to debate. One of these issues is the proof that there is only one PT (up to global spatial rotations, of course) that satisfies Eqs.~(\ref{01062019bb}) and (\ref{01062019b2}). (Furthermore, we must also test the consistence of IFs in more general spacetimes.) Another issue is whether we must work only with frames that are pure gravitational systems. If that is the case, then we will probably have to give up at least one of the conditions in Eqs.~(\ref{01062019bb}) and (\ref{01062019b2}) for some spacetimes.
 
\chapter{The meaning of torsion in teleparallel theories}\label{paper2}

The ambiguity of the Weitzenb\"{o}ck connection is investigated from the perspective of frames. In doing so, 
the pure-tetrad formalism and the formalism with an affine connection are compared with each other; it is shown that the ambiguity in the affine connection is equivalent to that of the frame. The role played by the regularization procedure is discussed, and it is argued that the frame approach gives better results than any regularization procedure, including those that make use of a suitable choice of the affine connection.  It is also argued that, for consistency, the Weitzenb\"{o}ck torsion must be either manifesting only gravitational effects or gravitational effects plus accelerations. Two possible interpretations for the torsion tensor are present, and an application to a frame that accelerates along the same direction as that of a $pp$-wave is used to show the consistence of these interpretations.  Finally, some possible solutions to well-known problems of the $f(T)$ theories are proposed.

\section{Introduction}
The use of torsion in theories of gravity dates back to Einstein's attempt to unify electromagnetic and gravitational interactions \cite{SAUER2006399,Gasperini}.  In general, torsion is either related to the intrinsic angular momentum of matter or is used to account for a distant parallelism. The most well-known theories with torsion are the Einstein-Cartan and the Teleparallel Equivalent of General Relativity (TEGR). (The former relates the torsion tensor to the spin density, while in the latter torsion is used to establish an absolute parallelism.)

The relevance of teleparallel theories is undeniable. These theories have attracted the attention of several researchers for many different reasons. They have been used to study the localization of  the gravitational energy \cite{ANDP:ANDP201200272}, to explain the late-time accelerated expansion of the universe without resorting to negative pressure \cite{PhysRevD.85.124007,Cai_2016,PhysRevD.79.124019}, to solve the particle horizon problem without resorting to an inflation field \cite{PhysRevD.75.084031}, to analyze the black hole entropy \cite{PhysRevD.100.124040,Miao_2011}; because of the expectation that the conformal invariance was important at early stages of the Universe and still is on small scales,  they have also been used to formulate conformal theories of gravity \cite{ANDP:ANDP201200037,Momeni2014,Silva2016,0264-9381-34-11-115012,doi:10.1142/S0217732317501139,PhysRevD.99.064047}.  There is also the claim that they are more suitable to quantization and  unification \cite{aldrovandi2012teleparallel}, since they  can ``dispense'' with the weak equivalence principle and can  be viewed as a gauge theory of the translation group \cite{PhysRevD.14.2521,PhysRevD.14.3335,HAYASHI1977441,PhysRevD.19.3524} (for a recent discussion on the gauge approach, see Refs.~\cite{PhysRevD.99.064006,universe5060139,PhysRevD.101.024059}).

However, in teleparallel theories, it is nor clear what torsion really is, physically speaking. Although it is argued that the torsion tensor describes gravity, the ambiguity of this description prevent us from finding a definite meaning for the Weitzenb\"{o}ck torsion.  In Ref.~\cite{PhysRevD.80.044036}, for example, Maluf et al. argued that the Weitzenb\"{o}ck torsion is related to the gravitational acceleration. Other authors, such as Aldrovandi and Pereira \cite{aldrovandi2012teleparallel}, focus more on the separation between inertia and gravitation, i.e., the torsion (more precisely, the contorsion) tensor represents the purely gravitational force. However, this separation is not unique because the Weitzenb\"{o}ck connection is not unique. In fact, the problem is worse: the theory allows one to choose  Weitzenb\"{o}ck connections that leads to  meaningless torsions. Some attempts to solve this problem have been made (see, e.g., the regularization procedures in Refs.~\cite{ANDP:ANDP201200272,PhysRevD.73.124017,Kr_k_2019}), but so far no definite solution has been given. In this article we discuss the meaning of torsion in teleparallel theories by analyzing the role played by the frame where the Weitzenb\"{o}ck  connection vanish. 

We begin with a brief review of Riemann-Cartan and Weitzenb\"{o}ck spacetimes in Sec.~\ref{14122017a} and \ref{22032020a}. The teleparallel frame problem is discussed in Sec.~\ref{05082019c} with the help of a notation that allows us to use both the pure-tetrad formalism and the so-called Metric-Affine Gravity (MAG). We revisit the relation between these two approaches in Sec.~\ref{22032020b} and prove that the ambiguity of the Weitzenb\"{o}ck connection is equivalent to that of the teleparallel frame. The regularization procedure used by Obukhov and Guillermo in Ref.~\cite{PhysRevD.73.124017} is discussed.

Section \ref{24082019a} is devoted to the meaning of torsion in teleparallel theories. There, we will argue that there are only two different ways to  interpret   the teleparallel theories based on the Weitzenb\"{o}ck connection, otherwise these theories will be inconsistent. We show that these two interpretations predict the same gravitational energy-momentum tensor if the frame satisfies some constraints. In Sec.~\ref{03072021a} we analyze  a $+$ polarized gravitational wave in a frame that accelerates along the direction of propagation of the wave. The consistence of the two interpretations when the aforementioned constraints are imposed on the frame is demonstrated. It is also shown that: the gravitational energy and momentum are shifted in a manner analogous to the Doppler effect of a electromagnetic wave in Minkowski; the density that is interpreted as the angular momentum density vanishes; the energy density cannot be made to vanish in the accelerated fame if it does not vanish in the freely falling one.

We devote Sec.~\ref{23032020b} to a general discussion about  the $f(T)$  theories. There, we discuss the possible limitations and even the viability of $f(T)$ theories. We also discuss possible solutions to the problems faced by these theories.

Our notation is as follows: The metric components in a coordinate basis is denoted by $g_{\mu\nu}$, while in a tetrad basis is denoted by $\eta_{ab}$. We denote the frame  and the coframe by $e_a$,  $\teta{^a}$, respectively; they satisfy the relation $\bracket{\teta{^a},\e{_b}}\equiv\teta{^a}(\e{_b})=\delta^a_b$. The spacetime signature is $(+,-,-,-)$. Greek letters represent spacetime indices, and  Latin letters represent tangent space ones, except for Latin letters in the middle of the alphabet ($i,j,k,\ldots$), which stand for spatial coordinate indices. The components of the frame $e_a$ and the coframe $\teta{^a}$ in a coordinate basis are represented by $\tensor{e}{_a^\mu}$ and $\tensor{e}{^a_\mu}$, respectively. We distinguish the tangent space indices from the coordinate ones  by using the former between  parentheses: $\tensor{e}{_{(0)}^0}$, $\tensor{e}{_{(1)}^2}$ etc. We use the convention $A_{[ab]}\equiv(1/2)(A_{ab}-A_{ba})$.

\section{Riemann-Cartan spacetime}\label{14122017a}
In this section we review some properties of the Riemann-Cartan spacetimes and present the main aspects of the formalism that will be used throughout this paper.

Let $M$ be a $n$-dimensional differentiable manifold, ${\cal F}(M)$ the set of smooth functions on $M$, and ${\cal V}(M)$ the set of the vector fields on $M$. An affine connection $\nablab$ is a map $\nablab:{\cal V}(M)\times{\cal V}(M)\mapsto {\cal V}(M) $, denoted by $(U,V)\mapsto\nablab_UV$, which satisfies the following conditions:
\begin{align}
\nablab_U(V+W)=\nablab_UV+\nablab_UW,\\
\nablab_{(U+V)}W=\nablab_UW+\nablab_VW,\\
\nablab_{(fU)}V=f\nablab_UV,\\
\nablab_U(fV)=U[f]V+f\nablab_UV,
\end{align}
where $U$, $V$, $W$~$\in {\cal V}(M)$, and $f$~$\in {\cal F}(M)$; by definition $U[f]\equiv U^\mu\pd{_\mu}f \in {\cal F}(M)$.

The affine connection of Riemann-Cartan spacetimes  possesses both a curvature and a torsion. In the index-free approach, they take the form  \cite{Nakahara}
\begin{equation}
\bm{R}(V,U)W\equiv\nablab_V\nablab_U W-\nablab_U\nablab_V W-\nablab_{[V,U]}W, \label{07112017b}
\end{equation}
\begin{equation}
\T(V,U)\equiv\nablab_V U-\nablab_U V- [V,U], \label{07112017d}
\end{equation}
where $[V,U]$ is the Lie bracket of $V$ and $U$, which is defined by $[V,U]f=V[U[f]]-U[V[f]]$.

In a coordinated basis $\pd{_\mu}$ (and its dual basis $dx^\mu$) , the affine connection coefficients are denoted by  $\connection{^\lambda_\mu_\nu}\equiv \bracket{dx^{\lambda},\nablab_{\mu}\partial_{\nu}}$, while in a tetrad one we use
\begin{equation}
\sconnection{^a_b_c}\equiv \bracket{\teta{^a},\nablab_b e_c }. \label{05082019a}
\end{equation}

For the curvature and torsion components, we use the conventions $\tensor{R}{^\alpha _\mu_\beta_\nu}\equiv\bracket{dx^{\alpha},\bm{R}(\pd{_\beta},\pd{_\nu})\pd{_\mu}}$ and $\torsion{^a_b_c}\equiv\bracket{\teta{^a},\T(e_b,e_c)}$\footnote{Note that $\tensor{R}{^a _b_c_d}=\bracket{\teta{^a},\bm{R}(\e{_c},\e{_d})\e{_b}}=\e{^a_\alpha}\e{_b^\mu}\e{_c^\beta}\e{_d^\nu}\tensor{R}{^\alpha _\mu_\beta_\nu}$, and $\torsion{^\alpha_\beta_\mu}=\bracket{dx^\alpha,\T(\pd{_\beta},\pd{_\mu})}=\e{_a^\alpha}\e{^b_\beta}\e{^c_\mu}\torsion{^a_b_c}$. }. For these conventions, Eqs.~(\ref{07112017b})-(\ref{07112017d}) yield
\begin{equation}
\tensor{R}{^\alpha _\mu_\beta_\nu}=\partial_{\beta}\connection{^\alpha_\nu_\mu}-\partial_{\nu}\connection{^\alpha_\beta_\mu}+\connection{^\phi_\nu_\mu}\connection{^\alpha_\beta_\phi}-\connection{^\phi_\beta_\mu}\connection{^\alpha_\nu_\phi}, \label{07112017c}
\end{equation}
\begin{equation}
\torsion{^a_b_c}=2\sconnection{^a_{[bc]}}+\nonholon{^a_b_c}, \label{16032019i}
\end{equation}
where
\begin{equation}
\connection{^\lambda_\mu_\nu}=\chr{^\lambda_\mu_\nu}+\contorsion{^\lambda_\mu_\nu}, \label{17112017a}
\end{equation}
\begin{equation}
\contorsion{^\lambda_\mu_\nu}\equiv-\frac{1}{2}\left(\torsion{_\mu_\nu^\lambda}+\torsion{_\nu_\mu^\lambda}-\torsion{^\lambda_\mu_\nu} \right),\label{17112017c}
\end{equation}
$\tensor{\Omega}{^a_b_c}\equiv-\bracket{\teta{^a},[e_b,e_c]}$ is the object of anholonomity (the structure functions of $\e{_a}$), $\chr{^\lambda_\mu_\nu}$ the Christoffel symbols, $\contorsion{^\lambda_\mu_\nu}$ the contorsion,   and $\sconnection{^a_{[bc]}}\equiv(\sconnection{^a_{bc}}-\sconnection{^a_{cb}})/2$. Furthermore, one can prove that
\begin{equation}
\sconnection{^a_b_c}=\frac{1}{2}\left(\tensor{\Omega}{_b_c^a}+\tensor{\Omega}{_c_b^a}-\tensor{\Omega}{^a_b_c} \right)+\contorsion{^a_b_c}. \label{07082019a}
\end{equation} 

Using the convention $\tensor{R}{_\mu_\nu}=\tensor{R}{^\alpha_\mu_\alpha_\nu}$, it is straightforward to check the validity of the well-known identity
\begin{align}
R_{\mu\nu}=&\ \R{_\mu_\nu}+\rcd{_\alpha}\tensor{K}{^\alpha_\nu_\mu}-\rcd{_\nu}\tensor{K}{^\alpha_\alpha_\mu}+\tensor{K}{^\alpha_\alpha_\lambda}\tensor{K}{^\lambda_\nu_\mu}
\nonumber\\
&-\tensor{K}{^\alpha_\nu_\lambda}\tensor{K}{^\lambda_\alpha_\mu}, \label{05082019b}
\end{align}
where $\rcd{_\lambda}\tensor{K}{^\alpha_\nu_\mu}$ are the components of the covariant derivative of $\bm{K}\equiv \tensor{K}{^\alpha_\nu_\mu}\pd{_\alpha}\otimes dx^\nu\otimes dx^\mu$ with respect to the Levi-Civita connection $\leviconnection{}$; in a coordinate basis, this connection becomes the Christoffel symbols: $\chr{^\lambda_\mu_\nu}\equiv \bracket{dx^\lambda,\leviconnection{_\mu}\pd{_\nu}}$.

\subsection{Differential forms}
The torsion components given by Eq.~(\ref{16032019i}) are commonly used by authors such as Maluf\footnote{Maluf's definition of the affine connection coefficients is slightly different: denoting Maluf's version as $\sconnection[\widehat]{_\mu^a_c}$, we have the relation $\sconnection{^a_\mu_c}=\sconnection[\widehat]{_\mu^a_c}$.} \cite{ANDP:ANDP201200272}. On the other hand, authors such as Obukhov and Rubilar \cite{PhysRevD.73.124017} use the torsion $2$-form
\begin{equation}
\torsionhehl{^a}=\frac{1}{2}\torsion{^a_b_c}\teta{^b}\wedge\teta{^c}.\label{18052019w}
\end{equation}
(Note that we use a slight different letter for the torsion $2$-form.) The relation of $\torsionhehl{^a}$ with Eq.~(\ref{07112017d})  is $\T(V,U)=\torsionhehl{^a}(V,U)\e{_a}$. So, the  torsion $2$-form is nothing but the vector component of $\T(V,U)$. 

 In terms of this component, Eq.~(\ref{16032019i}) can be recast as
\begin{equation}
\torsionhehl{^a}=d\teta{^a}+\connection{_b^a}\wedge\teta{^b},
\end{equation}
where $\connection{_b^a}$ is the connection $1$-form, which is related to the affine connection coefficients through
\begin{equation}
\connection{_a^b}=\sconnection{^b_c_a}\teta{^c}.\label{21052019a}
\end{equation}
We will use these two different approaches together in a consistent way.

\subsection{Lorentz transformation}
 As is well known, there are many different noncoordinate bases which yield the same metric. These bases are related to each other by a local orthogonal rotation, which in our case corresponds to the $SO(3,1)$ group. An element of this group [in fact $SO^+(3,1)$] is called a proper orthochronous Lorentz transformation and will be denoted by $\lorentz{^a_b}$. 

Given two tetrads $\teta{^a}$ and $\tetabar{^a}$ (their frames are denote by $\e{_a}$ and $\ebar{_a}$, respectively), we have
\begin{align}
\teta{^a}=\lorentz{^a_b}\tetabar{^b},\ \tetabar{^b}=\lorentz{_c^b}\teta{^c},\ \e{_c}=\lorentz{_c^b}\ebar{_b},\ \ebar{_c}=\lorentz{^d_c}\e{_d}, \label{25082017b}
\end{align}
where $\lorentz{_c^b}=\tensor{(\Lambda^{-1})}{^b_c}$. Note that $\lorentz{_c^a}\lorentz{^c_b}=\delta^a_b$ ($\Lambda^T\Lambda=\mathbb{I}$).

Using  $\scbar{^a_b_c}= \bracket{\tetabar{^a},\nablab_{\bar{b}} \ebar{_c} }$ and  Eqs.~(\ref{25082017b}) and (\ref{05082019a}), we obtain the transformations
\begin{align}
\scbar{^a_b_c}=\lorentz{_d^a}\lorentz{^f_b}\e{_f^\mu}\pd{_\mu}\lorentz{^d_c}+\lorentz{_g^a}\lorentz{^f_b}\lorentz{^d_c}\sconnection{^g_f_d}, \label{02082019a}
\\
\sconnection{^a_b_c}=\lorentz{^a_d}\lorentz{_b^f}\ebar{_f^\mu}\pd{_\mu}\lorentz{_c^d}+\lorentz{^a_g}\lorentz{_b^f}\lorentz{_c^d}\scbar{^g_f_d}. \label{25082017e}
\end{align} 
These are the well-known transformations of the affine connection coefficients in noncoordinate bases.
 
\section{Teleparallelism}\label{22032020a}
Given a manifold $M$ and an affine connection $\nablab$, we say that we have teleparallelism when we can transport a vector from one point to any other and the resulting vector does not depend on the path taken.  This means that if we have a vector $V$ defined at a point $p$, we can construct a unique vector field over $M$ by parallely propagating $V$. In other words, we demand that $\nablab_{d/ds}V=(dx^{\mu}/ds)\nablab_{\mu}V=0$ hold for an arbitrary curve $x^{\mu}(s)$, which implies $\nablab_{\mu}V=0$. Therefore, in teleparallel theories we can always find a tetrad field that satisfies the equation
\begin{equation}
\nablab_{\mu}\e{_a}=0\quad (\sconnection{^a_b_c}=0) \label{14022019bc}
\end{equation}
everywhere. This connection is called a {\it Weitzenb\"{o}ck connection}.

Equation (\ref{14022019bc}) is not invariant under a local Lorentz transformation (LLT). As a consequence, it depends on the  frame that is chosen to satisfy it.  Once we have chosen a particular basis $\e{_a}$, we cannot go to another basis $\ebar{_a}$ through a LLT without altering the form of Eq.~(\ref{14022019bc}): the affine connection coefficients cannot vanish in two different frames related by a LLT ``simultaneously''. 

\begin{definition}
A frame $\e{_a}$ (or its coframe $\teta{^a}$) will be called a teleparallel frame (TF) if it satisfies Eq.~(\ref{14022019bc}).
\end{definition}

In a TF, the torsion tensor becomes
\begin{align}
\torsion{^a_\mu_\nu}=\pd{_\mu}\e{^a_\nu}-\pd{_\nu}\e{^a_\mu}. \label{04102019p}
\end{align}
But, in general, it is given by Eq.~(\ref{16032019i}).

Let us now obtain some very interesting identities that are a consequence of teleparallelism. In doing so, we use the tensor (usually called {\it superpotential})
\begin{equation}
\potential{^\lambda^\mu^\nu}\equiv\frac{1}{4}\left( \torsion{^\lambda^\mu^\nu}+2\torsion{^{[\mu|}^\lambda^{|\nu]}}\right)+g^{\lambda[\nu}T^{\mu]}, \label{10112017lc}
\end{equation}
where $\torsion{_a}\equiv\torsion{^b_b_a}$. Note that  this tensor is antisymmetric in the last two indices. (It is also worth noting that we are not restricting ourselves to TFs here.) 

Using Eq.~(\ref{14022019bc}) in Eq. (\ref{07112017b}) we see that the curvature tensor must vanish, i.e., $\tensor{R}{^a_b_c_d}=\bracket{\teta{^a},\bm{R}(\e{_c},\e{_d})\e{_b}}=0$ for all basis,  including coordinate ones. Hence, we must also have $R_{\mu\nu}=0$.  Using this result in Eq.~(\ref{05082019b}), we can isolate $\R{_\mu_\nu}$ to obtain, with the help of Eq.~(\ref{10112017lc}), the identity (see appendix \ref{12042020f})
\begin{align}
\R{^\mu^\nu}=&\ 2\e{_b^\nu}\left( \pd{_\alpha}\potential{^b^\mu^\alpha}+\chr{^\lambda_\lambda_\alpha}\potential{^b^\mu^\alpha} \right)-2\potential{^a^b^\mu}\torsion{_a_b^\nu}
\nonumber\\
&+g^{\mu\nu}\rcd{_\alpha}\torsion{^\alpha}+2\sconnection{^c_\alpha_b}\e{_c^\nu}\potential{^b^\mu^\alpha}. \label{29032019a}
\end{align}
Multiplying it by $e=\det(\e{^a_\mu})=\pm\sqrt{-\det g}$ and using the identity $\pd{_\alpha}e=e\chr{^\lambda_\lambda_\alpha}$, we find that
\begin{align}
e\R{^\mu^\nu}=&2\e{_b^\nu} \pd{_\alpha}\left(e\potential{^b^\mu^\alpha} \right)-2e\potential{^a^b^\mu}\torsion{_a_b^\nu}
\nonumber\\
&+g^{\mu\nu}\pd{_\alpha}\left(e\torsion{^\alpha}\right)+2e\sconnection{^c_\alpha_b}\e{_c^\nu}\potential{^b^\mu^\alpha}. \label{29032019b} 
\end{align}
Contracting with $g_{\mu\nu}$, we get
\begin{equation}
e\R{}=2\pd{_\mu}\left(e\torsion{^\mu}\right)-eT, \label{29032019d}
\end{equation}
where $T\equiv\potential{^a^b^c}\torsion{_a_b_c}$ is the {\it torsion scalar}. Thus, we see that the Einstein tensor $\G{^\mu^\nu}$ of the Levi-Civita connection can be written in the form
\begin{equation}
e\G{^\mu^\nu}=2\e{_b^\nu}\pd{_\alpha}\left( e\potential{^b^\mu^\alpha}\right)-\frac{e}{2k}\psenergy{^\mu^\nu}, \label{29032019f} 
\end{equation}
where
\begin{equation}
\psenergy{^\mu^a}=t^{\mu a}-4k\sconnection{^a_\alpha_b}\potential{^b^\mu^\alpha}, \label{29032019g}
\end{equation}
\begin{equation}
\energy{^\mu ^a}=k(4\potential{^b^c^\mu}\torsion{_b_c^a}-\e{^a^\mu}T), \label{29032019h}
\end{equation}
and $k=1/(16\pi)$ in natural units. (Keep in mind that $\energy{^\mu^\nu}=\e{_a^\nu}\energy{^\mu ^a}$ and $\psenergy{^\mu^\nu}=\e{_a^\nu}\psenergy{^\mu^a}$.)

The first and the second terms in the right-hand side of Eq.~(\ref{29032019f}) transform as tensors under coordinate transformations, but not under LLTs. Note, nevertheless, that their sum transforms as a tensor under both transformations. Furthermore, $\energy{^\mu^\nu}$ also behaves as a tensor for both cases\footnote{In the pure-tetrad approach,   $\energy{^\mu^a}$ is not covariant under LLTs (see, e.g., Ref.~\cite{Cai_2016}).}, but it is not on the same footing as $\G{^\mu^\nu}$, as will become clear in Sec.~\ref{05082019c}. 

\subsection{The field equations of the TEGR}
The most famous teleparallel theory is the so-called Teleparallel Equivalent of General Relativity (TEGR). The reason why it is so famous is because its field equations are equivalent to that of GR. In this theory  one uses Eq.~(\ref{29032019d}) to go from the Einstein-Hilbert Lagrangian density $L=ke\R{}$ to $L=-keT$ by neglecting the surface term $2\pd{_\mu}\left(e\torsion{^\mu}\right)$. Defining the total Lagrangian by $L_\mathrm{total}\equiv L-L_M$ and varying the action with respect to $\e{_a^\mu}$ (the fundamental field), one obtains\footnote{For a discussion about the variational principle in teleparallel theories, see Refs.~\cite{0264-9381-34-14-145013,Maluf_2020}.} 
\begin{equation}
\G{^\mu^\nu}=\frac{1}{2k}T^{\mu\nu}, \label{29032019j} 
\end{equation}
where 
\begin{equation}
T_{\mu\nu}\equiv\frac{1}{e}\e{^a_\mu}\frac{\delta L_M}{\delta \e{^a^\nu}}=\frac{2}{e}\frac{\delta L_M}{\delta g^{\mu\nu}} \label{29032019i}
\end{equation}
is the energy-momentum of matter.

With the help of Eq.~(\ref{29032019f}), Eq.~(\ref{29032019j}) can be recast as
\begin{equation}
\pd{_\alpha}\left( e\potential{^a^\mu^\alpha}\right)=\frac{e}{4k}\left(\psenergy{^\mu^a}+T^{\mu a}\right). \label{29032019k}
\end{equation}
Applying $\pd{_\mu}$ to this equation  and using the identity $\potential{^a^\mu^\alpha}=-\potential{^a^\alpha^\mu}$, we obtain  $\pd{_\mu}\left[e\left(\psenergy{^\mu^a}+T^{\mu a}\right)\right]=0$. 

In the case of the pure-tetrad formalism we have $\psenergy{^\mu^a}=\energy{^\mu^a}$, where $\energy{^\mu^a}$ is identified as  the energy-momentum tensor of the gravitational field. Then,  the conservation law becomes $\pd{_\mu}\left[e\left(\energy{^\mu^a}+T^{\mu a}\right)\right]=0$.  In turn, the the energy-momentum contained within a three dimensional volume $V$ is \cite{ANDP:ANDP201200272}
\begin{equation}
P^a=\int_V d^3x e\energy{^0^a}+\int_V d^3x eT^{0a}, \label{10082019a}
\end{equation}
where the first integral is the gravitational energy-momentum. The energy-momentum (in the pure-tetrad formalism) can also be given by \cite{ANDP:ANDP201200272}
\begin{equation}
P^a=4k\oint_S dS_i e\Sigma^{a0i}.\label{22072018g}
\end{equation}

The total angular momentum  can be defined as
\begin{align}
L^{ab}\equiv -\int_V d^3xM^{ab},\label{30122019c}
\end{align}
where 
\begin{align}
M^{ab}=-4ke\left(\potential{^a^0^b}-\potential{^b^0^a} \right) \label{30122019d}
\end{align}
is the angular momentum density.  The quantity $L^{(0)b}$ may be interpreted as the center of mass moment, while $L^{(i)(j)}$ is the angular momentum \cite{ANDP:ANDP201200272}. The question whether Eqs.~(\ref{10082019a})-(\ref{30122019d}) hold beyond the pure-tetrad formalism will be discussed later.

The quantities given by Eqs.~(\ref{29032019h}) and (\ref{10082019a}) have yield interesting and consistent results for some cases \cite{ANDP:ANDP201200272,doi:10.1002/andp.201800320,PhysRevD.99.064047}. One example is the gravitational energy of the Universe within a three dimensional volume, which for a spatially flat Universe is given by $E_g=-c^2H^2a^3r^3/(2G)$ [see, e.g., Eq.~(170) of Ref.~\cite{PhysRevD.99.064047}], where $H$ is the Hubble parameter,  $a$ the scalar factor, and $r$ is a radial coordinate. Notice that the density is $-3c^2H^2/(8\pi G)$, which is constant  for a constant $H$. The total energy of the Universe within a finite volume for this case vanishes, while the one for a closed Universe predicted by Eq.~(\ref{10082019a}) is the same as that predicted by the Einstein gravitational pseudo-tensor [compare Eq.~(161) of Ref.~\cite{PhysRevD.99.064047} with Eq.~(14) of Ref.~\cite{Rosen1994} for $\phi=0$ and $k=1$.].

\section{The teleparallel frame problem}\label{05082019c}
In order to tackle the teleparallel frame problem,
we adopt the following notation. The components of the torsion tensor in a basis $\ebar{_a}$ when $e_a$ is taken as the TF will be denoted by $\torsionbare{^a_b_c}$. On the other hand, when the components are written in the basis $\ebar{_a}$ and at the same time this basis is the TF, we use $\torsionbarbar{^a_b_c}$. Note that these definitions can be applied to any object that depend on the choice of the TF. 
 
One should be aware that $\torsionbare{^a_b_c}$ and $\torsionbarbar{^a_b_c}$ may represent torsion components of different connections. To be more precise, if $\ebar{_a}$ and  $e_a$ are not related to each other by a global Lorentz transformation, then the Weitzenb\"{o}ck connections $\nablab_{\bf e}$  and  $\nablab_{\bf \bar{e}}$ are not the same, where $\nablab_{\bf e}$ ($\nablab_{\bf \bar{e}}$) is the Weitzenb\"{o}ck connection that satisfies $\nablab_{\!{\bf e}\mu}\e{_a}=0$ ($\nablab_{{\!\bf \bar{e}}\mu}\ebar{_a}=0$). Hence, we will be dealing with a manifold endowed with an infinite number of Weitzenb\"{o}ck connections, one for each set of possible TFs that are not related to each other by global transformations.

By definition, we must have $\scbarbar{^a_b_c}=0$, that is, when the Weitzenb\"{o}ck connection $\nablab_{\bf \bar{e}}$ is expanded in terms of $\ebar{_a}$, the connection coefficients vanish. So, from Eq.~(\ref{16032019i}) we see that
\begin{equation}
\torsionbarbar{^a_b_c}=\nonholonbar{^a_b_c}, \label{16032019e}
\end{equation}
where 
\begin{equation}
\nonholonbar{^a_b_c}=-\bracket{\tetabar{^a},[\ebar{_b},\ebar{_c}]}=2\ebar{_b^\mu}\ebar{_c^\nu}   \pd{_{[\mu|}}\ebar{^a_{|\nu]}}. \label{16032019h}
\end{equation}
Note that Eq.~(\ref{04102019p}) corresponds to $\torsione{^a_\mu_\nu}$.

From Eq.~(\ref{16032019i}) we  see that
\begin{equation}
\torsionbare{^a_b_c}=2\scbare{^a_{[bc]}} +\torsionbarbar{^a_b_c}, \label{16032019f}
\end{equation}
where we have used  Eq.~(\ref{16032019e}) in the second term of Eq.~(\ref{16032019i}). In case $\nablab_{\bf e}$ and $\nablab_{\bf \bar{e}}$ do not represent the same connection, Eq.~(\ref{16032019f}) gives the relation between their torsions.

From Eq.~(\ref{02082019a}), we see that the connection coefficients of $\nablab_{\bf e}$ expanded in $\ebar{_a}$, that is $\scbare{^a_b_c}$,  is given by
\begin{equation}
\scbare{^a_b_c}=\lorentz{_d^a}\ebar{_b^\alpha}\pd{_\alpha}\lorentz{^d_c}. \label{09042020a}
\end{equation}
 It is clear that if the matrix $\Lambda$ is not constant, then the description of the torsion tensor in the basis $\ebar{_a}$ when $\ebar{_a}$ is the TF is not equivalent to that where $\e{_a}$ is the TF: note that both $\torsionbare{^a_b_c}$ and $\torsionbarbar{^a_b_c}$ are written in the same basis, the only difference is the choice of the TF. Therefore, we  conclude that  {\it the Weitzenb\"{o}ck torsion cannot be made invariant under an arbitrary change of the TF (the torsion of $\nablab_{\bf e}$ is not the same as that of $\nablab_{\bf \bar{e}}$ when $\Lambda$ is not constant)}. Note that this problem also affects all quantities defined solely by combinations of the Weitzenb\"{o}ck torsion tensor, such as the superpotential and the torsion scalar.

It is easy to show that the relation between $\scbare{^a_b_c}$ and $\scebar{^a_b_c}$ is given by
\begin{align}
\scebar{^a_b_c}=-\lorentz{^a_f}\lorentz{_b^h}\lorentz{_c^g}\scbare{^f_h_g}, \label{06032020a}
\\ 
 \scbare{^a_b_c}=-\lorentz{_f^a}\lorentz{^h_b}\lorentz{^g_c}\scebar{^f_h_g}. \label{12012020a}
\end{align}

An interesting aspect of the notation used here is that, to ``invert'' the identities that relate the quantities defined with both $\scebar{^a_b_c}$ and $\scbare{^a_b_c}$, we only need to change overbars to ``no overbars'' (and vice versa) where it is appropriate, and  change the order of the indices of the Lorentz matrices, as in the expressions above.  

Since $\torsionbare{^a_b_c}$ and $\torsionbarbar{^a_b_c}$, in general, are not the same torsion, which of them, if any, has physical meaning? To pursue this question, we will need the following definitions. 

\begin{definition}\label{03042020a}
Let \E be the set of all frames (coframes) related to a particular basis $\e{_a}$ (or $\teta{^a}$) by a global Lorentz transformation. This set will be called the global set of $\e{_a}$.
\end{definition}

\begin{definition}\label{27052019b}
Let \E and \Ebar be the global sets of $\e{_a}$ and $\ebar{_a}$, respectively. Assume that \E and \Ebar are not the same. The transformation of an object $A$, which may or may not be a tensor, from $\tensor{A}{_{\bf e}}$ to $\tensor{\bar{A}}{_{\bf \bar{e}}}$, and vice versa, will be called a  {\it teleparallel transformation} (TT).
\end{definition}

\begin{definition}\label{27052019c}
Quantities that are invariant under LLTs and TTs will be called  absolute invariant (AI).
\end{definition}

\begin{definition}\label{27052019d}
Quantities that are invariant under LLTs, but not under TTs will be called  relative invariant (RI).
\end{definition}
Let $A$ and $B$ be  a RI and an AI, respectively. Assume that we can expand them in the frame $\e{_a}$ as $A=A^a\e{_a}$ and $B=B^a\e{_a}$. Under a LLT given by 
$\bar{e}_c=\tensor{\Lambda}{^d_c}e_d$, $A^a$ and $B^a$  change as  $\tensor{\bar{A}}{_{\bf e}^a}=\lorentz{_b^a}\tensor{A}{_{\bf e}^b}$ ($\tensor{\bar{A}}{_{\bf \bar{e}}^a}=\lorentz{_b^a}\tensor{A}{_{\bf \bar{e}}^b}$) and  $\tensor{\bar{B}}{_{\bf e}^a}=\lorentz{_b^a}\tensor{B}{_{\bf e}^b}$ ($\tensor{\bar{B}}{_{\bf \bar{e}}^a}=\lorentz{_b^a}\tensor{B}{_{\bf \bar{e}}^b}$). As a result, we have $A_{\bf\bar{e}}=\bar{A}_{\bf\bar{e}}$ and $B_{\bf\bar{e}}=\bar{B}_{\bf\bar{e}}$. Nevertheless, under a TT  we have $\tensor{B}{_{\bf e}}=\tensor{\bar{B}}{_{\bf \bar{e}}}$, while $\tensor{A}{_{\bf e}}\neq \tensor{\bar{A}}{_{\bf \bar{e}}}$. As an example of a RI, we have the torsion tensor, whose components in the TF $\e{_a}$ is $\torsione{^a_b_c}=\nonholon{^a_b_c}$, while in the TF $\ebar{_a}$ is $\torsionbarbar{^a_b_c}=\nonholonbar{^a_b_c}\neq \lorentz{_d^a}\lorentz{^e_b}\lorentz{^f_c}\nonholon{^d_e_f}$. So, we conclude that  $\T_{\bf e}\neq\bar{\T}_{\bf \bar{e}}$ (different Weitzenb\"{o}ck connections). Nevertheless, we still have $\T_{\bf e}=\bar{\T}_{\bf e}$ and $\T_{\bf \bar{e}}=\bar{\T}_{\bf \bar{e}}$ (the same Weitzenb\"{o}ck connection). [Recall that $\T$ is given by Eq.~(\ref{07112017d}).]

 In general, we interpret the components of a tensor as being related to some measurable quantity. For instance,  the component $T^{(0)(0)}$ of the energy-momentum tensor of a mater field is interpreted as being the energy density measured by an observer with a $4$-velocity $\e{_{(0)}}$; when we change to a frame $\ebar{_{(0)}}$, then we have $\bar{T}^{(0)(0)}$, which is the energy density measured by another observer whose $4$-velocity is $\ebar{_{(0)}}$. As will be shown later, this tensor is necessarily an AI. How about the components of RIs? Again, we need more definitions to deal with this question:

\begin{definition}\label{27052019e}
The components of a tensor written in the TF will be called a teleparallel component.
\end{definition}

Weitzenb\"{o}ck torsion can be invariant under LLTs with a fixed TF, but not under  TTs. To understand what is happening in teleparallelism, we need to differentiate the ordinary meaning of a LLT from a mere mathematical procedure. In general, when we change the components of a tensor from one frame $\e{_a}$ to another $\ebar{_a}$ by means of a LLT, we say that we changed the observer and the new components are the quantities measured by this new observer. However, this view does not necessarily hold for a RI.

\subsection{Torsion scalar}
Let us see the relation between the torsion scalars of two distinct Weitzenb\"{o}ck torsions.

Equation (\ref{29032019d}) can be recast as $\R{}+T-2\rcd{_\mu} \torsion{^\mu}=0$. Thus, it is clear that $\R{}+\torsione{}-2\rcd{_\mu} \torsione{^\mu}=0$. (Because $\R{}$ is an AI, we do not need to write $\R{_{\bf e}}$.) Since $\torsione{}=\torsionbare{}$ and $\torsione{^\mu}=\torsionbare{^\mu}$, we must also have
\begin{equation}
\R{}+\torsionbare{}-2\rcd{_\mu} \torsionbare{^\mu}=0. \label{26032019c}
\end{equation}

It is obvious that the following identity also holds
\begin{equation}
\R{}+\torsionbarbar{}-2\rcd{_\mu} \torsionbarbar{^\mu}=0. \label{26032019d}
\end{equation}
Subtracting Eq.~(\ref{26032019d}) from Eq.~(\ref{26032019c}), we obtain
\begin{align}
\torsionbare{}=&\ \torsionbarbar{}+2\rcd{_\mu}\left(\torsionbare{^\mu} -\torsionbarbar{^\mu}\right)
\nonumber\\
=&\ \torsionbarbar{}+\frac{2}{e}\pd{_\mu}\left(e\ebar{^b^\mu}\scbare{^a_a_b}\right),
\label{26032019e}
\end{align}
where Eq.~(\ref{16032019f}) and the definition $\torsion{_a}\equiv\torsion{^b_b_a}$ were used in the second line.
Thus, changing the TF, but keeping the frame where the torsion scalar is written, induces a surface term.  This motivates the following definition.
\begin{definition}\label{27052019f}
Let $A$ be an object that may or may not be a tensor. The transformation from  $A_{\bf e}$ to $A_{\bf\bar{e}}$, or vice versa, will be called an affine teleparallel transformation (ATT).
\end{definition}

Note from definitions \ref{27052019b} and \ref{27052019f} that the differences between a TT and an ATT is that the latter does not change the basis where the components are written, only the TF or, equivalently, the affine connection, which does not necessarily  vanish for all frames. (It is only a change in the choice of the affine connection.)

\begin{theorem}
An object is invariant under LLTs and ATTs if and only if it is an AI.
\end{theorem}
{\it Proof.} If  $A$ is invariant under a LLT, then $\bar{A}_{\bf e}=A_{\bf e}$. If, in addition, $A$ is also invariant under an ATT, then $\bar{A}_{\bf e}=\bar{A}_{\bf \bar{e}}$. Hence, we have $\bar{A}_{\bf \bar{e}}=A_{\bf e}$ (invariant under TTs). On the other hand, if $A$ is an AI, then both $\bar{A}_{\bf e}=A_{\bf e}$ and $\bar{A}_{\bf \bar{e}}=A_{\bf e}$ hold (see definition \ref{27052019c}), which leads to $\bar{A}_{\bf e}=\bar{A}_{\bf \bar{e}}$.

\subsection{Energy-momentum tensors}
Now we search for the relation between the stress-energy tensors of two different Weitzenb\"{o}ck connections.

From Eq.~(\ref{29032019j}) we see that, in the TEGR, the matter energy-momentum tensor must be an AI, that is $\tensor{T}{_{\bf e}^\mu^\nu}=\tensor{T}{_{\bf \bar{e}}^\mu^\nu}=\tensor{\bar{T}}{_{\bf e}^\mu^\nu}=\tensor{\bar{T}}{_{\bf \bar{e}}^\mu^\nu}$. However, neither $\psenergy{^\mu^\nu}$ nor $\energy{^\mu^\nu}$ are.

Using the fact that Eq.~(\ref{29032019f}) holds for any basis $\ebar{_a}$, regardless of whether it is a TF or not, we have
\begin{equation}
e\G{^\mu^\nu}=2\ebar{_b^\nu}\pd{_\alpha}\left( e\potentialbarbar{^b^\mu^\alpha}\right)-\frac{e}{2k}\psenergybarbar{^\mu^\nu}, \label{29032019l}
\end{equation}
\begin{equation}
e\G{^\mu^\nu}=2\ebar{_b^\nu}\pd{_\alpha}\left( e\potentialbare{^b^\mu^\alpha}\right)-\frac{e}{2k}\psenergybare{^\mu^\nu}. \label{29032019m}
\end{equation}
Subtracting Eq.~(\ref{29032019m}) from Eq.~(\ref{29032019l}) and contracting the result with $\ebar{^a_\nu}$, we obtain
\begin{equation}
\psenergybare{^\mu^a}-\psenergybarbar{^\mu^a}=\frac{8k}{e}\pd{_\alpha}\left( e\pifactorbar{^a^\mu^\alpha}\right), \label{29032019n}
\end{equation}
with
\begin{equation}
\pifactorbar{_a_b_c}=\frac{1}{4}\left(\scbare{_b_a_c}+\scbare{^d_d_b}\eta_{ac}-\scbare{^d_d_c}\eta_{ab} \right), \label{11012020a}
\end{equation}
where Eqs.~(\ref{10112017lc}), (\ref{16032019f}) and (\ref{16032019e}) were used to write $\potentialbarbar{^b^\mu^\alpha}-\potentialbare{^b^\mu^\alpha}$ in terms of the affine connection coefficients.

The tensor $\psenergybare{^\mu^a}\pd{_\mu}\otimes \e{_a}$ is neither a RI nor an AI. In fact, as we can see from Eq.~(\ref{29032019g}), it is invariant only under coordinate changes\footnote{$\sconnection{^a_\alpha_c}$ transforms as a tensor under coordinate transformations.}. (Keep in mind that, throughout this paper, the word ``tensor'' refers to any quantity that is invariant under coordinate transformations, not necessarily LLTs. )

Using Eqs.~(\ref{29032019g})-(\ref{29032019h}) in Eq.~(\ref{29032019n}), we find that
\begin{align}
\energybare{^\mu^a}=\energybarbar{^\mu^a}+\frac{8k}{e}\pd{_\alpha}\left( e\pifactorbar{^a^\mu^\alpha}\right)+4k\scbare{^a_\alpha_b}\potentialbare{^b^\mu^\alpha}. \label{29032019p}
\end{align}
(Note that $\psenergybarbar{^\mu^\nu}=\energybarbar{^\mu^\nu}$.) It is clear that $\energy{^\mu^\nu}\pd{_\mu}\otimes\pd{_\nu}$ is not invariant under ATTs. Nevertheless, it is invariant under LLTs and coordinate transformations, which means that  it is a RI.

\section{Pure-tetrad formulation versus MAG}\label{22032020b}
At the level of the field equations, we have basically two approaches to teleparallel theories: setting the connection to always vanish or using an affine connection. To better understand these two approaches, we have dealt with a notation that have allowed us to work with both simultaneously. Now we focus a little bit more on the metric-affine gravity (MAG) and adapt the notation used in Refs. \cite{PhysRevD.73.124017,Obukhov_2009,PhysRevD.80.064043} to ours.  

The volume $4$-form is defined as $\eta\equiv\teta{^{(0)}}\wedge\teta{^{(1)}}\wedge\teta{^{(2)}}\wedge\teta{^{(3)}}$, while the bases\footnote{We use boldface letters to distinguish the basis $\baseeta{_a_b}$ from the metric in the tetrad basis, $\eta_{ab}$.} for $3$-, $2$-, and $1$- forms are defined as  $\baseeta{_a}\equiv\e{_a}\rfloor\eta$ , $\baseeta{_a_b}\equiv\e{_b}\rfloor\baseeta{_a}$, $\baseeta{_a_b_c}\equiv\e{_c}\rfloor\baseeta{_a_b}$, respectively. (The symbol $\rfloor$ denotes the interior product.)  The object $\baseeta{_a_b_c_d}\equiv\e{_d}\rfloor\baseeta{_a_b_c}$ is the Levi-Civita tensor density. The connection $1$-form defined in Refs.~\cite{PhysRevD.73.124017,Obukhov_2009,PhysRevD.80.064043} is exactly the one in Eq.~(\ref{21052019a}).

The Lagrangian density for the TEGR in the MAG approach can be written as
\begin{equation}
V=-\frac{\eta}{2\kappa}\torsion{^a^b^c}\potential{_a_b_c}=-\frac{\eta}{2\kappa}\torsion{},\label{18052019ae}
\end{equation}
where $\kappa=1/(2k)$. In this approach, one usually works with the translational gauge field momentum $2$-form and the canonical energy-momentum $3$-form:
\begin{equation}
H_a=\frac{1}{\kappa}\potential{_a_b_c}\baseeta{^b^c},\label{21052019b}
\end{equation}
\begin{equation}
E_a=\energy{^b_a}\baseeta{_b}. \label{21052019cc}
\end{equation}

\subsection{Pure-tetrad formulation}
Since the quantities (2.1)-(2.2) and (2.6)-(2.7) in Ref.~\cite{PhysRevD.73.124017} are written in the TF, we have 
\begin{equation}
\widetilde{H}_a=\frac{1}{\kappa}\potentiale{_a_b_c}\baseeta{^b^c},\label{20052019c}
\end{equation}
\begin{equation}
\widetilde{V}=-\frac{\eta}{2\kappa}\torsione{}.\label{24052019d}
\end{equation}
The same goes for $F^{\alpha}$ and other quantities there with a ``tilde''.

An interesting result shown  by Obukhov and Rubilar \cite{PhysRevD.73.124017} is that the main pure-tetrad objects depend only on the Levi-Civita  connection, which we denote by $\rsconnection{^\alpha_\beta_\gamma}$. Translating Eqs.~(2.27), (2.28), and (2.31) there into components, we get
\begin{equation}
\potentiale{_a_b_c}=\frac{1}{2}\rsconnection{_c_a_b}+\rsconnection{^d_d_{[c}}\tensor{\eta}{_{b]a}},\label{20052019fc}
\end{equation}
\begin{equation}
\widetilde{V}=\frac{\eta}{2\kappa}\left(\rsconnection{^a_a_c}\rsconnection{^b_b^c}-\rsconnection{^a_b_c}\rsconnection{^b_a^c} \right),\label{20052019g}
\end{equation}
\begin{align}
\energye{^b_a}=&\ \frac{1}{\kappa}\biggl(2\rsconnection{^c_{[ad]}}\rsconnection{^b_c^d}-2\rsconnection{^b_{[ad]}}\rsconnection{^c_c^d}
\nonumber\\
&-\rsconnection{^c_c_a}\rsconnection{^d_d^b}+\delta^{b}_{a}\rsconnection{^c_{[c|f}}\rsconnection{^d_{|d]}^f}\biggr), \label{20052019hc}
\end{align} 
respectively. Note that, from Eq.~(\ref{20052019fc}), we can recast (\ref{30122019d}) as
\begin{align}
M_{\bf e}^{ab}=-4ke\left(\rsconnection{^{[ba]}_c}\e{^c^0}+\rsconnection{^d_d^{[b}}\e{^{a]}^0} \right). \label{30122019bc}
\end{align}

Equations~(2.16)-(2.18)  can be easily translated into
\begin{equation}
\torsione{}=\torsionbarbar{}+\frac{2}{e}\pd{_\mu}\left[e\lorentz{_b^a}(\pd{_\nu}\lorentz{^b_c})\ebar{_a^\nu}\ebar{^c^\mu}\right], \label{23052019a}
\end{equation}
\begin{align}
\energye{^b_a}=\lorentz{^b_c}\lorentz{_a^d}\energybarbar{^c_d}+\frac{2}{\kappa}\lorentz{^b_c}(\pd{_\mu}\lorentz{_a^d})\potentialbarbar{_d^c^\mu}
\nonumber\\
-\frac{4}{\kappa e}\e{^b_\mu}\pd{_\nu}\left(e\lorentz{_a^c}\pifactorbar{_c^\nu^\mu}\right),\label{23052019b}
\end{align}
\begin{equation}
\potentiale{_a_b_c}=\lorentz{_a^d}\lorentz{_b^f}\lorentz{_c^g}\left( \potentialbarbar{_d_f_g}+2\pifactorbar{_d_f_g}\right), \label{23052019d}
\end{equation}
where
\begin{align}
\pifactorbar{_a^\mu^\nu}=\ebar{^b ^\mu}\ebar{^{c \nu}}\pifactorbar{_a_b_c},\ \widetilde{V}(\tetabar{})=-k\eta\torsionbarbar{},
\nonumber\\
 \widetilde{V}(\teta{})=-k\eta\torsione{} \label{23052019c}
\end{align}
and we have changed the prime notation used there: $\teta{^\prime^a}\to\teta{^a}$ and $\teta{^a}\to\tetabar{^a}$. 

From the perspective of the pure-tetrad formalism, Eqs.~(\ref{23052019a})-(\ref{23052019d}) correspond to the transformations of the torsion scalar, the gravitational energy-momentum tensor, and the superpotential under the change $\teta{^a}=\lorentz{^a_b}\tetabar{^b}$. On the other hand, from the viewpoint of the theory with an affine connection, this is just a TT.

Using Eqs.~(\ref{11012020a}) and (\ref{12012020a}), we find that $\pifactorbar{_a_b_c}=-\lorentz{^f_a}\lorentz{^r_b}\lorentz{^g_c}\pifactorebar{_f_r_g}$. Thus, we can recast Eq.~(\ref{23052019d}) in the form
\begin{align} 
\potentiale{_a_b_c}=\potentialebar{_a_b_c}-2\pifactorebar{_a_b_c}, \label{13012020f}
\end{align}
where
\begin{equation}
\pifactorebar{_a_b_c}=\frac{1}{4}\left(\scebar{_b_a_c}+\eta_{ac}\scebar{^d_d_b}-\eta_{ab}\scebar{^d_d_c} \right).\label{21052019d}
\end{equation}
 This identity will be very important because it relates the superpotentials of the two main interpretations for the gravitational energy that will be presented in this paper.

A remark is in order here regarding global transformations. To avoid the undesirable derivatives of $\lorentz{}$ in Eqs.~(\ref{23052019a})-(\ref{23052019d}), we could consider only  global $SO(3,1)$ transformations. However, the ``problem'' is that the set of physical acceptable frames is not limited to frames which are  related to each other by global transformations, and it is impossible to have two different frames related by a LLT that are simultaneously TFs. Think, for example, of the frame of an observer that is in free fall and the frame of a static observer, both in Schwarzschild spacetime. We just cannot have a unified description of them using only global transformations.   Therefore, if we want to be able to describe gravity in more than one global set (see definition \ref{03042020a}), we have to use LLTs.

\subsection{Affine connection ambiguity}\label{11032020a}
The connection for the TEGR is completely undetermined in the MAG frame work\footnote{In the pure-tetrad approach this ambiguity is hidden in the choice of the TF.} \cite{PhysRevD.73.124017}. As a matter of fact, this is true for any approach to the TEGR that considers  an affine connection. Furthermore, this ambiguity is equivalent to the ambiguity in the choice of the TF, as will be demonstrated below.

In terms of components, the transformations  (for more details about these transformations, see section III A of Ref.\cite{PhysRevD.73.124017})
\begin{equation}
\eta^\prime_{ab}=\eta_{ab},\ \teta{^\prime^a}=\teta{^a},\ \connection{^\prime_b^a}=\connection{_b^a}+\tensor{\Psi}{_b^a}, \label{27052019a}
\end{equation}
 become
\begin{equation}
\sconnection{^\prime^a_b_c}=\sconnection{^a_b_c}+\tensor{\Psi}{_c_b^a}, \label{20052019i}
\end{equation}
where $\tensor{\Psi}{_b^a}=\tensor{\Psi}{_b_c^a}\teta{^c}$. The above connection coefficients are written  in the frame $\e{_a}$. To know how the above transformation affects the connection coefficients in the frame $\ebar{_a}$, which is related to $\e{_a}$ through $\ebar{_a}=\lorentz{^b_a}\e{_b}$, we can use Eq.~(\ref{02082019a}). From this equation we see that $\scbar{^\prime^a_b_c}=\lorentz{_d^a}\ebar{_b^\mu}\pd{_\mu}\lorentz{^d_c}+\lorentz{_d^a}\lorentz{^f_b}\lorentz{^g_c}\sconnection{^\prime^d_f_g}$, which combined with Eq.~(\ref{20052019i}) gives  $\scbar{^\prime^a_b_c}=\scbar{^a_b_c}+\lorentz{_d^a}\lorentz{^f_b}\lorentz{^g_c}\tensor{\Psi}{_g_f^d}$. Since, in teleparallelism, there always exists a TF, we can assume (without loss of generality) that
$\nablab_{\bar{b}}\ebar{_a}=0$ $(\scbar{^c_b_a}=0)$
where $\bar{b}\equiv\ebar{_b}$. In this case, we have
\begin{equation}
\scbar{^\prime^a_b_c}=\lorentz{_d^a}\lorentz{^f_b}\lorentz{^g_c}\tensor{\Psi}{_g_f^d}. \label{20052019n}
\end{equation}
In other words, in the new affine connection $\nablab^\prime$ the frame $\ebar{_a}$ is no longer a TF: $\nablab^\prime_{\bar{b}}\ebar{_c}=\lorentz{_d^a}\lorentz{^f_b}\lorentz{^g_c}\tensor{\Psi}{_g_f^d}\ebar{_a}.$
By definition, this new affine connection is still a connection in a geometry with teleparallelism. Hence, there must exist a third frame  $\e[\widehat]{_a}$, related to $\ebar{_a}$ by
$\ebar{_b}=\lorentz[\widehat]{^a_b}\e[\widehat]{_a}$, such that $\nablab^\prime_{\widehat{b}}\e[\widehat]{_a}=0$. From (\ref{25082017e}), we find that $\sconnection[\widehat]{^\prime^a_b_c}=\lorentz[\widehat]{^a_d}\lorentz[\widehat]{_b^f}\ebar{_f^\mu}\pd{_\mu}\lorentz[\widehat]{_c^d}+\lorentz[\widehat]{^a_d}\lorentz[\widehat]{_b^f}\lorentz[\widehat]{_c^g}\scbar{^\prime^d_f_g}=0$, which yields $\scbar{^\prime^a_b_c}=\lorentz[\widehat]{_d^a}\lorentz[\widehat]{^f_b}\e[\widehat]{_f^\mu}\pd{_\mu}\lorentz[\widehat]{^d_c}$. Combining this  expression with (\ref{20052019n}),  we see that $\tensor{\Psi}{_g_f^d}=\lorentz{^d_a}\lorentz{_f^b}\lorentz{_g^c}\lorentz[\widehat]{_h^a}\ebar{_b^\mu}\pd{_\mu}\lorentz[\widehat]{^h_c}$. Therefore, the transformations given by Eq.~(\ref{27052019a}) are equivalent to a change in the choice of the TF, and the affine connection is undetermined because the TF is undetermined.

\subsection{MAG and pure-tetrad objects}
In this section we show the relation between MAG and tetrad objects in terms of their components and from the perspective of the TF.

Using Eqs.~(\ref{21052019a}), (\ref{21052019b}), (\ref{20052019c}) in Eq.~(3.18) of Ref.~\cite{PhysRevD.73.124017}, one can easily check that Eq. (3.18)  is equivalent to Eq.~(\ref{13012020f}). Note that $H_{{\bf\bar{e}}a}=H_{a}$ and $H_{{\bf e}a}=\widetilde{H}_{a}$. In turn, using Eq.~(\ref{21052019a}), (\ref{21052019b}), and (\ref{21052019cc}) in Eq.~(3.19) there, we obtain
\begin{align}
\energyebar{^\lambda^a}=\energye{^\lambda^a}+\frac{8k}{e}\pd{_\mu}\left( e\pifactorebar{^a^\lambda^\mu}\right)+4k\scebar{^a_\mu_b}\potentialebar{^b^\lambda^\mu}. \label{21052019e}
\end{align}
Using the same procedure for Eq.~(3.21), we find that it is equivalent to
\begin{equation}
\torsionebar{}=\torsione{}+2\rcd{_\mu}\left(\e{^b^\mu}\scebar{^a_a_b} \right).\label{21052019f}
\end{equation}

It is clear that the relation between MAG and pure-tetrad objects can be seen as being the same as writing the relation between the same components of an object when different frames are taken as the TF. In other words, Eqs.~(3.18), (3.19) and (3.21) of Ref.~\cite{PhysRevD.73.124017}  is a result of an ATT (see definition \ref{27052019f}).

\subsection{Regularization}
When calculating the energy-momentum of the spacetime we may obtain a divergent result, even in the cases where the spacetime is asymptotically flat. Nonetheless, since the affine connection is not unique (the choice of the TF is arbitrary), we can regularize the result. 

In Sec. IV of Ref.~\cite{PhysRevD.73.124017}, the regularization is applied for asymptotically flat spacetimes and works basically as follows. Let $\connectioninfinity{_a^b}$  be the values of $\connectiontilde{_a^b}$ (Levi-Civita connection) at spatial infinity. Then, to regularize we can use a relocalization of the type
\begin{equation}
\hat{V}=\widetilde{V}+d\hat{\Psi},\label{23052019e}
\end{equation} 
where $\hat{\Psi}=\frac{1}{2\kappa}\connectioninfinity{^a^b}\wedge\baseeta{_a_b}$. Of course, this procedure is not unique and is not necessary when the Levi-Civita connection $\rsconnection{^a_b_c}$ goes to zero fast enough at spatial infinity [see, e.g., Eqs.~(\ref{20052019fc})-(\ref{20052019hc})].

A more fundamental way to avoid divergences is by choosing a TF that already gives the right behavior for the Levi-Civita connection coefficient $\rsconnection{^a_b_c}$, since in this way we can calculate and interpret the results in finite regions such as the event horizons of the Schwarzschild and Ker spacetimes \cite{PhysRevD.85.044050}. To see how this approach connects to the previous one, let us use Eq.~(\ref{23052019a}). Multiplying this equation  by $-\eta/(2\kappa)$ and using  $\widetilde{V}(\tetabar{})=-\eta/(2\kappa)\torsionbarbar{}$, $\widetilde{V}(\teta{})=-\eta/(2\kappa)\torsione{}$, and\footnote{Note from Eq.~(\ref{25082017e}) that $\scebar{^a_b_c}=\e{_b^\mu}\lorentz{^a_d}\pd{_\mu}\lorentz{_c^d}$.}  $\connection{_{\bf\bar{e}}_a^b}=\scebar{^b_c_a}\teta{^c}$, we obtain
\begin{equation}
\widetilde{V}(\tetabar{})=\widetilde{V}(\teta{})-\frac{1}{2\kappa}d\left( \connection{_{\bf\bar{e}}^a^b}\wedge\baseeta{_a_b}\right).\label{23052019f}
\end{equation}
For a convenient choice of  $\ebar{_a}$ (the new TF), we can have  $\connection{_{\bf\bar{e}}_a^b}=\connectioninfinity{_a^b}$ at spatial infinity, ensuring that Eq.~(\ref{23052019f}) is equivalent to Eq.~(\ref{23052019e}) asymptotically.

The regularization (\ref{23052019e}) is easier to apply, but it is more limited and arbitrary. The procedure of finding the best frame to interpret the gravitational energy-momentum may be harder, nevertheless it can be applied to any spacetime, including nonasymptotically flat and nonstatic ones. Another advantage is that this procedure allows us to interpret the gravitational energy-momentum density as measured by a set of observers.

While the only principle guiding us for the regularization procedure is to find a finite value, which is clear an ambiguous procedure, the principle that would lead to the right frame for a given set of observers would naturally lead to a unique result for each set of observers, with a clear interpretation of the result. (A 
promising set of restrictions on the frames used to calculate RIs is proposed\footnote{The content of the present paper can also be found in this preprint; the content in Ref.~\cite{formiga2020meaning} related to the ``ideal'' frame will eventually be submitted elsewhere.} in Ref.~\cite{formiga2020meaning})

\section{Teleparallel versus non-teleparallel components }\label{24082019a}
In Ref.~\cite{Obukhov_2009}, Obukhov et. al  choose a particular TF $\e{_a}$  and arrive at a vanishing energy for gravitational waves. Then they conclude that, since the result is covariant, it remains null in all reference frames. This could be true only if we assume that the observers which use the frame $\ebar{_a}$ measure $\energybare{^a_b}$, rather than $\energybarbar{^a_b}$.  In this case, for consistency, there should be a fundamental principle to justify why the particular tetrad $\e{_a}$ should always be the TF for gravitational waves. This raises the question of whether  non-teleparallel components have any physical significance at all.

\subsection{Possible interpretations}\label{15092019a}
We have seen that the torsion tensor in teleparallel theories  is not an AI (see definition \ref{27052019c}) and that different TFs will lead to different energies and momenta. So, the teleparallel theories in the way it has been used is ambiguous. How can we solve that problem? One step towards the solution of this problem is to remove the spurious effects from the TF. 

To deal with those spurious effects, we use the following types of frames:
\begin{definition}\label{20032021a}
A frame that becomes the Lorentz frame ($\pd{_t}$,$\pd{_x}$,$\pd{_y}$,$\pd{_z}$) in the absence of gravity and satisfies the conditions
\begin{equation}
\e{_{(i)j}}=\e{_{(j)i}},\ \e{_{(i)}^0}=0\ (i,j=1,2,3), \label{01062019bc}
\end{equation}
in a particular coordinate system,  will be called the ``{\bf fundamental frame}'' (FF).
\end{definition}
\begin{definition}\label{15022020b2}
Let $e{_a}$ be the FF of a certain spacetime. The frame given by $\lorentz{_a^b}e{_b}$ (or $\lorentz{^b_a}e{_b}$) will be called an ideal frame (IF) if   the transformation $\lorentz{^b_a}$  does not include artificial rotations and depends only on the time coordinate of the coordinate system that allows us to write the FF in the form given by Eq.~(\ref{01062019bc}).
\end{definition}
(for more details about these definitions, see the discussion in Ref.~\cite{formiga2020meaning}). We will assume that an IF and, of course, the FF are free from spurious effects.

Although an IF solves the problem of spurious effects, it does not fix the TF. To be more precise, we still have the following possible interpretations: 
\begin{description}
\item[Teleparallel component (TC) interpretation]  The RIs are meaningful only when calculated in the TF and the TF is not fixed. 
\item[Non-teleparallel component (NTC) interpretation] All non-teleparallel components of the RIs are meaningful. (A fixed TF is necessary.)
\item[Fundamental frame (FF) interpretation] There is only one type of frame that possess only gravitational effects, namely, the fundamental frame. This frame must always be taken as the TF, when interpreting RIs, and some, but not all, non-teleparallel components may be meaningful.
\end{description}

Now we discuss these interpretations in detail.

\subsubsection{TC interpretation}
This interpretation seems to be the most common in the literature. It is basically the essence of the pure-tetrad formalism. 

In this interpretation, there is no fixed TF: if we want to calculate the components of a RI in an ideal accelerated frame, then we take this frame as the TF. (Note that, unlike what has  been   done in the pure-tetrad formalism, here, we use only IFs.)

The meaning of torsion in the TC interpretation is two-fold: in the absence of gravity, it is attached to the nonclosure of infinitesimal parallelograms due to accelerations and rotations only (see the next paragraph); in the presence of a gravitational field, it can measure both gravity and acceleration. (The association of torsion with the nonclosure of parallelograms in the context of the Pound-Rebka experiment was discussed by Maluf et al.  in Ref.~\cite{PhysRevD.80.044036}.)

In the absence of gravity, we have the Minkowski spacetime and the FF  is simply $\ebar{^a_\mu}=\delta^a_\mu$. 
 Thus, a general IF in Minkowski will be given by $\e{^a_\mu}=\lorentz{^a_b}(t)\delta^b_\mu$. (See definition \ref{20032021a}.) By substituting this expression into Eq.~(\ref{04102019p}), we obtain $\torsione{^a_\mu_\nu}=\left(d\lorentz{^a_b}/dt \right)\left( \delta^0_\mu\delta^b_\nu-\delta^0_\nu\delta^b_\mu \right)$, which clearly does not vanish in general. Although the torsion is not zero, it is possible to show that both $M^{ab}$ and $\energy{^\mu ^a}$ are; therefore, the TC interpretation gives consistent results in Minkowski if we use suitable frames. (This includes the proper reference frame of an observer \cite{formiga2020meaning}.)

With respect to the Minkowski spacetime, an interesting point should be emphasized here. Unlike an Euclidean space, where you can use a holonomic frame even in a rocket, in Minkowski, an accelerated  frame is necessarily  anholonomic, and this property is not just a matter of choice; it is due to the physical interaction which produces the acceleration. Therefore, the association of torsion with the nonclosure of infinitesimal parallelograms due to acceleration should not be seen as meaningless: the acceleration affects the proper coordinates and these coordinates have physical meaning.


\subsubsection{NTC interpretation}
This does not seem to be a good interpretation. To see why, let us rewrite Eq.~(\ref{29032019f}) for the case $\G{^\mu^\nu}=0$ (no matter field):
\begin{align}
\pd{_\alpha}\left( e\potentialebar{^a^\mu^\alpha}\right)=\frac{e}{4k}\psenergyebar{^\mu^a}
\nonumber\\
\psenergyebar{^\mu^a}=\energyebar{^\mu^a}-4k\scebar{^a_\alpha_b}\potentialebar{^b^\mu^\alpha}. \label{05012020a}
\end{align}
It is clear in these expressions that we are assuming that $\ebar{_a}$  is the TF while  $\e{_a}$  is not. So,
 $\scebar{^a_\alpha_b}$ is not zero, in general. Since this connection is not always zero, the integral  $\int_Vd^3x e\energyebar{^0^a}$ cannot always be transformed into a surface integral.

If $\e{_a}$ is an IF and $\ebar{_a}$ is the FF, then 
\begin{align}
\scebar{^a_\alpha_b}=\lorentz{^a_d}\frac{d\lorentz{_b^d}}{dt}\delta^0_\alpha, \label{28042020a}
\end{align}
which leads to $\energyebar{^0^a}=\psenergyebar{^0^a}$; this means that $\int_Vd^3x e\energyebar{^0^a}$ can be turned into a surface integral for IFs. However, the problem persists for the conservation of $\energyebar{^\mu^a}$:  in an IF, the conservation equation $\pd{_\mu}\left(e \psenergy{^\mu^a}\right)=0$ yields  $\pd{_\mu}\left(e\energyebar{^\mu^a}\right)=4k\frac{d\lorentz{^a_d}}{dt}\pd{_j}\left( e\potentialbarbar{^d^0^j}\right)$, and the equation for the gravitational energy-momentum flux becomes  $\frac{d}{dt}\int_Vd^3x e\energyebar{^0^a}=-\oint_SdS_j e \energyebar{^j^a}+\frac{d\lorentz{^a_d}}{dt}\bar{P}_{\bar{e}}^d$. (Note that the last term does not vanish if the frame is accelerated relatively to the FF.)  
  
In any case, if the non-teleparallel component $\torsionebar{^a_b_c}$ has a meaning, then it is natural to assume that they measure gravitational effects only, combined with the Lorentz transformations, of course.

\subsubsection{FF interpretation}
In this interpretation we set the FF as the TF, as in the NTC interpretation. However, we do not insist that all  non-teleparallel components of  RIs are meaningful. Furthermore, since $\energyebar{^\mu^a}$ does not yield a consistent gravitational energy-momentum flux, we will assume that the gravitational energy-momentum density in the FF interpretation is given by $\psenergyebar{^\mu^a}$. (We need this assumption to be able to calculate the gravitational energy in more general frames, otherwise we will be limited to the FF only.).

The reader might think that $\psenergyebar{^\mu^a}$ is not a good choice because of the affine connection coefficients appearing in its definition, see Eq.~(\ref{29032019g}). Nevertheless, those coefficients, i.e., $\sconnection{^a_\alpha_b}$, transform as tensor components under both a general coordinate transformation and global $SO(3,1)$. So, the calculations will not depend on the coordinate system. In addition, $\psenergyebar{^\mu^a}$ is also trace free, conserved, $\pd{_\mu}\left(e\psenergyebar{^\mu^a} \right)=0$ in the absence of matter, and vanishes in Minkowski spacetime. The latter property can be inferred from the fact that in Minkowski both $\potentialbarbar{^a^b^c}$ and $\energybarbar{^\mu^a}$ vanishes (note that this implies $\potentialebar{^a^b^c}=\energyebar{^\mu^a}=0$) if $\ebar{_a^\mu}$ is the FF.

We define the gravitational energy-momentum in the FF interpretation as
\begin{align}
P_\epsilon^a\equiv\int_V d^3x e\psenergyebar{^0^a},  \label{26022020a}
\end{align}
while the total energy of the spacetime will be defined as
\begin{equation}
{\cal P}^a\equiv P_\epsilon^a+\int_V d^3x eT^{0a}, \label{26022020b}
\end{equation} 
Equation (\ref{29032019k}) can be used to recast ${\cal P}^a$ in the form
\begin{equation}
{\cal P}^a=4k\oint_S dS_i e\potentialebar{^a^0^i}.\label{22072018gb2}
\end{equation}

The fact that $\potentialebar{^a^0^i}$ appears in the integrand of Eq.~(\ref{22072018gb2}) shows that, if the FF interpretation is right, then this non-teleparallel component will have an important role in the calculation of energy.

Since $\potentialebar{^a^0^i}=\lorentz{^a_b}\potentialbarbar{^b^0^i}$, we see from Eqs.~(\ref{22072018gb2}) and (\ref{22072018g}) that ${\cal P}^a=\lorentz{^a_b}(t)\bar{P}^b_{\bar{e}}$; in turn, since   $\psenergyebar{^0^a}=\energyebar{^0^a}=\lorentz{^a_b}\energybarbar{^0^b}$, we see  from Eqs~(\ref{26022020a}) and the first integral in Eq.~(\ref{10082019a})  that  $P_\epsilon^a=\lorentz{^a_b}(t)\bar{P}^b_{\bar{e},\textrm{grav}}$, where  $\bar{P}^b_{\bar{e},\textrm{grav}}$ is the gravitational energy-momentum tensor in the FF.

\subsubsection{Comparing the interpretations}
Here we show that, using only IFs, the three interpretations discussed above yield the same predictions for the gravitational angular momentum density and the gravitational energy-momentum tensor.

From Eq.~(\ref{28042020a}), we find $\scebar{^0^a^\alpha}=\left(d\lorentz{_b^\bzero}/dt \right)\ebar{_{(0)}^0}\e{^b^\alpha}\e{^a^0}$, $\e{^a^\alpha}\scebar{^b_b^0}=-\left( d\lorentz{_b^\bzero}/dt\right)\ebar{_{(0)}^0}\e{^b^0}\e{^a^\alpha}$, and $-\e{^a^0}\scebar{^b_b^\alpha}=-\left( d\lorentz{_e^\bzero}/dt\right)\ebar{_{(0)}^0}\e{^e^\alpha}\e{^a^0}$, where we have used the second constraint in Eq.~(\ref{01062019bc}), and the definition $\scebar{^0^a^\alpha}\equiv \e{_b^0}\e{^c^\alpha} \scebar{^b^a_c}$ and so on. Substituting these expressions into Eq.~(\ref{21052019d}), we get $\pifactorebar{^a^0^\alpha}=-(1/4)\left( d\lorentz{^b^\bzero}/dt\right)\lorentz{_b^\bzero}\left( \ebar{_{(0)}^0}\right)^2\e{^a^\alpha}$; by differentiating $\lorentz{^b^\bzero}\lorentz{_b^\bzero}=\eta^{\bzero\bzero}$, one can easily show that $\pifactorebar{^a^0^\alpha}$ vanishes. Using this result in Eqs.~(\ref{13012020f}) and (\ref{29032019n}), we arrive at $\potentiale{^a^0^\alpha}=\potentialebar{^a^0^\alpha}$ and $\psenergyebar{^0^a}=\energye{^0^a}$; therefore, the FF and the TC interpretations predict the same gravitational (and also  spacetime) energy-momentum tensor. It is natural to assume that the angular momentum density in the FF interpretation is given by Eq.~(\ref{30122019d}) with the superpotential $\potentialebar{}$; hence, these interpretations will also give the same result for $M^{ab}$. Note that, since the only difference between the NTC and the FF interpretations is that all non-teleparallel components have a meaning in the NTC interpretation, then, once we assume that $M^{ab}$ in the FF interpretation is calculated with $\potentialebar{}$, a non-teleparallel component, these two interpretation will predict the same angular momentum density. Besides, we already know that they predict the same energy-momentum tensor because $\psenergyebar{^0^a}=\energyebar{^0^a}$.

Unfortunately, it has not been possible to find a unique interpretation. But, as long as the TC and the FF interpretations yield consistent predictions, there is nothing to worry about. If these interpretations are equivalent, then the meaning of torsion in teleparallel theories is ambiguous: {\it acceleration plus gravitational effects for TC interpretation} or {``only'' gravitational effects for FF interpretation}. 

\section{Worked example}\label{03072021a}
As an example of how the TC and the FF interpretation work, we analyze the case of a plane-fronted $+$ polarized gravitational wave traveling along a $z$ axis. First we review some of the properties of the fundamental frame. Then we evaluate and analyze the energy-momentum tensor in a frame that accelerates along the $z$ axis and has the form 
$\e{_a}=\lorentz{_a^b}(t,z)\ebar{_b}$, where $\ebar{_b}$ is the FF.  (We let $\lorentz{_a^b}$ depend on $z$ in order to exhibit the problems that may arise.)
 
\subsection{The fundamental frame}
The metric of the $pp$-wave spacetimes for a $+$ polarized gravitational wave can be written as 
\begin{equation}
ds^2=dt^2-f(u)^2dx^2-g(u)^2dy^2-dz^2, \label{30062021a}
\end{equation}
where $u=t-z$, and $f$ and $g$ are functions that satisfy the equation $-(1/f)(d^2f/du^2)=(1/g)(d^2g/du^2)$. The fundamental frame (coframe) of this spacetime is
\begin{equation}
\tetabar{^a}=(dt,f(u)dx,g(u)dy,dz),\label{02062019d2}
\end{equation}
where $f$ and $g$ must be constant in the absence of gravity \cite{formiga2020meaning}. It turns out that this frame is a freely ``falling'' frame. 

The tensor $\energybarbar{^a^\mu}$ was calculated for the first time in Ref.~\cite{doi:10.1002/andp.201800320}; it corresponds to Eq.~(25) there. The value of $\bar{P}_{\bf \bar{e}}^a$ has not been given explicitly yet, but it will be given here implicitly latter on.

\subsection{Accelerated observers}\label{20210602a}

Now we calculate $P^a$   for observers that are accelerated along the $z$ direction. 

To obtain the accelerated frame 
 $\e{_a}$ from the relation  $\e{_a}=\lorentz{_a^b}(t,z)\ebar{_b}$, where $\ebar{_b}$ is the frame associated with the coframe (\ref{02062019d2}), we use
\begin{align}
\lorentz{_a^b}=(\gamma\hat{t}_a-\alpha\hat{z}_a)\hat{t}^b+(\alpha\hat{t}_a-\gamma\hat{z}_a)\hat{z}^b-\hat{x}_a\hat{x}^b-\hat{y}_a\hat{y}^b, \label{13012019a}
\end{align}
where $\gamma=\gamma(t,z)$ is the Lorentz factor and $\alpha=\alpha(t,z)=\pm\sqrt{\gamma^2-1}$; we are also using  the definition $\hat{t}^a=\delta^a_{(0)}$, $\hat{x}^a=\delta^a_{(1)}$, $\hat{y}^a=\delta^a_{(2)}$, $\hat{z}^a=\delta^a_{(3)}$, where the indices are lowered with $\eta_{ab}$. From Eq.~(\ref{02062019d2}), we see that we can write the frame $\ebar{_a}$ in the form $\ebar{_b^\mu}=\hat{t}_b\delta_0^\mu-(1/f)\hat{x}_b\delta^\mu_1-(1/g)\hat{y}_b\delta^\mu_2-\hat{z}_b\delta^\mu_3$. Thus, from $\e{_a}=\lorentz{_a^b}\ebar{_b}$, we find that
\begin{align}
\e{_a^\mu}=\left(\gamma\hat{t}_a-\alpha\hat{z}_a\right)\delta^\mu_0-\frac{\hat{x}_a}{f}\delta^\mu_1-\frac{\hat{y}_a}{g}\delta^\mu_2+\left( \alpha\hat{t}_a-\gamma\hat{z}_a\right)\delta^\mu_3. \label{04032020a}
\end{align}

Let us first calculate the gravitational energy-momentum tensor for the FF interpretation. In this case, we need to know $\potentialebar{^a^b^c}$.  To evaluate this tensor, we can use $\potentialbarbar{^a^b^c}$ given in Eq.~(23) of Ref.~\cite{doi:10.1002/andp.201800320}. Writing that equation in terms of $\hat{t}^a$, $\hat{x}^a$, $\hat{y}^a$, and $\hat{z}^a$, we get
\begin{align}
\potentialbarbar{^a^b^c}=&\ \left( \hat{t}^{[b|}+\hat{z}^{[b|}\right)\left[(\ln g)'\hat{x}^a\hat{x}^{|c]}+(\ln f)'\hat{y}^a\hat{y}^{|c]} \right]
\nonumber\\
&+\left[\ln(fg) \right]'\left(\hat{t}^a+\hat{z}^a \right)\hat{t}^{[b}\hat{z}^{c]}, \label{08032020a}
\end{align}
where the prime denotes the derivative with respect to $u$. From Eq.~(\ref{13012019a}) and $\potentialebar{^a^b^c}=\lorentz{^a_d}\lorentz{^b_e}\lorentz{^c_f}\potentialbarbar{^d^e^f}$, we find that
\begin{align}
\potentialebar{^a^b^c}=&\ (\gamma-\alpha)\Bigl[[\ln (fg) ]'(\hat{t}^a+\hat{z}^a)\hat{t}^{[b}\hat{z}^{c]}+(\ln g)'(\hat{t}^{[b}\hat{x}^{c]}
\nonumber\\
&-\hat{x}^{[b}\hat{z}^{c]})\hat{x}^a+(\ln f)'(\hat{t}^{[b}\hat{y}^{c]}-\hat{y}^{[b}\hat{z}^{c]})\hat{y}^a \Bigr]. \label{08012020a}
\end{align}

Now we use the definitions $\hat{t}^\lambda\equiv \e{_a^\lambda}\hat{t}^a$,  $\hat{x}^\lambda\equiv \e{_a^\lambda}\hat{x}^a$, $\hat{y}^\lambda\equiv \e{_a^\lambda}\hat{y}^a$, and $\hat{z}^\lambda\equiv \e{_a^\lambda}\hat{z}^a$ to evaluate $\potentialebar{^a^0^\nu}$. (For more details about the use of these definitions, see Ref.~\cite{formiga2020meaning}.) From Eq.~(\ref{04032020a}), we get
\begin{align}
\hat{t}^\lambda=\gamma\delta^\lambda_0+\alpha\delta^\lambda_3,\ \hat{x}^\lambda=\frac{\delta^\lambda_1}{f},\ \hat{y}^\lambda=\frac{\delta^\lambda_2}{g},
\nonumber\\ 
\hat{z}^\lambda=\alpha\delta^\lambda_0+\gamma\delta^\lambda_3. \label{08012020b}
\end{align}
It is clear that 
\begin{align}
\hat{t}^{[0}\hat{x}^{c]}=\frac{1}{2}\gamma\hat{x}^c,\ \hat{t}^{[0}\hat{y}^{c]}=\frac{1}{2}\gamma\hat{y}^c,\ \hat{x}^{[0}\hat{z}^{c]}=-\frac{1}{2}\alpha\hat{x}^c,
\nonumber\\
\hat{y}^{[0}\hat{z}^{c]}=-\frac{1}{2}\alpha\hat{y}^c,\ \hat{t}^{[0}\hat{z}^{c]}=\frac{1}{2}\left(\gamma\hat{z}^c-\alpha\hat{t}^c\right). \label{09032020a}
\end{align}
Using these expressions and Eq.~(\ref{08012020a}), we find that
\begin{align}
\potentialebar{^a^0^c}=&\ \frac{(\gamma-\alpha)}{2}\left[\ln\left(fg\right) \right]'\left(\hat{t}^a+\hat{z}^a \right)\left(\gamma\hat{z}^c-\alpha\hat{t}^c \right)
\nonumber\\
&+\frac{1}{2}\left( \ln g\right)'\hat{x}^a\hat{x}^c+\frac{1}{2}\left(\ln f \right)'\hat{y}^a\hat{y}^c. \label{09032020b}
\end{align} 
By changing $c$ to $\nu$, using Eq.~(\ref{08012020b}), and multiplying the result by $e=f(u)g(u)$, we obtain
\begin{align}
e\potentialebar{^a^0^\nu}=\frac{(\gamma-\alpha)}{2}\left(fg\right)'(\hat{t}^a+\hat{z}^a)\delta^\nu_3+\frac{1}{2}(g'\hat{x}^a\delta^\nu_1+f'\hat{y}^a\delta^\nu_2). \label{08012020c}
\end{align}

To calculate the energy-momentum of the spacetime, consider the rectangular parallelepiped  $x_-\leq x\leq x_+$,  $y_-\leq y\leq y_+$, and  $z_< \leq z\leq z_>$, where $x_\pm=x_0\pm a/2$, $y_\pm=y_0\pm b/2$,  $z_>=z_0+l/2$, and $z_<=z_0-l/2$. To calculate $P_\epsilon^a$ inside this region, we use  Eqs.~(\ref{08012020c}) and (\ref{22072018gb2}). (Note that, in our case, $P^a_\epsilon={\cal P}^a$.) The result is
\begin{align}
P_\epsilon^a=&2kab\sqrt{\frac{1-\beta(t,z_>)}{1+\beta(t,z_>)}}\left[\left(fg\right)'(z_>)-\left(fg\right)'(z_<) \right]
\nonumber\\
&\times (\hat{t}^a+\hat{z}^a),\label{10012020b}
\end{align}
where $\alpha=\beta\gamma$ ($\beta$ is the velocity of the observers with respect to the frame $\ebar{_a}$), and we have assumed that $\gamma(t,z_>)=\gamma(t,z_<)$ and $\alpha(t,z_>)=\alpha(t,z_<)$. (See the motivation for this assumption in section VI C of Ref~\cite{formiga2020meaning}.)

As expected from a field that propagates at the speed of light, we have $P_\epsilon^aP_{\epsilon a}=0$. It is also interesting to note that, for observers moving in the same direction as the wave, the magnitudes of $P_\epsilon^{(0)}$ and $P_\epsilon^{(3)}$ decrease as  the observers' speed increases and, in the limit $\beta\to 1$, we have $P_\epsilon^a=0$. On the other hand, when the observers move in the opposite direction, we get an increase of the magnitudes of the energy and the momentum; for this case, $P_\epsilon^a$ diverges as $\beta\to -1$. Note that this is the same behavior as that of  the frequency of a light wave in Minkowski: if $f_0$ is the frequency of the wave in the rest frame of the source and $f$ the frequency measured by a detector with relative velocity $\beta$, then the relativistic Doppler effect is $f=f_0\sqrt{(1-\beta)/(1+\beta)}$.

The question whether the energy is positive or negative will not be discussed here. Instead, we show that Eq.~(\ref{10012020b}) is consistent with Eq.~(25) in Ref.~\cite{doi:10.1002/andp.201800320}. 

By assuming that $l$ is small, one can verify  that $\left(fg\right)'(z_>)-\left(fg\right)'(z_<)\approx-l d^2(fg)/du^2|_{u_0}$, where $u_0=t-z_0$. The next step is to realize that, since $f''/f+g''/g=0$, we have $d^2(fg)/du^2=2f'g'$. Hence, for $\beta=0$, Eq.~(\ref{10012020b}) becomes  $P_\epsilon^a \approx -4kabl f'(u_0)g'(u_0)(\hat{t}^a+\hat{z}^a)$. Now, assuming that $a$ and $b$ are also small and using the proper volume $V\approx abl e(u_0)=ablf(u_0)g(u_0)$, we arrive at $P_\epsilon^a/V\approx -4k(\hat{t}^a+\hat{z}^a)f'g'/(fg)$.

\subsubsection{TC interpretation}
Let's now consider the calculation of the gravitational energy-momentum tensor in the context of the TC interpretation. We begin by calculating $\pifactorebar{_a_b_c}$.

From Eq.~(\ref{13012019a}) we find that $\partial_\mu\lorentz{_c^d}=\left[\hat{t}_c\partial_\mu\gamma-\hat{z}_c\partial_\mu\alpha \right]\hat{t}^d+\left[\hat{t}_c\partial_\mu\alpha-\hat{z}_c\partial_\mu\gamma \right]\hat{z}^d$. Using this expression in $\scebar{^a_\mu_c}=\lorentz{^a_d}\pd{_\mu}\lorentz{_c^d}$ [this can be inferred from Eq.~(\ref{09042020a})], we obtain
\begin{align}
\scebar{^a_\mu_c}=&\ \left(\gamma\hat{t}^a-\alpha\hat{z}^a\right)\left(\hat{t}_c\partial_\mu\gamma-\hat{z}_c\partial_\mu\alpha\right)
\nonumber\\
&-\left(\alpha\hat{t}^a-\gamma\hat{z}^a\right)\left(\hat{t}_c\partial_\mu\alpha-\hat{z}_c\partial_\mu\gamma\right). \label{13012020b}
\end{align}
By using Eq.~(\ref{04032020a}) and the identities $\partial_\mu\gamma=\delta^0_\mu\partial_t\gamma+\delta^3_\mu\partial_z\gamma$ and $\partial_\mu\alpha=\delta^0_\mu\partial_t\alpha+\delta^3_\mu\partial_z\alpha$, we see that $\e{_b^\mu}\partial_\mu\gamma=(\gamma\hat{t}_b-\alpha\hat{z}_b)\partial_t\gamma+(\alpha\hat{t}_b-\gamma\hat{z}_b)\partial_z\gamma$ and $\e{_b^\mu}\partial_\mu\alpha=(\gamma\hat{t}_b-\alpha\hat{z}_b)\partial_t\alpha+(\alpha\hat{t}_b-\gamma\hat{z}_b)\partial_z\alpha$. Using these expressions and the relation $\gamma\partial_\mu\gamma=\alpha\partial_\mu\alpha$ in Eq.~(\ref{13012020b}) we find, after a lengthy calculation,  that
\begin{align}
\scebar{^a_b_c}=2\Biggl[-\left(\partial_t\alpha+\frac{\alpha}{\gamma}\partial_z\alpha \right)\hat{t}_b+\left(\frac{\alpha}{\gamma}\partial_t\alpha+\partial_z\alpha \right)\hat{z}_b \Biggr]\hat{t}^{[a}\hat{z}_{c]}. \label{13012020bb}
\end{align}
Contracting  $a$ with $b$, we find $ \scebar{^a_a_c}=[(\alpha/\gamma)\partial_t\alpha+\partial_z\alpha ]\hat{t}_c-[\partial_t\alpha+(\alpha/\gamma)\partial_z\alpha]\hat{z}_c$. Substituting this and Eq.~(\ref{13012020bb}) into Eq.~(\ref{21052019d})  yields
\begin{align}
\pifactorebar{_a_b_c}=&-\frac{1}{2}\Bigl[ (\partial_t\gamma+\partial_z\alpha)(\hat{x}_a\hat{t}_{[b}\hat{x}_{c]}+\hat{y}_a\hat{t}_{[b}\hat{y}_{c]})
\nonumber\\
&+(\partial_t\alpha+\partial_z\gamma)(\hat{x}_a\hat{x}_{[b}\hat{z}_{c]}+\hat{y}_a\hat{y}_{[b}\hat{z}_{c]})\Bigr]. \label{13012020e}
\end{align}

Finally, from this equation and Eq.~(\ref{08012020b}), we obtain $-2\pifactorebar{^a^0^\nu}=(\partial_z\alpha)/(2\gamma)\left(\delta^\nu_1\hat{x}^a/f+\delta^\nu_2\hat{y}^a/g\right)$. Since this expression does not depend on $x$ and  $y$, its integral over the boundary of the parallelepiped vanishes.  Thus, although  $\potentiale{^a^0^\nu}$ and $\potentialebar{^a^0^\nu}$ are different,  they yield the same gravitational energy-momentum tensor.  But, like the case of Eqs.~(122) and (143) of Ref.~\cite{formiga2020meaning}, the  FF interpretation has provided a more convenient result: there is no ``strange'' term that disappears only after integration. This term vanishes for $\alpha=\alpha(t)$ anyway, i.e., it vanishes when the accelerated frame is an IF (remember the definition \ref{15022020b2}). In short, both interpretations yield the same prediction for the energy and momentum.

\subsection{Angular momentum density}
For concreteness, let's evaluate the gravitational angular momentum densities measured by the static and the accelerated observers. We begin with the static ones, i.e, we calculate $\bar{M}^{ab}_{\bar{e}}$ first. 

From Eqs.~(\ref{08032020a}) and (\ref{09032020a}) (with $\gamma=1$ and $\alpha=0$), we find that $ 2\potentialbarbar{^{[a|}^0^{|b]}}= \left[\ln\left(fg\right)\right]'\hat{t}^{[a}\hat{z}^{b]}$. Substituting into Eq.~(\ref{30122019d}) and using $e=fg$, we obtain
\begin{align}
 \bar{M}^{ab}_{\bar{e}}=-4k\left( fg\right)'\hat{t}^{[a}\hat{z}^{b]}. \label{14032020a}
\end{align}

To calculate the angular momentum for the accelerated observers, we have to decide which interpretation  will be used. Let us first use the TC interpretation. In this interpretation, we use Eq.~(\ref{30122019d}) with the superpotential $\potentiale{^a^b^c}$. Since we have already evaluated $\potentialebar{^a^0^c }$ and $\pifactorebar{^a^b^c}$, we use the identity (\ref{13012020f})  to rewrite  Eq.~(\ref{30122019d}) in the form $M_e^{ab}=M_{\bar{e}}^{ab}+16ke\pifactorebar{^{[a|}^0^{|b]}}$, where $M_{\bar{e}}^{ab}$ is written in terms of $\potentialebar{^a^0^b}$. By using Eqs.~(\ref{13012020e}) and (\ref{09032020a}), it is easy to show that $\pifactorebar{^{[a|}^0^{|b]}}=0$. So, we have $M_e^{ab}=M_{\bar{e}}^{ab}$. Using Eq.~(\ref{09032020b}) to calculate $M_{\bar{e}}^{ab}$, one finds that $M_{\bar{e}}^{ab}$ gives the same result as $\bar{M}^{ab}_{\bar{e}}$. Therefore, the accelerated observers ``measure'' the same gravitational angular momentum density. This is probably a consequence of the fact that the observers' trajectory is in the same direction as that of the wave. (Note that we have only a nonvanishing center of mass moment; the angular momentum $L^{(i)(j)}$ vanishes.)

In the case of the FF interpretation, the situation is more complicated: this interpretation says nothing about the angular momentum density because there is no surface term. However, it seems natural to assume that this density is given by $M_{\bar{e}}^{ab}$. If this is the case, the FF interpretation also gives the same result as $\bar{M}^{ab}_{\bar{e}}$.

Since Eq.~(25) of Ref.~\cite{doi:10.1002/andp.201800320} reproduces the energy-momentum tensor of gravitational waves that is usually defined in the literature for GR (see Ref.~\cite{doi:10.1002/andp.201800320}), the reader might be tempted to see if Eq.~(\ref{14032020a}) can also reproduce the angular momentum density [see, e.g., Eq.~(2.50) of Ref.~\cite{MicheleMaggiore1}]. However, this does not seem to be possible   because the definition used in Ref.~\cite{MicheleMaggiore1} is not local. The fact that this definition is not local  can be seen by noticing that the first term in Eq.~(2.50) is a type of orbital angular momentum. On the other hand, $M^{ab}$ is clearly defined at a point $x^\alpha$, which seems to be incompatible with 
the idea that it represents the angular momentum density around $x_{\cal O}^\alpha=(0,0,0,0)$; perhaps we could interpret $M^{ab}$ as being associated to the angular momentum density around $x^\alpha$ (local definition). Of course, after integration, we obtain  $L^{(i)(j)}$  and  interpret it as the total angular momentum around the ``center of mass''.

\subsection{Energy Density }
Let us  prove that, if $\e{_a^\mu}$  is an IF, then the energy density does not vanish along the curve of any observer that accelerates along the $z$ direction. 

We can write Eq.~(25) in Ref.~\cite{doi:10.1002/andp.201800320} as $\energybarbar{^\mu^b}=\rho_g\left(\delta^\mu_0+\delta^\mu_3\right)\left(\hat{t}^b+\hat{z}^b\right)$, where $\rho_g=-4kf'g'/(fg)$. Using Eq.~(\ref{13012019a}) in $\energyebar{^\mu ^a}=\lorentz{^a_b}\energybarbar{^\mu^b}$, we find that
\begin{align}
\energyebar{^\mu^a}=(\gamma-\alpha)\rho_g\left(\delta^\mu_0+\delta^\mu_3\right)\left(\hat{t}^a+\hat{z}^a\right). \label{18032020a}
\end{align}
From Eqs.~(\ref{13012020b}), (\ref{08012020a}), and (\ref{08012020b}), we obtain
\begin{align}
\scebar{^a_\alpha_b}\potentialebar{^b^\mu^\alpha}=&\frac{(\gamma-\alpha)}{2\gamma}\left[\ln\left(fg\right) \right]'\left(\hat{t}^a+\hat{z}^a \right)
\nonumber\\
&\times\left(\pd{_z}\alpha\delta^\mu_0-\pd{_t}\alpha\delta^\mu_3 \right). \label{18032020b}
\end{align}
Finally, by substituting Eqs.~(\ref{18032020a}) and (\ref{18032020b}) into Eq.~(\ref{05012020a}), we arrive at
\begin{align}
\psenergyebar{^\mu^a}=&\ \sqrt{\frac{1-\beta}{1+\beta}}\left(\hat{t}^a+\hat{z}^a \right)\Bigr[\rho_g\left(\delta^\mu_0+\delta^\mu_3 \right) 
\nonumber\\
& -\frac{2k\left[\ln(fg)\right]'}{1-\beta^2}\left(\pd{_z}\beta\delta^\mu_0-\pd{_t}\beta\delta^\mu_3 \right)\Bigl], \label{18032020c}
\end{align}
where we have used $\alpha=\beta\gamma$ and $\gamma=1/\sqrt{1-\beta^2}$.

As we can see from Eq.~(\ref{18032020c}), the energy density  $\psenergyebar{^0^{(0)}}$ does not vanish if $\beta=\beta(t)$. Furthermore, since $\pifactorebar{^a^0^\nu}=0$ in this case [see, e.g., the paragraph below Eq.~(\ref{13012020e})], then $\energye{^0^{(0)}}$ equals $\psenergyebar{^0^{(0)}}$ [see Eq.~(\ref{29032019n})]; therefore, the TC interpretation is also consistent.

On the other hand, if we let $\beta$ depend on $z$ as well, we can find a set of congruences where the energy density vanishes. It is straightforward to verify that, for
\begin{align}
\beta(t,z)=\frac{\xi(t,z)-1}{\xi(t,z)+1},\quad \xi=\chi^2(t)\left(\frac{dfg}{du} \right)^2,
\end{align}
where $\chi$ is an arbitrary function of $t$, the energy density is zero. This result may be interpreted in the following manner: different observers in the neighborhood of $z$ have different velocities with respect to the FF; they ``measure'' the gravitational energy in such a synchronized way that, in an instant $t$,  some of them measure positive energy, others negative energy, and the net effect is a zero energy density. (This is a clear example of why we should not use an arbitrary congruence of curves to interpret RIs.)

\section{$f(T)$ theories}\label{23032020b}
When dealing with geometric invariants, we have to distinguish between two types: those that are related to  physical invariants and those that are not. In mathematics, we can define plenty of geometric invariants: metric, torsion, nonmetricity etc. However, in physics, we have very feel. For instance, the only independent physical invariant that is used in GR (and TEGR) is the  infinitesimal interval between two events, which is associated with the local value of the speed of light. We associate the pseudo-Riemannian metric tensor to this interval  and build all other geometric invariants from it: curvature tensor, Ricci tensor, scalar tensor etc. In general, we search for physical invariants, associate them to some geometric object, and then build the Lagrangian density of the theory from them. But physics is not limited to invariants: energy, momentum, temperature, and heat are just a few examples. However, we do not use quantities that are not invariant in the Lagrangian density because we want the field equations to reflect this invariance.

Since the Lagrangian density of the TEGR differs from that of GR only by a surface term, the resultant field equations are invariant under the same transformation group. Furthermore, the TEGR does not have any additional physical invariant.  Therefore, in this approach, the new geometric entity, the torsion tensor, gives us only an additional substructure to help us describe the gravitational field at the level of quantities that are not invariant under LLTs. In other words, the Weitzenb\"{o}ck torsion is not a new geometric entity associated to a new physical invariant (under LLTs), unlike in the Einstein-Cartan theory, where torsion is related to the spin density.

 We could just use the Einstein-Hilbert action to obtain Einstein field equations by taking variations with respect to $g_{\mu\nu}$. Then, after obtaining the field equations,  we would use the identity (\ref{29032019f}) to write these equations in the form given by Eq.~(\ref{29032019k}). We could even write the Einstein tensor in terms of the nonmetricity tensor and see if this new tensor could helps us to improve the description of gravity, without altering the invariance of the field equations under LLTs.

Now, when we consider a $f(T)$ and take variations with respect to the tetrad field, the resultant field equations are not necessarily invariant under LLTs \cite{PhysRevD.83.064035,PhysRevD.83.104030}, which means that the field equations might yield different solutions for different choices of the TF. It seems that there are only two ways to overcome this difficulty: If the field equations of a particular  $f(T)$ yield the same metric as a solution for all IFs, i.e, the lack of local Lorentz invariance happens only with ``artificial frames'', then we may take any IF as a TF and use it to solve the field equations; this means that, in principle, the TC interpretation could be applied to this theory. On the other hand, if the field equations of a $f(T)$  do not give the same spacetime for different IFs, then we should use only the FF  to solve the field equations.

It is worth noting that the TEGR, i.e, $f(T)=T$, is not the only teleparallel theory with local Lorentz invariance, because we can always add the Weitzenb\"{o}ck connection to any spacetime. For instance, if we want to build a teleparallel theory with high order derivatives and local Lorentz invariance, we can take any $f(\R{})$ and add to it the Weitzenb\"{o}ck connection. By doing so, we can substitute the identities (\ref{29032019b}) and (\ref{29032019d}) in the field equations of the $f(\R{})$ theory and study the role played by the Weitzenb\"{o}ck torsion in this theory. (The definition of the gravitational energy-momentum tensor of the $f(\R{})$ theories is still an open question.)

In the $f(T)$ theories, the ``main'' geometry is the Weitzenb\"{o}ck version of the Riemann-Cartan geometry. Perhaps, the best approach to teleparallel theories is to avoid using this geometry to construct the  action of the theory. Instead, one could construct a Lagrangian density based on a geometry with the Levi-Civita connection, such as $f(\R{})$ theories, obtain the field equations by varying the action with respect to the metric, then, only after that, use the Weitzenb\"{o}ck connection to build a substructure with the help of identities such as Eqs.~(\ref{29032019b}) and (\ref{29032019d}). In this approach,  the teleparallel structure would server only to extract some important information about the gravitational field.

A natural and interesting extension of both Einstein-Cartan and teleparallel theories would be a theory with two affine connections with torsions: the torsion of the ``Einstein-Cartan connection'' would account for the spin, while the Weitzenb\"{o}ck torsion would allow us to identify the gravitational energy.

\section{Discussion and conclusions}
In this article we have used a notation that not only clarifies the relation between the pure-tetrad formulation of teleparallel theories and the MAG approach, but also clarifies the teleparallel frame problem. In this context, we have shown that the torsion tensor of the teleparallel theories cannot be invariant under an arbitrary change of the teleparallel frame. (It is not a universal quantity, unless one fixes the TF.) We have also proved that the ambiguity of the Weitzenb\"{o}ck affine connection in the MAG, or any other theory with local Lorentz invariance, is equivalent to the ambiguity in the choice of the TF (see Sec.~\ref{11032020a}).

In Sec.~\ref{24082019a} we have discussed two possible interpretations for teleparallel theories: the TC and the FF interpretations. We have argued that there are no other interpretations that will not lead to contradictions and ambiguities. We have also argued that, in the TC interpretation, the torsion tensor is measuring both gravitational effects and the observers' acceleration. ( In Minkowski, of course, it measures only the nonclosure of infinitesimal parallelograms made with the observer's  proper coordinates, which is caused by the acceleration.) On the other hand, we have argued that, in the FF interpretation, torsion is measuring only the gravitational effects. (These gravitational effects, however, can be changed by the accelerations of the observers in  much the same way that the energy of an electron changes when we change the acceleration of the frame.) In both interpretations, the torsion tensor may become meaningless in frames with artificial  properties (not related to the observer's motion).

We have shown that the TC and the FF yield the same predictions for the energy and momentum when using the IFs. As an example, we have analyzed the gravitational energy-momentum tensor of a $+$ polarized gravitational wave in a frame that accelerates along the same direction as that of the wave. The results were all consistent with our expectations.

It is worth noting that the ambiguity of the TF disappears in the non-local gravity  approach adopted by Mashhoon \cite{Mashhoonbook2017}. It would be interesting to see if the analysis made here could be extended to this theory.

\appendix

\chapter{Applying the hybrid machinery to  Eqs.~(\ref{15122019l})-(\ref{15122019n})}\label{03092019f}
In this appendix we give the torsion tensor, the superpotential, the torsion scalar, and the quantity $\potential{^\lambda^\mu^\nu}\torsion{_\lambda_\mu_\alpha}$ in terms of the tetrad given by Eq.~(\ref{14122019e}) with $\{\hat{t}_\lambda,\hat{s}_\lambda,\hat{\phi}_\lambda,\hat{z}_\lambda\}$ given by  Eqs.(\ref{15122019l})-(\ref{15122019n}). 

By inverting Eqs.~(\ref{15122019l})-(\ref{15122019n}), we get
\begin{align}
\delta_\lambda^0=\frac{1}{A}(f\hat{t}_\lambda-g\hat{z}_\lambda), \label{15122019q}
\\
\delta_\lambda^1=\frac{1}{B}(g\cos\theta\hat{t}_\lambda-\sin\theta\hat{s}_\lambda-f\cos\theta\hat{z}_\lambda), \label{15122019r} 
\\ 
\delta_\lambda^2=\frac{1}{\rho B}(-g\sin\theta\hat{t}_\lambda-\cos\theta\hat{s}_\lambda+f\sin\theta\hat{z}_\lambda), \label{15122019s}
\\ 
\delta_\lambda^3=-\frac{\hat{\phi}_\lambda,}{\rho B\sin\theta}. \label{15122019t}
\end{align}
We can simplify the calculations by using the following definitions:
\begin{align}
h_1\equiv A\pd{_z}f+B\pd{_t}g,\ h_2\equiv h_1\cos\theta+ f\pd{_\rho}A,
\nonumber\\ 
h_3\equiv A\pd{_z}g+B\pd{_t}f,\ h_4\equiv h_3\cos\theta+g\pd{_\rho}A, 
\nonumber\\ 
h_5\equiv \frac{\pd{_\rho}A}{AB},\ h_6\equiv \frac{\pd{_\rho}B}{B^2},
\nonumber\\ 
h_7\equiv (f^2h_5-g^2h_6)\sin\theta,\ h_8\equiv \frac{h_1}{AB}+fh_5\cos\theta,
\nonumber\\ 
h_9\equiv (h_5-h_6)fg\sin\theta,\ h_{10}\equiv\frac{h_3}{AB}+gh_5\cos\theta,
\nonumber\\ 
h_{11}\equiv (f^2h_6-g^2h_5)\sin\theta,\ h_{12}\equiv h_6g\cos\theta,
\nonumber\\ 
h_{13}\equiv h_6\sin\theta,\ h_{14}\equiv h_6f\cos\theta,
\nonumber\\ 
h_{15}\equiv h_{11}+h_{13},\ h_{16}\equiv h_{10}+h_{12},
\nonumber\\ 
h_{17}\equiv h_7+h_{11},\ h_{18}\equiv h_8+h_{14},
\nonumber\\ 
h_{19}\equiv h_7+h_{13},\ f\partial_{\mu}f=g\partial_\mu g. \label{15122019p}
\end{align}

Substituting Eqs.~(\ref{15122019l})-(\ref{15122019n}) and (\ref{15122019t}) into Eqs.~(\ref{15122019b})-(\ref{15122019e}), we obtain
\begin{align}
\torsion{^\bzero_\mu_\nu}=&-2h_2\delta^0_{[\mu}\delta^1_{\nu]}+2h_1\rho\sin\theta\delta^0_{[\mu}\delta^2_{\nu]}
\nonumber\\
&+2g\rho(\pd{_\rho}B)\sin\theta\delta^1_{[\mu}\delta^2_{\nu]}, \label{27122019a}
\\
\torsion{^\bbs_\mu_\nu}=&-2\rho(\pd{_\rho}B)\cos\theta\delta^1_{[\mu}\delta^2_{\nu]}, \label{15122019u}
\\
\torsion{^\bbphi_\mu_\nu}=&-2\pd{_\rho}(\rho B)\sin\theta\delta^1_{[\mu}\delta^3_{\nu]}-2\rho B\cos\theta\delta^2_{[\mu}\delta^3_{\nu]}
\nonumber\\
&+2\delta^3_{[\mu}\hat{s}_{\nu]},  \label{15122019v}
\\
\torsion{^\bbz_\mu_\nu}=&-2h_4\delta^0_{[\mu}\delta^1_{\nu]}+2h_3\rho\sin\theta\delta^0_{[\mu}\delta^2_{\nu]}
\nonumber\\
&+2(\pd{_\rho}B)f\rho\sin\theta\delta^1_{[\mu}\delta^2_{\nu]}. \label{15122019x}
\end{align}
To write the torsion components in terms of $\{\hat{t}_\lambda,\hat{s}_\lambda,\hat{\phi}_\lambda,\hat{z}_\lambda\}$, we use Eqs.~(\ref{15122019q})-(\ref{15122019t}). From these equations, one can check that
\begin{align}
\delta^0_{[\mu}\delta^1_{\nu]}=&\ \frac{-1}{AB}( f\sin\theta\hat{t}_{[\mu}\hat{s}_{\nu]}+\cos\theta\hat{t}_{[\mu}\hat{z}_{\nu]}
\nonumber\\
&+g\sin\theta\hat{s}_{[\mu}\hat{z}_{\nu]}), \label{16122019a}
\\ 
\delta^0_{[\mu}\delta^2_{\nu]}=&\ \frac{1}{\rho AB}( -f\cos\theta\hat{t}_{[\mu}\hat{s}_{\nu]}+\sin\theta\hat{t}_{[\mu}\hat{z}_{\nu]}
\nonumber\\
&-g\cos\theta\hat{s}_{[\mu}\hat{z}_{\nu]}), \label{16122019b}
\\ 
\delta^1_{[\mu}\delta^2_{\nu]}=&\ \frac{-1}{\rho B^2}(g\hat{t}_{[\mu}\hat{s}_{\nu]}+f\hat{s}_{[\mu}\hat{z}_{\nu]} ), \label{16122019c}
\\ 
\delta^1_{[\mu}\delta^3_{\nu]}=&\ \frac{1}{\rho B^2\sin\theta}( -g\cos\theta\hat{t}_{[\mu}\hat{\phi}_{\nu]}+\sin\theta\hat{s}_{[\mu}\hat{\phi}_{\nu]}
\nonumber\\
&-f\cos\theta\hat{\phi}_{[\mu}\hat{z}_{\nu]}), \label{16122019d}
\\ 
\delta^2_{[\mu}\delta^3_{\nu]}=&\ \frac{1}{\rho^2 B^2\sin\theta}( g\sin\theta\hat{t}_{[\mu}\hat{\phi}_{\nu]}+\cos\theta\hat{s}_{[\mu}\hat{\phi}_{\nu]}
\nonumber\\
&+f\sin\theta\hat{\phi}_{[\mu}\hat{z}_{\nu]} ), \label{16122019e}
\\ 
\delta^3_{[\mu}\hat{s}_{\nu]}=&\ \frac{\hat{s}_{[\mu}\hat{\phi}_{\nu]}}{\rho B\sin\theta}. \label{16122019f}
\end{align}
By substituting Eqs.~(\ref{16122019a})-(\ref{16122019f}) into  Eqs.~(\ref{27122019a})-(\ref{15122019x}) and simplifying, we find that
\begin{align}
\torsion{^\bzero_\mu_\nu}=&\ 2h_7\hat{t}_{[\mu}\hat{s}_{\nu]}+2h_8\hat{t}_{[\mu}\hat{z}_{\nu]}+2h_9\hat{s}_{[\mu}\hat{z}_{\nu]}, \label{23122019c}
\\ 
\torsion{^\bbs_\mu_\nu}=&\ 2h_6\cos\theta\left(g\hat{t}_{[\mu}\hat{s}_{\nu]}+f\hat{s}_{[\mu}\hat{z}_{\nu]} \right),\label{23122019d}
\\ 
\torsion{^\bbphi_\mu_\nu}=&\ 2h_6(g\cos\theta\hat{t}_{[\mu}\hat{\phi}_{\nu]}-\sin\theta\hat{s}_{[\mu}\hat{\phi}_{\nu]}
\nonumber\\
&+f\cos\theta\hat{\phi}_{[\mu}\hat{z}_{\nu]} ), \label{23122019e}
\\ 
\torsion{^\bbz_\mu_\nu}=&\ 2h_9\hat{t}_{[\mu}\hat{s}_{\nu]}+2h_{10}\hat{t}_{[\mu}\hat{z}_{\nu]}-2h_{11}\hat{s}_{[\mu}\hat{z}_{\nu]}. \label{16122019g}
\end{align}
Substituting these equations into Eq.~(\ref{15122019a}) and manipulating the result  yield
\begin{align}
\torsion{^\lambda^\mu^\nu}=&\ 2(h_7\hat{t}^\lambda-h_{12}\hat{s}^\lambda-h_9\hat{z}^\lambda)\hat{t}^{[\mu}\hat{s}^{\nu]}
\nonumber\\
&-2h_{12}\hat{\phi}^\lambda\hat{t}^{[\mu}\hat{\phi}^{\nu]}+2(h_8\hat{t}^\lambda-h_{10}\hat{z}^\lambda)\hat{t}^{[\mu}\hat{z}^{\nu]}
\nonumber\\
&+2h_{13}\hat{\phi}^\lambda\hat{s}^{[\mu}\hat{\phi}^{\nu]}+2(h_9\hat{t}^\lambda-h_{14}\hat{s}^\lambda+h_{11}\hat{z}^\lambda)\hat{s}^{[\mu}\hat{z}^{\nu]}
\nonumber\\
&-2h_{14}\hat{\phi}^\lambda\hat{\phi}^{[\mu}\hat{z}^{\nu]}. \label{27122019d}
\end{align}
Rearranging the terns on the right-hand side of Eq.~(\ref{27122019d}),  one can verify that $ 2\torsion{^{[\mu|}^\lambda^{|\nu]}}=\torsion{^\lambda^\mu^\nu}$.

From Eqs.~(\ref{23122019c})-(\ref{16122019g}), it is possible to show that 
\begin{align}
\torsion{^\bzero _a_\nu}\hat{t}^a=h_7\hat{s}_\nu+h_8\hat{z}_\nu,
\\ 
\torsion{^\bbs_a_\nu}\hat{s}^a=h_{12}\hat{t}_\nu-h_{14}\hat{z}_\nu,
\\  
\torsion{^\bbphi_a_\nu}\hat{\phi}^a=h_{12}\hat{t}_\nu-h_{13}\hat{s}_\nu-h_{14}\hat{z}_\nu,
\\ 
\torsion{^\bbz_a_\nu}\hat{z}^a=h_{10}\hat{t}_\nu-h_{11}\hat{s}_\nu.
\end{align}
Using these equations,  Eq.~(\ref{15122019a}), and the definition $T_{\nu}=\torsion{^a_a_\nu}$, we obtain
\begin{align}
\torsion{^\nu}=&-(h_{10}+2h_{12})\hat{t}^\nu+(h_{17}+h_{13})\hat{s}^\nu
\nonumber\\
&+(h_8+2h_{14})\hat{z}^\nu.
\end{align}
From this expression and  Eq.~(\ref{14122019g}), we find that
\begin{align}
g^{\lambda[\nu}T^{\mu]}=&\ \left[-(h_{17}+h_{13})\hat{t}^\lambda+(h_{10}+2h_{12})\hat{s}^\lambda \right]\hat{t}^{[\mu}\hat{s}^{\nu]}
\nonumber\\
&+(h_{10}+2h_{12})\hat{\phi}^\lambda\hat{t}^{[\mu}\hat{\phi}^{\nu]}
\nonumber\\
&+\left[-(h_8+2h_{14})\hat{t}^\lambda+(h_{10}+2h_{12})\hat{z}^\lambda \right]\hat{t}^{[\mu}\hat{z}^{\nu]}
\nonumber\\
&-(h_{17}+h_{13})\hat{\phi}^\lambda\hat{s}^{[\mu}\hat{\phi}^{\nu]}
\nonumber\\
&+\left[ (h_8+2h_{14})\hat{s}^\lambda-(h_{17}+h_{13})\hat{z}^\lambda\right]\hat{s}^{[\mu}\hat{z}^{\nu]}
\nonumber\\
&+(h_8+2h_{14})\hat{\phi}^\lambda\hat{\phi}^{[\mu}\hat{z}^{\nu]}. \label{27122019b}
\end{align}
Finally, by substituting Eqs.~(\ref{27122019b}) and (\ref{27122019d}) into Eq.~(\ref{10112017la})  (remember that $ 2\torsion{^{[\mu|}^\lambda^{|\nu]}}=\torsion{^\lambda^\mu^\nu}$), we arrive at
\begin{align}
\potential{^\lambda^\mu^\nu}=&\ \left(-h_{15}\hat{t}^\lambda+h_{16}\hat{s}^\lambda-h_9\hat{z}^\lambda \right)\hat{t}^{[\mu}\hat{s}^{\nu]} +h_{16}\hat{\phi}^\lambda\hat{t}^{[\mu}\hat{\phi}^{\nu]}
\nonumber\\
&+2\left(-h_{14}\hat{t}^\lambda+h_{12}\hat{z}^\lambda\right)\hat{t}^{[\mu}\hat{z}^{\nu]}-h_{17}\hat{\phi}^\lambda\hat{s}^{[\mu}\hat{\phi}^{\nu]}
\nonumber\\
&+\left(h_9\hat{t}^\lambda+h_{18}\hat{s}^\lambda-h_{19}\hat{z}^\lambda \right)\hat{s}^{[\mu}\hat{z}^{\nu]}+h_{18}\hat{\phi}^\lambda\hat{\phi}^{[\mu}\hat{z}^{\nu]}. \label{27122019c}
\end{align}

To calculate the gravitational angular momentum density, we need the component $\potential{^a^0^b}$. From Eqs.~(\ref{27122019c}), (\ref{15122019p}), and Eqs.~(\ref{15122019g})-(\ref{15122019i}), we obtain
\begin{align}
\potential{^a^0^b}=&-\frac{h_6}{A}(f\hat{t}^a-g\hat{z}^a)\Bigl[\sin\theta\hat{s}^b-\cos\theta(g\hat{t}^b-f\hat{z}^b) \Bigr]
\nonumber\\
&+\frac{(\partial_zg)}{2ABf}(\hat{s}^a\hat{s}^b+\hat{\phi}^a\hat{\phi}^b), \label{27122019e}
\end{align}
where we have used the identities  $ fh_{16}-gh_{18}=(\pd{_z}g)/(fB)$, $ fh_{15}+gh_9=2fh_6\sin\theta$ and $ gh_{19}-fh_9=2gh_6\sin\theta$.

To evaluate $\potential{^\lambda^\mu^\nu}\torsion{_\lambda_\mu_\alpha}$, it is convenient to proceed in the following manner: Let $X$, $Y$, and $Z$ be elements in $\{\hat{t}^a,\hat{s},\hat{\phi},\hat{z}\}$; assume that $X$ is different from both $Y$ and $Z$. From the orthonormality condition, we see that
\begin{align}
X^{[\mu}Y^{\nu]}X_{[\mu}Z_{\alpha]}=\frac{1}{4}(X^aX_a)Y^\nu Z_\alpha+\frac{1}{4}(Y^aZ_a)X^\nu X_\alpha. \label{20122019b}
\end{align}
We can use this identity to obtain, after a lengthy calculation, the expression
\begin{align}
\potential{^\lambda^\mu^\nu}\torsion{_\lambda_\mu_\alpha}=&\ (\frac{h_9^2}{2}+\frac{h_7h_{15}}{2}+h_8h_{14}-h_{12}^2-2h_{10}h_{12})\hat{t}^\nu\hat{t}_\alpha
\nonumber\\
&+\frac{1}{2}(-2h_9^2-h_7h_{15}-h_{14}h_{18}+h_{12}h_{16}
\nonumber\\
&-h_{13}h_{17}-h_{11}h_{19} )\hat{s}^\nu\hat{s}_\alpha+\frac{1}{2}(h_{12}h_{16}
\nonumber\\
&-h_{13}h_{17}-h_{14}h_{18} )\hat{\phi}^\nu\hat{\phi}_\alpha-(\frac{h_9^2}{2}+\frac{h_{11}h_{19}}{2}
\nonumber\\
&+h_{14}^2+2h_8h_{14}-h_{10}h_{12} )\hat{z}^\nu\hat{z}_\alpha+(\frac{h_{13}h_{16}}{2}
\nonumber\\
&+h_9h_{14}+h_{11}h_{12})\hat{t}^\nu\hat{s}_\alpha+\frac{1}{2}(h_{12}h_{17}-h_8h_9
\nonumber\\
&+h_{10}h_{19})\hat{s}^\nu\hat{t}_\alpha+(h_{14}h_{16}-\frac{h_9h_{13}}{2})\hat{t}^\nu\hat{z}_\alpha
\nonumber\\
&+(h_{12}h_{18}-\frac{h_9h_{13}}{2})\hat{z}^\nu\hat{t}_\alpha-\frac{1}{2}(h_8h_{15}+h_{10}h_9
\nonumber\\
&+h_{14}h_{17})\hat{s}^\nu\hat{z}_\alpha+(-h_7h_{14}+h_9h_{12}
\nonumber\\
&-\frac{h_{13}h_{18}}{2})\hat{z}^\nu\hat{s}_\alpha  \label{20122019c}
\end{align}

Contracting  $\nu$ with $\alpha$ and using (\ref{15122019p}) to eliminate $h_{18}$ and $h_{16}$, we find that
\begin{align}
T=&\ 2h_9^2+h_7h_{15}+4h_8h_{14}+2h_{14}^2-2h_{12}^2
\nonumber\\
&-4h_{10}h_{12}+h_{13}h_{17}+h_{11}h_{19}. \label{20122019d}
\end{align}

\chapter{Proof of Eq.~(\ref{29032019a})}\label{12042020f}
From Eqs.~(\ref{17112017c}) and (\ref{10112017lc}) we obtain
\begin{align}
\contorsion{^\alpha^\nu^\mu}=-2\potential{^\nu^\mu^\alpha}+g^{\nu\alpha}\torsion{^\mu}-g^{\mu\nu}\torsion{^\alpha},\label{12042020a}\\
\contorsion{^\alpha_\alpha_\mu}=\torsion{_\mu}.  \label{12042020b}
\end{align}
Substituting Eqs.~(\ref{12042020a})-(\ref{12042020b}) into the second and third terms on the right hand side of Eq.~(\ref{05082019b}), respectively, and taking $R_{\mu\nu}=0$, we find that
\begin{align}
\R{^\mu^\nu}=2 \rcd{_\alpha}\potential{^\nu^\mu^\alpha}+g^{\mu\nu}\rcd{_\alpha}\torsion{^\alpha}+V^{\mu\nu}, \label{12042020c}
\end{align}
where $V^{\mu\nu}=-\contorsion{^\alpha_\alpha_\lambda}\contorsion{^\lambda^\nu^\mu}+\contorsion{^\alpha^\nu_\lambda}\contorsion{^\lambda_\alpha^\mu}$.

We know that 
\begin{align}
2 \rcd{_\alpha}\potential{^\nu^\mu^\alpha}=2\left(\pd{_\alpha}\potential{^\nu^\mu^\alpha}+\chr{^\nu_\alpha_\lambda}\potential{^\lambda^\mu^\alpha} +\chr{^\alpha_\alpha_\lambda}\potential{^\nu^\mu^\lambda}\right). \label{12042020d}
\end{align}
(Note that $\chr{^\mu_\alpha_\lambda}\potential{^\nu^\lambda^\alpha}=0$ because $\potential{^\nu^\lambda^\alpha}=-\potential{^\nu^\alpha^\lambda}$.) Using the identity  $\pd{_\alpha}\e{_b^\nu}+\connection{^\nu_\alpha_\lambda}\e{_b^\lambda}-\sconnection{^c_\alpha_b}\e{_c^\nu}=0$ in the identity   $\pd{_\alpha}\potential{^\nu^\mu^\alpha}=\left(\pd{_\alpha}\e{_b^\nu} \right)\potential{^b^\mu^\alpha}+\e{_b^\nu}\pd{_\alpha}\potential{^b^\mu^\alpha}$ to eliminate $\pd{_\alpha}\e{_b^\nu}$, we can write $\pd{_\alpha}\potential{^\nu^\mu^\alpha}=\e{_b^\nu}\pd{_\alpha}\potential{^b^\mu^\alpha}-\connection{^\nu_\alpha_\lambda}\potential{^\lambda^\mu^\alpha}+\sconnection{^c_\alpha_b}\e{_c^\nu}\potential{^b^\mu^\alpha}$. Thus, we can recast Eq.~(\ref{12042020d}) as
\begin{align}
2 \rcd{_\alpha}\potential{^\nu^\mu^\alpha}=&\ 2\Bigl[ \e{_b^\nu}\pd{_\alpha}\potential{^b^\mu^\alpha}-\contorsion{^\nu_\alpha_\lambda}\potential{^\lambda^\mu^\alpha}+\chr{^\alpha_\alpha_\lambda}\potential{^\nu^\mu^\lambda}
\nonumber\\
&+\sconnection{^c_\alpha_b}\e{_c^\nu}\potential{^b^\mu^\alpha}\Bigr], \label{12042020e}
\end{align}
where we have used $\chr{^\nu_\alpha_\lambda}-\connection{^\nu_\alpha_\lambda}=-\contorsion{^\nu_\alpha_\lambda}$ [see Eq.~(\ref{17112017a})].

We can write the term with the contorsion tensor in Eq.~(\ref{12042020e}) as $-\contorsion{^\nu_\alpha_\lambda}\potential{^\lambda^\mu^\alpha}=-\torsion{_a_b^\nu}\potential{^a^b^\mu}+\tilde{V}^{\mu\nu}$, where\\ $\widetilde{V}^{\mu\nu}=(1/2)\left(-\torsion{_\alpha_\lambda^\nu}+\torsion{_\lambda_\alpha^\nu}+\torsion{^\nu_\alpha_\lambda}\right)\potential{^\lambda^\alpha^\mu}$. (It is straightforward to show that $2\widetilde{V}^{\mu\nu}=-V^{\mu\nu}$.) By substituting the expression with $-\contorsion{^\nu_\alpha_\lambda}\potential{^\lambda^\mu^\alpha}$ into Eq.~(\ref{12042020e}) and then substituting the result into Eq.~(\ref{12042020c}), one can easily prove that Eq.~(\ref{29032019a}) holds.


\begin{thebibliography}{10}

\bibitem{Einstein:1915by}
Albert Einstein.
\newblock {Zur Allgemeinen Relativit\"atstheorie}.
\newblock {\em Sitzungsber. Preuss. Akad. Wiss. Berlin (Math. Phys. )},
  1915:778--786, 1915.

\bibitem{10.2307/20488488}
A. Papapetrou.
\newblock Einstein's theory of gravitation and flat space.
\newblock {\em Proc. R. Irish Acad. (Sect. A)}, 52:11--23, 1948.

\bibitem{PhysRev.89.400}
Peter G. Bergmann and Robb Thomson.
\newblock Spin and angular momentum in general relativity.
\newblock {\em Phys. Rev.}, 89:400--407, Jan 1953.

\bibitem{MOLLER1961118}
C {M{\o}ller}.
\newblock Further remarks on the localization of the energy in the general
  theory of relativity.
\newblock {\em Annals of Physics}, 12(1):118 -- 133, 1961.

\bibitem{Landaufourthv2}
L. D. Landau and E. M. Lifschitz.
\newblock {\em {The Classical Theory of Fields}}, volume 2.
\newblock Pergamon Press, Oxford, 4 edition, 1975.

\bibitem{Weinberg:1972kfs}
Steven Weinberg.
\newblock {\em {Gravitation and Cosmology}: {Principles and Applications of the
  General Theory of Relativity}}.
\newblock John Wiley and Sons, New York, 1972.

\bibitem{aldrovandi2012teleparallel}
R. Aldrovandi and J.G. Pereira.
\newblock {\em Teleparallel Gravity: An Introduction}.
\newblock Springer, London, 2012.

\bibitem{ANDP:ANDP201200272}
Jos\' e W. Maluf.
\newblock The teleparallel equivalent of general relativity.
\newblock {\em Ann. Phys. (Berlin)}, 525(5):339--357, 2013.

\bibitem{PhysRevD.19.3524}
Kenji Hayashi and Takeshi Shirafuji.
\newblock New general relativity.
\newblock {\em Phys. Rev. D}, 19:3524--3553, Jun 1979.

\bibitem{Cai_2016}
Yi-Fu Cai, Salvatore Capozziello, Mariafelicia De Laurentis, and Emmanuel N
  Saridakis.
\newblock f(t) teleparallel gravity and cosmology.
\newblock {\em Rep. Prog. Phys.}, 79(10):106901, sep 2016.

\bibitem{0264-9381-34-11-115012}
Chao-Qiang Geng and Ling-Wei Luo.
\newblock Teleparallel conformal invariant models induced by kaluza-klein
  reduction.
\newblock {\em Class. Quantum Grav.}, 34(11):115012, 2017.

\bibitem{Silva2016}
J. G. Silva, A. F. Santos, and S. C. Ulhoa.
\newblock On friedmann--robertson--walker model in conformal teleparallel
  gravity.
\newblock {\em Eur. Phys. J. C}, 76(3):167, Mar 2016.

\bibitem{PhysRevD.47.1407}
J. David Brown and James W. York.
\newblock Quasilocal energy and conserved charges derived from the
  gravitational action.
\newblock {\em Phys. Rev. D}, 47:1407--1419, Feb 1993.

\bibitem{PhysRevD.85.044050}
J. W. Maluf, S. C. Ulhoa, and J. F. da Rocha-Neto.
\newblock Gravitational pressure on event horizons and thermodynamics in the
  teleparallel framework.
\newblock {\em Phys. Rev. D}, 85:044050, Feb 2012.

\bibitem{PhysRevD.100.124040}
F. Hammad, D. Dijamco, A. Torres-Rivas, and D. B\'erub\'e.
\newblock Noether charge and black hole entropy in teleparallel gravity.
\newblock {\em Phys. Rev. D}, 100:124040, Dec 2019.

\bibitem{PhysRevD.78.047502}
J. W. Maluf and S. C. Ulhoa.
\newblock The energy-momentum of plane-fronted gravitational waves in the
  teleparallel equivalent of gr.
\newblock {\em Phys. Rev. D}, 78:047502, Aug 2008.

\bibitem{Obukhov_2009}
Yuri N Obukhov, J G Pereira, and Guillermo F Rubilar.
\newblock On the energy transported by exact plane gravitational-wave
  solutions.
\newblock {\em Class. Quantum Grav.}, 26(21):215014, oct 2009.

\bibitem{doi:10.1002/andp.201800320}
Jansen B. Formiga.
\newblock The energy-momentum tensor of gravitational waves, wyman spacetime,
  and freely falling observers.
\newblock {\em Ann. Phys. (Berlin)}, 530(12):1800320, 2018.

\bibitem{Xulu2000}
S. S. Xulu.
\newblock {Total Energy of the Bianchi Type I Universes}.
\newblock {\em {Int. J. Theor. Phys.}}, 39:1153 -- 1161, 04 2000.

\bibitem{Vargas2004}
T. Vargas.
\newblock The energy of the universe in teleparallel gravity.
\newblock {\em Gen. Rel. Grav.}, 36(6):1255--1264, Jun 2004.

\bibitem{SOUSA2010}
A. A. Sousa, J. S. Moura, and R. B. Pereira.
\newblock {Energy in an expanding universe in the teleparallel geometry}.
\newblock {\em {Braz. J. Phys}}, 40:1 -- 8, 03 2010.

\bibitem{doi:10.1142/S021827181001813X}
S. C. Ulhoa, J. F. da Rocha Neto, and J. W. Maluf.
\newblock The gravitational energy problem for cosmological models in
  teleparallel gravity.
\newblock {\em Int. J. Mod. Phys. D}, 19(12):1925--1935, 2010.

\bibitem{ABEDI201954}
Habib Abedi, Amir M. Abbassi, and Salvatore Capozziello.
\newblock Cosmological perturbations in gravitational energy-momentum
  complex.
\newblock {\em Ann. Phys.}, 405:54--68, 2019.

\bibitem{doi:10.1002/andp.201900507}
J. B. Formiga.
\newblock The gravitational energy-momentum density of radially accelerated
  observers in schwarzschild spacetime.
\newblock {\em Ann. Phys. (Berlin)}, 532(3):1900507, 2020.

\bibitem{formiga2020meaning}
J. B. Formiga.
\newblock The meaning of torsion in teleparallel theories, 2020.
\newblock arXiv: 2004.10788v2.

\bibitem{PhysRevD.14.2521}
Y. M. Cho.
\newblock Einstein lagrangian as the translational yang-mills lagrangian.
\newblock {\em Phys. Rev. D}, 14:2521--2525, Nov 1976.

\bibitem{PhysRevD.14.3335}
Y. M. Cho.
\newblock Gauge theory of poincar\'e symmetry.
\newblock {\em Phys. Rev. D}, 14:3335--3340, Dec 1976.

\bibitem{HAYASHI1977441}
Kenji Hayashi.
\newblock The gauge theory of the translation group and underlying geometry.
\newblock {\em Phys. Lett. B}, 69(4):441 -- 444, 1977.

\bibitem{PhysRevLett.84.4533}
V. C. de Andrade, L. C. T. Guillen, and J. G. Pereira.
\newblock Gravitational energy-momentum density in teleparallel gravity.
\newblock {\em Phys. Rev. Lett.}, 84:4533--4536, May 2000.

\bibitem{Arnowitt2008}
Richard Arnowitt, Stanley Deser, and Charles W. Misner.
\newblock Republication of: The dynamics of general relativity.
\newblock {\em Gen. Relativ. Gravit.}, 40:1997--2027, 2008.

\bibitem{0264-9381-9-7-009}
G Bergqvist.
\newblock Quasilocal mass for event horizons.
\newblock {\em Class. and Quantum Grav.}, 9(7):1753, 1992.

\bibitem{PhysRevLett.83.1897}
Chia-Chen Chang, James M. Nester, and Chiang-Mei Chen.
\newblock Pseudotensors and quasilocal energy-momentum.
\newblock {\em Phys. Rev. Lett.}, 83:1897--1901, Sep 1999.

\bibitem{PhysRevD.42.2045}
Friedrich W. Hehl and Wei-Tou Ni.
\newblock Inertial effects of a dirac particle.
\newblock {\em Phys. Rev. D}, 42:2045--2048, Sep 1990.

\bibitem{JansenRevistaMexicana}
J. B. Formiga.
\newblock The {F}renet-{S}erret description of {B}orn rigidity and its
  application to the {D}irac equation.
\newblock {\em Rev. Mex. Fis.}, 66(2):180, 2020.

\bibitem{PhysRevD.80.064043}
Tiago Gribl Lucas, Yuri N. Obukhov, and J. G. Pereira.
\newblock Regularizing role of teleparallelism.
\newblock {\em Phys. Rev. D}, 80:064043, Sep 2009.

\bibitem{PhysRevD.99.024022}
Rui-Hui Lin and Xiang-Hua Zhai.
\newblock New proper tetrad for teleparallel gravity in curved spacetimes.
\newblock {\em Phys. Rev. D}, 99:024022, Jan 2019.

\bibitem{0264-9381-34-14-145013}
Alexey Golovnev, Tomi Koivisto, and Marit Sandstad.
\newblock On the covariance of teleparallel gravity theories.
\newblock {\em Class. Quantum Grav.}, 34(14):145013, 2017.

\bibitem{Krssak2017}
Martin Kr{\v{s}}{\v{s}}{\'a}k.
\newblock Holographic renormalization in teleparallel gravity.
\newblock {\em The European Physical Journal C}, 77(1):44, Jan 2017.

\bibitem{Formiga2021Braz}
J. B. Formiga.
\newblock Revisiting the gravitational energy of the schwarzschild spacetime
  with a new approach to the calculations.
\newblock {\em Braz. J. Phys.}, 2021.

\bibitem{doi:10.1002/andp.201700175}
Combi Luciano and Romero Gustavo E.
\newblock Is teleparallel gravity really equivalent to general relativity?
\newblock {\em Ann. Phys.}, 530(1):1700175, 2018.

\bibitem{Gravitation}
C. W. Misner, K. S. Thorne, and J. A. Wheeler.
\newblock {\em Gravitation}.
\newblock W.H. Freeman and Company, New York, 1973.

\bibitem{doi:10.1119/1.10744}
Hans C. Ohanian.
\newblock What is the principle of equivalence?
\newblock {\em Am. J. Phys.}, 45(10):903--909, 1977.

\bibitem{MicheleMaggiore1}
Michele Maggiore.
\newblock {\em Gravitational Waves: Theory and Experiments}, volume 1.
\newblock Oxford University Press Inc, New York, 2008.

\bibitem{Maluf_2007}
J W Maluf, F F Faria, and S C Ulhoa.
\newblock On reference frames in spacetime and gravitational energy in freely
  falling frames.
\newblock {\em Class. Quantum Grav.}, 24(10):2743--2753, apr 2007.

\bibitem{Gonalves2021}
V. R. Gonalves and J. B. Formiga.
\newblock The black hole energy and the energy of the other universe.
\newblock {\em Eur. Phys. J. Plus}, 136:1063, 2021.

\bibitem{Maluf:2004vc}
Jos\' e Wadih Maluf.
\newblock {Accelerated observers and gravitational radiation}.
\newblock {\em Grav. Cosmol.}, 11:284--288, 2005.

\bibitem{Mashhoonbook2017}
Bahram Mashhoon.
\newblock {\em Nonlocal Gravity}.
\newblock Oxford University Press Inc, Oxford, 2017.

\bibitem{NORTON1985203}
John Norton.
\newblock What was {E}instein's principle of equivalence?
\newblock {\em Stud. Hist. Phil. Sci.}, 16(3):203--246, 1985.

\bibitem{Norton_1993}
J D Norton.
\newblock {\em Rep. Prog. Phys.}, 56(7):791--858, jul 1993.

\bibitem{PhysRev.130.1253}
Julian Schwinger.
\newblock Quantized gravitational field.
\newblock {\em Phys. Rev.}, 130:1253--1258, May 1963.

\bibitem{PhysRevD.65.124001}
J. W. Maluf, J. F. da Rocha-Neto, T. M. L. Tor\'{\i}bio, and K. H.
  Castello-Branco.
\newblock Energy and angular momentum of the gravitational field in the
  teleparallel geometry.
\newblock {\em Phys. Rev. D}, 65:124001, May 2002.

\bibitem{griffiths1991colliding}
J.B. Griffiths.
\newblock {\em Colliding plane waves in general relativity}.
\newblock Dover Publications, New York, 2016.

\bibitem{StephaniBook}
Hans Stephani, Dietrich Kramer, Malcolm Maccallum, Cornelius Hoenselaers, and
  Eduard Herlt.
\newblock {\em { Exact Solutions to Einstein's Field Equations}}.
\newblock Cambridge Monographs on Mathematical Physics. Cambridge University
  Press, Cambridge, 2nd edition, 2003.

\bibitem{AbrahamPais}
Abraham Pais.
\newblock {\em {Subtle is the lord}}.
\newblock Oxford University Press, New York, 2005.

\bibitem{SAUER2006399}
Tilman Sauer.
\newblock Field equations in teleparallel space-time: Einstein's
  fernparallelismus approach toward unified field theory.
\newblock {\em Historia Mathematica}, 33(4):399, 2006.
\newblock Special Issue on Geometry and its Uses in Physics, 1900-1930.

\bibitem{Gasperini}
Venzo de Sabbata and Maurizio Gasperini.
\newblock {\em {Introduction to Gravitation}}.
\newblock World Scientific, Singapore, 1985.

\bibitem{PhysRevD.85.124007}
Vincenzo F. Cardone, Ninfa Radicella, and Stefano Camera.
\newblock Accelerating $f(t)$ gravity models constrained by recent cosmological
  data.
\newblock {\em Phys. Rev. D}, 85:124007, Jun 2012.

\bibitem{PhysRevD.79.124019}
Gabriel R. Bengochea and Rafael Ferraro.
\newblock Dark torsion as the cosmic speed-up.
\newblock {\em Phys. Rev. D}, 79:124019, Jun 2009.

\bibitem{PhysRevD.75.084031}
Rafael Ferraro and Franco Fiorini.
\newblock Modified teleparallel gravity: Inflation without an inflaton.
\newblock {\em Phys. Rev. D}, 75:084031, Apr 2007.

\bibitem{Miao_2011}
Rong-Xin Miao, Miao Li, and Yan-Gang Miao.
\newblock Violation of the first law of black hole thermodynamics inf(t)
  gravity.
\newblock {\em J. Cosmol. Astropart. Phys.}, 2011(11):033--033, nov 2011.

\bibitem{ANDP:ANDP201200037}
Jos\' e W. Maluf and F F Faria.
\newblock Teleparallel gauge theory of gravity.
\newblock {\em Ann. Phys. (Berlin)}, 524(6-7):366--370, 2012.

\bibitem{Momeni2014}
Davood Momeni and Ratbay Myrzakulov.
\newblock Conformal invariant teleparallel cosmology.
\newblock {\em Eur. Phys. J. Plus}, 129(6):137, Jun 2014.

\bibitem{doi:10.1142/S0217732317501139}
J. G. da Silva and S. C. Ulhoa.
\newblock On gravitational energy in conformal teleparallel gravity.
\newblock {\em Mod. Phys. Lett. A}, 32(21):1750113, 2017.

\bibitem{PhysRevD.99.064047}
J. B. Formiga.
\newblock Conformal teleparallel theories and weyl geometry.
\newblock {\em Phys. Rev. D}, 99:064047, Mar 2019.

\bibitem{PhysRevD.99.064006}
M. Fontanini, E. Huguet, and M. Le Delliou.
\newblock Teleparallel gravity equivalent of general relativity as a gauge
  theory: Translation or cartan connection?
\newblock {\em Phys. Rev. D}, 99:064006, Mar 2019.

\bibitem{universe5060139}
Jos\' e G. Pereira and Yuri N. Obukhov.
\newblock Gauge structure of teleparallel gravity.
\newblock {\em Universe}, 5(6), 2019.

\bibitem{PhysRevD.101.024059}
M. Le Delliou, E. Huguet, and M. Fontanini.
\newblock Teleparallel theory as a gauge theory of translations: Remarks and
  issues.
\newblock {\em Phys. Rev. D}, 101:024059, Jan 2020.

\bibitem{PhysRevD.80.044036}
J. W. Maluf, S. C. Ulhoa, and F. F. Faria.
\newblock Pound-rebka experiment and torsion in the schwarzschild spacetime.
\newblock {\em Phys. Rev. D}, 80:044036, Aug 2009.

\bibitem{PhysRevD.73.124017}
Yuri N. Obukhov and Guillermo F. Rubilar.
\newblock Covariance properties and regularization of conserved currents in
  tetrad gravity.
\newblock {\em Phys. Rev. D}, 73:124017, Jun 2006.

\bibitem{Kr_k_2019}
M Kr{\v{s}}{\v{s}}{\'{a}}k, R J van den Hoogen, J G Pereira, C G B\" ohmer, and
  A A Coley.
\newblock Teleparallel theories of gravity: illuminating a fully invariant
  approach.
\newblock {\em Class. Quantum Gravit.}, 36(18):183001, aug 2019.

\bibitem{Nakahara}
Mikio Nakahara.
\newblock {\em {Geometry, topology and physics}}.
\newblock IOP Publishing Ltd, 2nd edition, 2003.

\bibitem{Maluf_2020}
J W Maluf, S C Ulhoa, J F da Rocha-Neto, and F L Carneiro.
\newblock Difficulties of teleparallel theories of gravity with local lorentz
  symmetry.
\newblock {\em Class. Quantum Gravit.}, 37(6):067003, feb 2020.

\bibitem{Rosen1994}
Nathan Rosen.
\newblock The energy of the universe.
\newblock {\em Gen. Rel. Grav.}, 26(3):319--321, Mar 1994.

\bibitem{PhysRevD.83.064035}
Baojiu Li, Thomas P. Sotiriou, and John D. Barrow.
\newblock $f(t)$ gravity and local lorentz invariance.
\newblock {\em Phys. Rev. D}, 83:064035, Mar 2011.

\bibitem{PhysRevD.83.104030}
Thomas P. Sotiriou, Baojiu Li, and John D. Barrow.
\newblock Generalizations of teleparallel gravity and local lorentz symmetry.
\newblock {\em Phys. Rev. D}, 83:104030, May 2011.

\end{thebibliography}

\end{document}